\documentclass[traditabstract]{aa} 

\usepackage{psfig}

\begin{document}

\title{Unveiling the nature of {\it INTEGRAL} objects through optical 
spectroscopy\thanks{Based on observations collected at the following 
observatories: Cerro Tololo Interamerican Observatory (Chile); 
Observatorio del Roque de los Muchachos of the Instituto de 
Astrof\'{\i}sica de Canarias (Canary Islands, Spain); ESO (La Silla, 
Chile) under programme 079.A-0171(A); Astronomical Observatory of 
Bologna in Loiano (Italy); Observatorio Astron\'omico Nacional (San Pedro 
M\'artir, M\'exico); Anglo-Australian Observatory (Siding Spring, 
Australia); Apache Point Observatory (New Mexico, USA).}}

\subtitle{VII. Identification of 20 Galactic and extragalactic hard X--ray 
sources}

\author{N. Masetti\inst{1},
P. Parisi\inst{1,2},
E. Palazzi\inst{1},
E. Jim\'enez-Bail\'on\inst{3},
L. Morelli\inst{4},
V. Chavushyan\inst{5},
E. Mason\inst{6}, \\
V.A. McBride\inst{7},
L. Bassani\inst{1},
A. Bazzano\inst{8}, 
A.J. Bird\inst{7},
A.J. Dean\inst{7},
G. Galaz\inst{9},
N. Gehrels\inst{10},
R. Landi\inst{1}, \\
A. Malizia\inst{1},
D. Minniti\inst{9,11},
F. Schiavone\inst{1},
J.B. Stephen\inst{1} and
P. Ubertini\inst{8}
}

\institute{
INAF -- Istituto di Astrofisica Spaziale e Fisica Cosmica di 
Bologna, Via Gobetti 101, I-40129 Bologna, Italy
\and
Dipartimento di Astronomia, Universit\`a di Bologna,
Via Ranzani 1, I-40127 Bologna, Italy
\and
Instituto de Astronom\'{\i}a, Universidad Nacional Aut\'onoma de M\'exico,
Apartado Postal 70-264, 04510 M\'exico D.F., M\'exico
\and
Dipartimento di Astronomia, Universit\`a di Padova,
Vicolo dell'Osservatorio 3, I-35122 Padua, Italy
\and
Instituto Nacional de Astrof\'{i}sica, \'Optica y Electr\'onica,
Apartado Postal 51-216, 72000 Puebla, M\'exico
\and
European Southern Observatory, Alonso de Cordova 3107, Vitacura,
Santiago, Chile
\and
School of Physics \& Astronomy, University of Southampton, Southampton, 
Hampshire, SO17 1BJ, United Kingdom  
\and
INAF -- Istituto di Astrofisica Spaziale e Fisica Cosmica di
Roma, Via Fosso del Cavaliere 100, I-00133 Rome, Italy
\and
Departamento de Astronom\'{i}a y Astrof\'{i}sica, Pontificia Universidad 
Cat\'olica de Chile, Casilla 306, Santiago 22, Chile
\and
NASA/Goddard Space Flight Center, Greenbelt, MD 20771, USA
\and
Specola Vaticana, V-00120 Citt\`a del Vaticano
}

\offprints{N. Masetti (\texttt{masetti@iasfbo.inaf.it)}}
\date{Received 10 November 2008; accepted 19 November 2008}

\abstract{
Within the framework of our program of assessment of the nature of 
unidentified or poorly known {\it INTEGRAL} sources, we present here 
spectroscopy of optical objects, selected through positional 
cross-correlation with soft X--ray detections (afforded with
satellites such as {\it Swift}, {\it ROSAT}, {\it Chandra} and/or 
{\it XMM-Newton}) as putative counterparts of hard X--ray sources 
detected with the IBIS instrument onboard {\it INTEGRAL}.
Using 6 telescopes of various sizes and archival data from two on-line 
spectroscopic surveys we are able to identify, either for the first time 
or independent of other groups, the nature of 20 {\it INTEGRAL} hard 
X--ray sources.
Our results indicate that: 11 of these objects are active galactic nuclei
(AGNs) at redshifts between 0.014 and 0.978, 7 of which display 
broad emission lines, 2 show narrow emission lines only, and 2 have 
unremarkable or no emission lines (thus are likely Compton thick AGNs); 
5 are cataclysmic variables (CVs), 4 of which are (possibly magnetic) 
dwarf novae and one is a symbiotic star; and 4 are Galactic X--ray 
binaries (3 with high-mass companions and one with a low-mass secondary).
It is thus again found that the majority of these sources are AGNs or 
magnetic CVs, confirming our previous findings.
When possible, the main physical parameters for these hard X--ray sources 
are also computed using the multiwavelength information available in the 
literature. These identifications support the importance of {\it 
INTEGRAL} in the study of the hard X--ray spectrum of all classes of
X--ray emitting objects, and the effectiveness of a strategy of 
multi-catalogue cross-correlation plus optical spectroscopy to securely 
pinpoint the actual nature of unidentified hard X--ray sources.}


\titlerunning{The nature of 20 Galactic and extragalactic IGR sources}
\authorrunning{N. Masetti et al.}

\maketitle

\section{Introduction}

One of the fundamental aims of the {\it INTEGRAL} mission (Winkler et al. 
2003) is the census of the whole sky in the hard X--ray band above 20 keV. 
This makes use of the unique imaging capability of the IBIS instrument 
(Ubertini et al. 2003) which allows the detection of sources at the mCrab 
level with a typical localization accuracy of less than 5 arcmin (Gros et 
al. 2003).

The IBIS surveys secured the detection of extragalactic sources in the 
so-called `Zone of Avoidance', which hampers observations in soft X--rays 
along the Galactic Plane due to the presence of gas and dust. Moreover, 
these surveys are expanding our knowledge about Galactic X--ray binaries, 
by showing the existence of a new class of heavily absorbed supergiant 
massive X--ray binaries (first suggested by Revnivtsev et al. 2003), by 
allowing the discovery and the study of supergiant fast X--ray transients 
(e.g., Sguera et al. 2006; Leyder et al. 2007; Sidoli et al. 2008), by 
doubling the number of known high-mass X--ray binaries (HMXBs; see Walter 
2007), and by detecting a substantial number of new magnetic cataclysmic 
variables (CVs; e.g., Barlow et al. 2006; Bonnet-Bidaud et al. 2007; Landi 
et al. 2008a).

Up to now, IBIS has detected more than 500 sources at hard X--rays between 
20 and 100 keV (see e.g., Bird et al. 2007; Krivonos et al. 2007; see also 
Bodaghee et al. 2007). If we consider the most complete compilation of 
IBIS sources available (Bodaghee et al. 2007; see also Bird et 
al. 2007), one can see that most of these sources are active galactic 
nuclei (AGNs, 33\% of the total number of detected objects) followed by 
known Galactic X--ray binaries (32\%) and CVs (4.5\%). However, a large 
number of the remaining objects (26\% of all IBIS detections, according to 
Bodaghee et al. 2007) has no obvious counterpart at other wavelengths and 
therefore cannot immediately be associated with any known class of 
high-energy emitting objects.  The multiwavelength study of these 
unidentified sources is thus critical to determine their nature.  

Therefore, in 2004 we started a multisite observational campaign devoted 
to the identification of these unidentified objects through optical 
spectroscopy (see Masetti et al. 2004, 2006a,b,c,d, 2008a; hereafter 
Papers I-VI). Our results showed that about half of these objects are 
nearby ($z \la$ 0.1) AGNs (Papers I-VI), while a non-negligible fraction 
($\sim$15\%) of objects belongs to the class of magnetic CVs (Paper VI and 
references therein).
The results of this identification program, along with those obtained by 
other groups, are collected as a service to the scientific community on
a web page\footnote{{\tt http://www.iasfbo.inaf.it/extras/IGR/main.html}} 
reporting information on {\it INTEGRAL} sources identified via optical or 
near-infrared (NIR) observations.

Continuing our effort to reveal the real nature of {\it INTEGRAL} 
sources, we present here optical spectroscopy of 20 objects that were 
detected by IBIS but are still unidentified, unclassified or poorly 
studied. The data reported here were obtained at 6 different telescopes 
around the world, or retrieved from two public spectroscopic archives.

A thorough description of the approach to the sample selection is reported 
in Papers I-VI, so we refer the reader to these papers for details on the 
procedure and caveats thereof. Here we simply state that all of the 
optical counterparts to the IBIS sources studied in this paper are 
positionally associated with a soft X--ray object detected with {\it 
ROSAT} (Voges et al. 1999; {\it ROSAT} Team 2000), 
{\it Swift}/XRT\footnote{XRT archival data are freely available at \\
{\tt http://www.asdc.asi.it/}}
(Malizia et al. 2007; Landi et al. 2007a,b, 2008b,c,d,e; Revnivtsev et al. 
2007; Rodriguez et al. 2008), {\it Chandra} (Jonker \& Kuiper 2007; 
Sazonov et al. 2008; Tomsick et al. 2008a,b), and/or {\it XMM-Newton} 
(Saxton et al. 2008; Watson et al. 2008; Ibarra et al. 2008a,b) within the 
corresponding {\it INTEGRAL} error circle (Bird et al. 2007; Krivonos et 
al. 2007; Sazonov et al. 2008).

We remark that this approach is highly effective, as we have already 
sucessfully identified more than 80 selected targets (see Papers 
I-VI; Masetti et al. 2007, 2008b).

For the high-energy sources with more than one 
optical candidate, we spectroscopically observed all objects with 
magnitude $R\la$ 18 within the corresponding soft X--ray arcsec-sized 
error circle. However, in the following we will report only on their firm 
or likely optical counterparts, recognized via their peculiar spectral 
features (basically, the presence of emission lines). All other candidates 
were excluded because their spectra did not show any peculiarity (in 
general they are recognized as Galactic stars) and will not be considered 
further.

We note that we included in our list of objects the hard X--ray source IGR 
J12415$-$5750, which is reported as a Seyfert 2 AGN in Bird et al. (2007)
but for which no optical spectrum is available in the literature (Winter 
et al. 2008). We also considered IGR J19267+1325, which has a soft X--ray 
source close to (but formally outside of) its 90\% IBIS error circle; it 
can nevertheless be considered the likely soft X--ray counterpart of this
{\it INTEGRAL} source on the basis of X--ray spectral comparisons 
(Tomsick et al. 2008b).

In our final sample of sources we also included the {\it INTEGRAL} 
objects mentioned by Negueruela et al. (2007), Bikmaev et al. (2008a,b), 
Revnivtsev et al. (2008) and Steeghs et al. (2008). These objects, 
although already identified elsewhere, still have fragmentary 
longer-wavelength information, or were independently studied by us before 
their identification was announced. Our observations are thus meant to 
confirm their nature and to improve their classification and known 
information.

Figures 1 and 2 report the optical finding charts of the 20 sources of the 
selected sample. The corresponding optical counterparts are indicated with 
tick marks. The list of identified {\it INTEGRAL} sources is reported in 
Table 1 (which we thoroughly describe in the next section).

The outline of the rest of the paper is the following: in Sect. 2 a 
description of the observations and of the employed telescopes is given.  
Sect. 3 reports and discusses the results, divided into three broad 
classes of sources (CVs, X-ray binaries and AGNs), together with an update 
of the statistical outline of the identifications of {\it INTEGRAL} 
sources obtained until now. Conclusions are drawn in Sect. 4. In this 
work, if not otherwise stated, errors and limits are reported at 
1$\sigma$ and 3$\sigma$ confidence levels, respectively.

\begin{figure*}[th!]
\hspace{-.1cm}
\centering{\mbox{\psfig{file=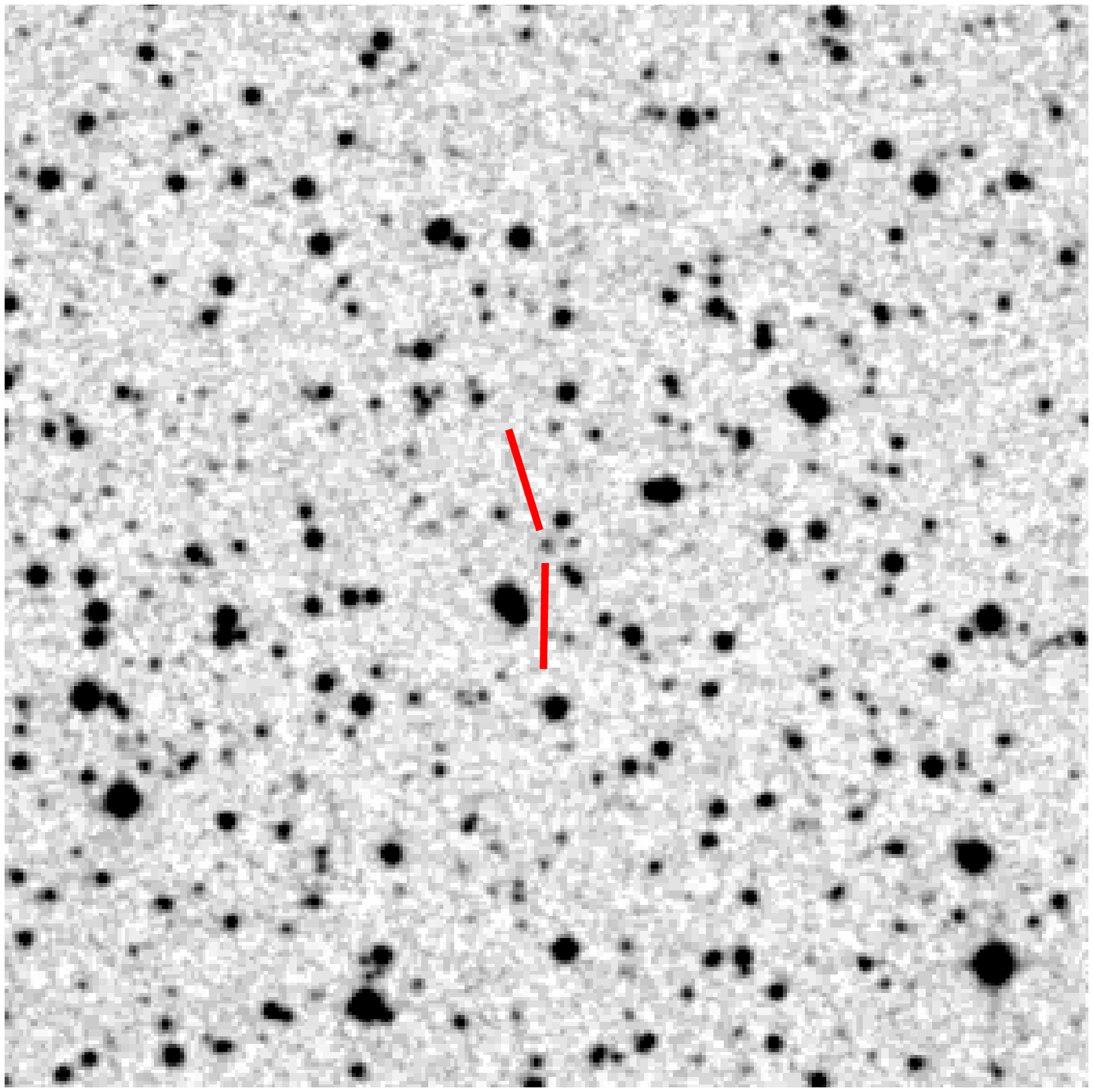,width=5.9cm}}}
\centering{\mbox{\psfig{file=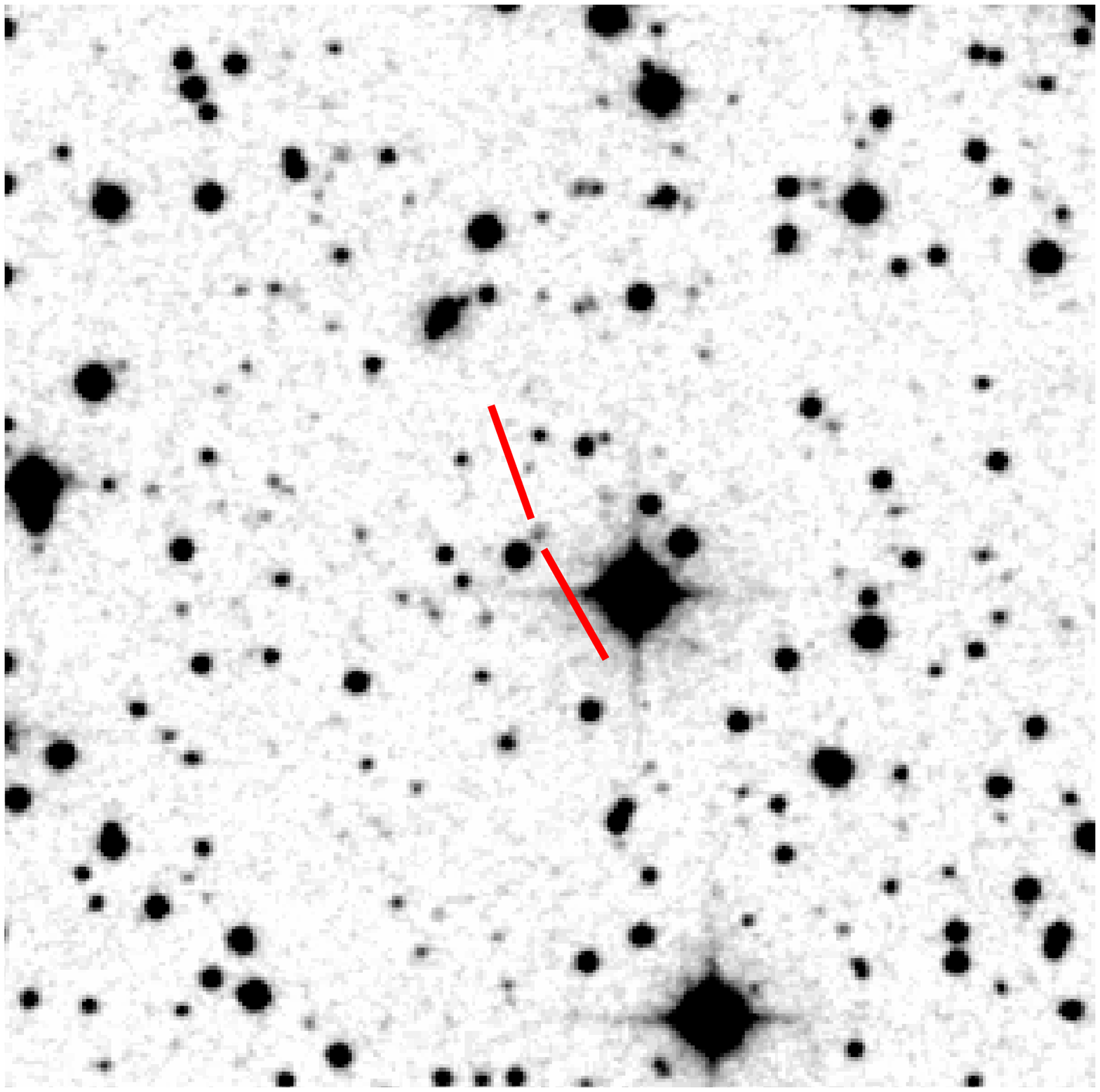,width=5.9cm}}}
\centering{\mbox{\psfig{file=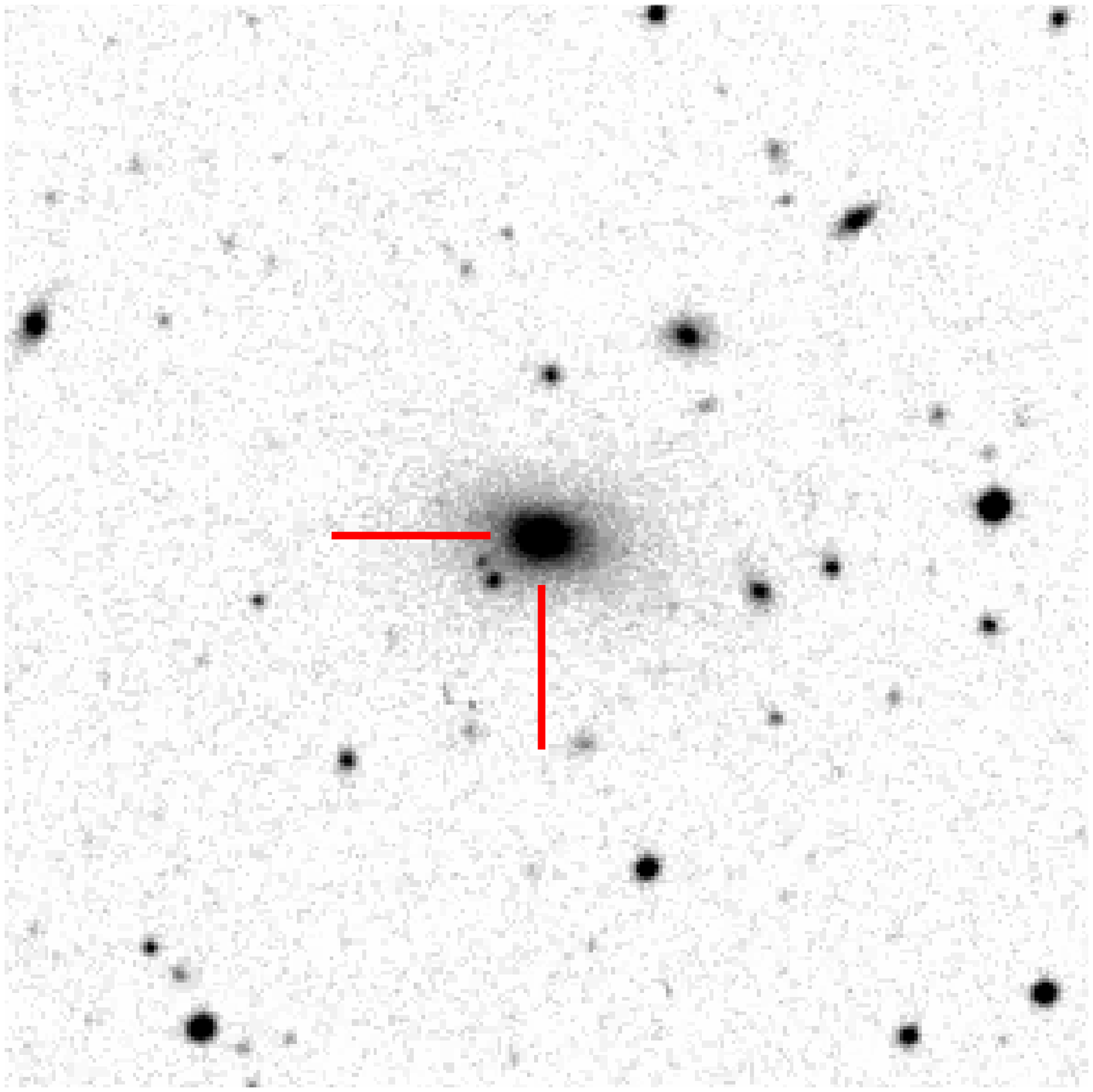,width=5.9cm}}}
\centering{\mbox{\psfig{file=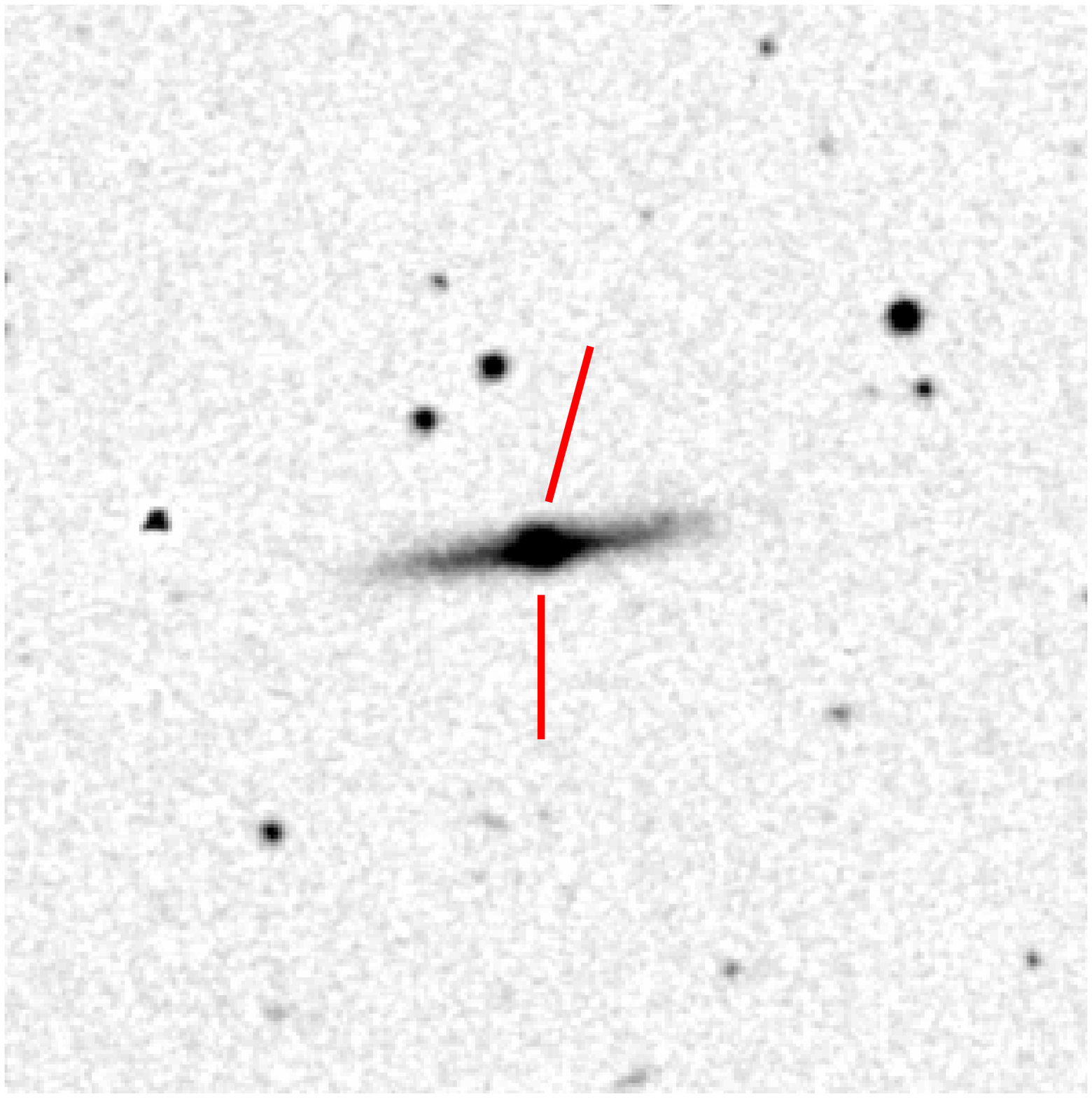,width=5.9cm}}}
\centering{\mbox{\psfig{file=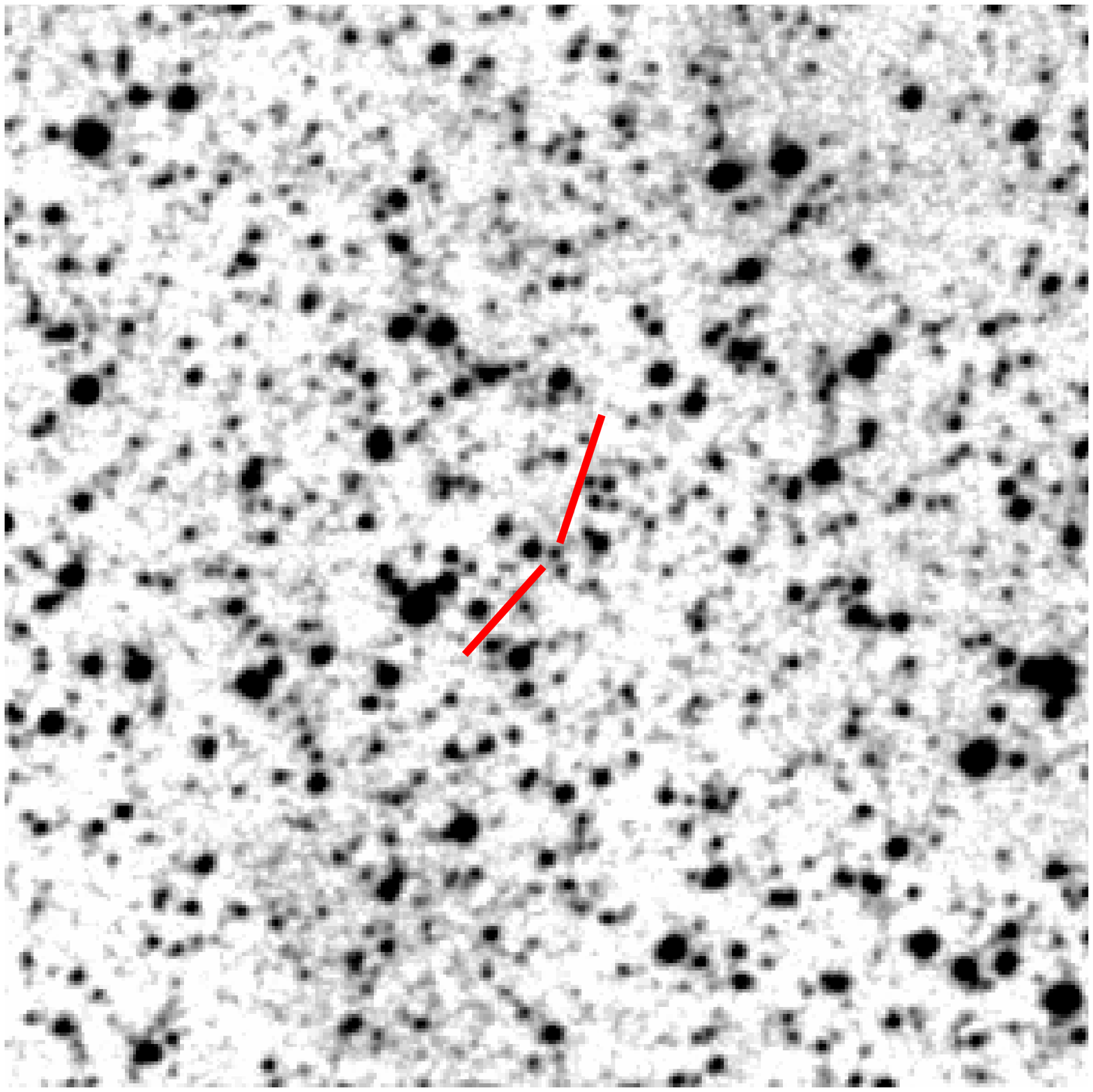,width=5.9cm}}}
\centering{\mbox{\psfig{file=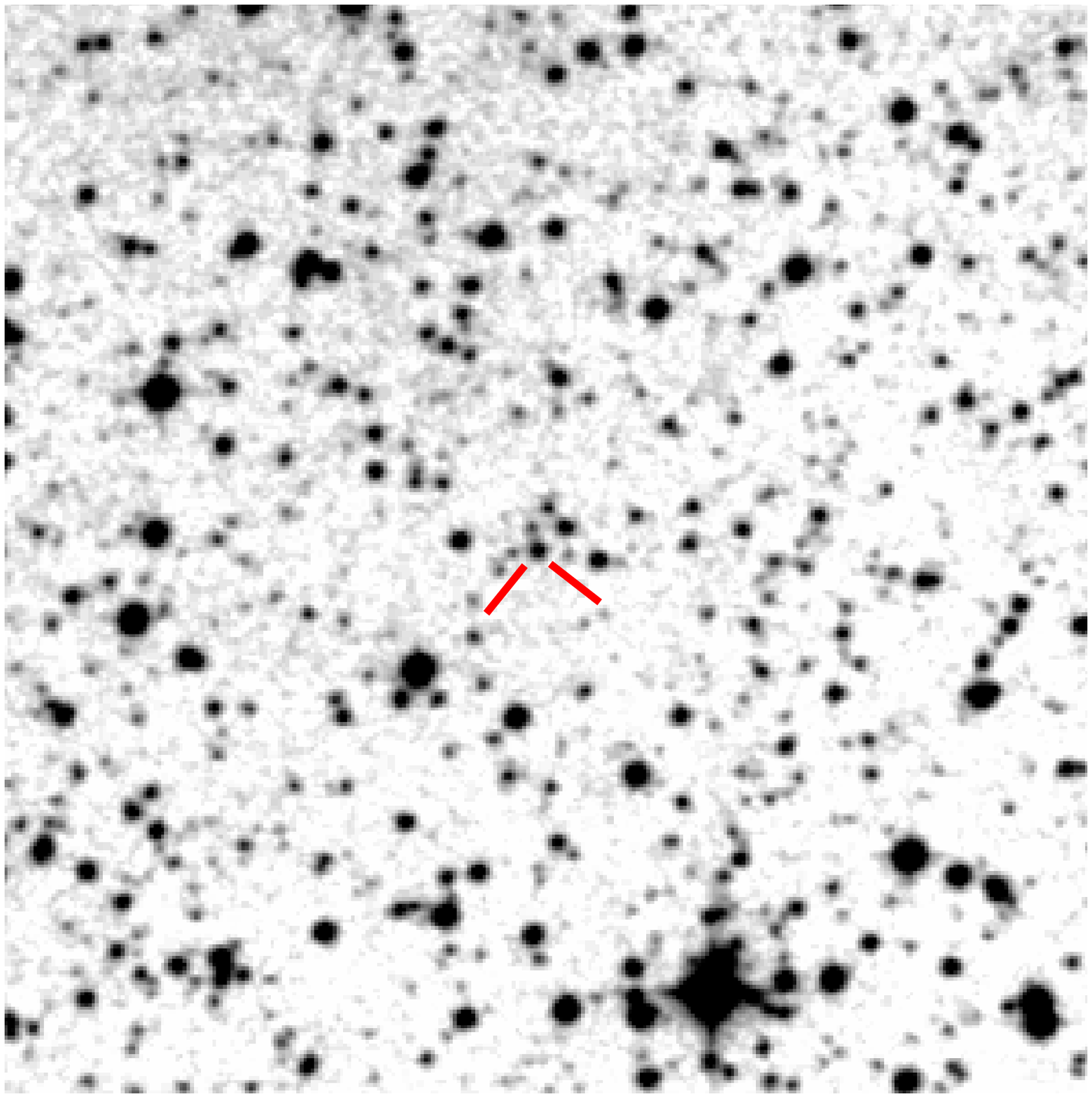,width=5.9cm}}}
\centering{\mbox{\psfig{file=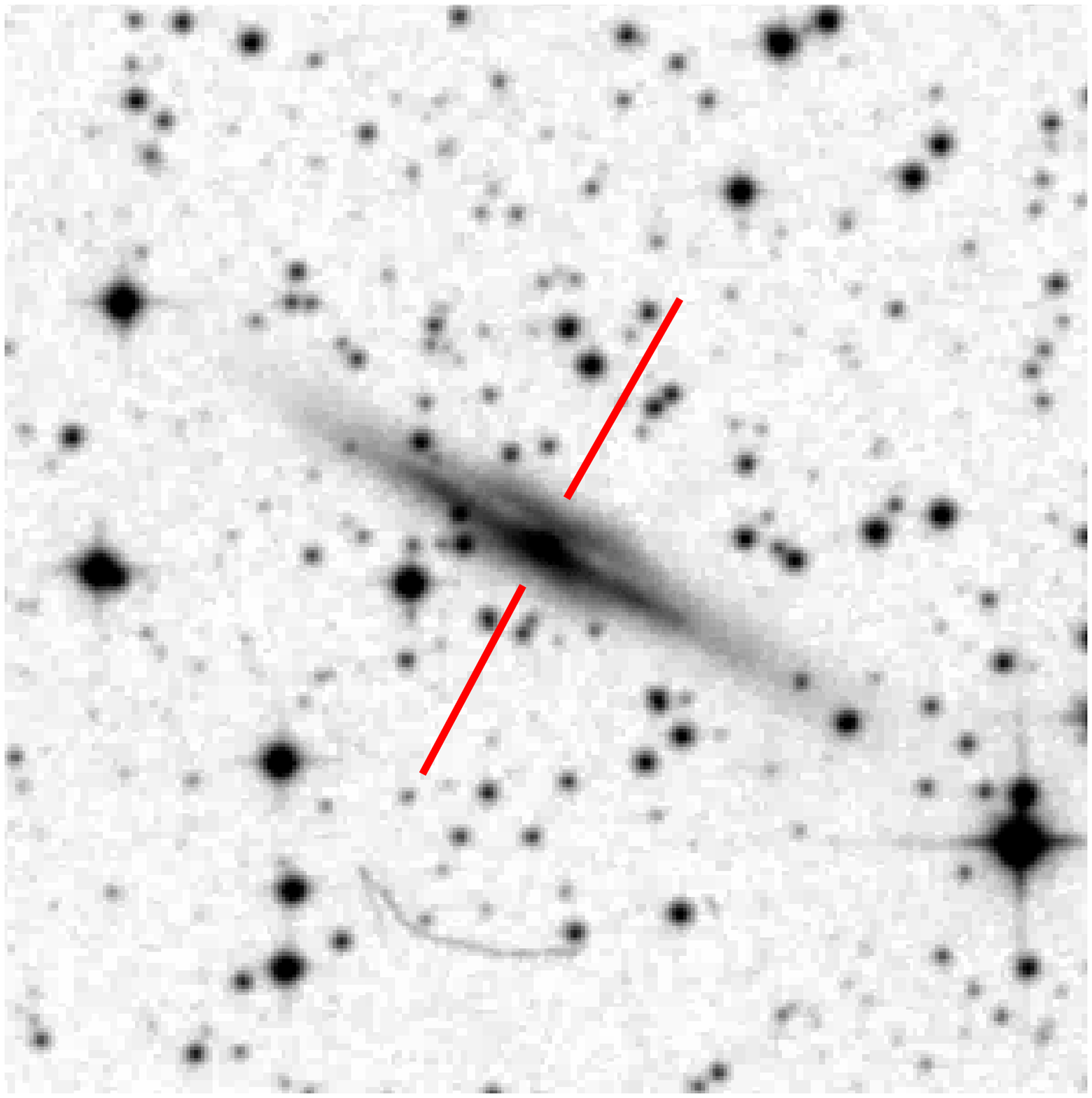,width=5.9cm}}}
\centering{\mbox{\psfig{file=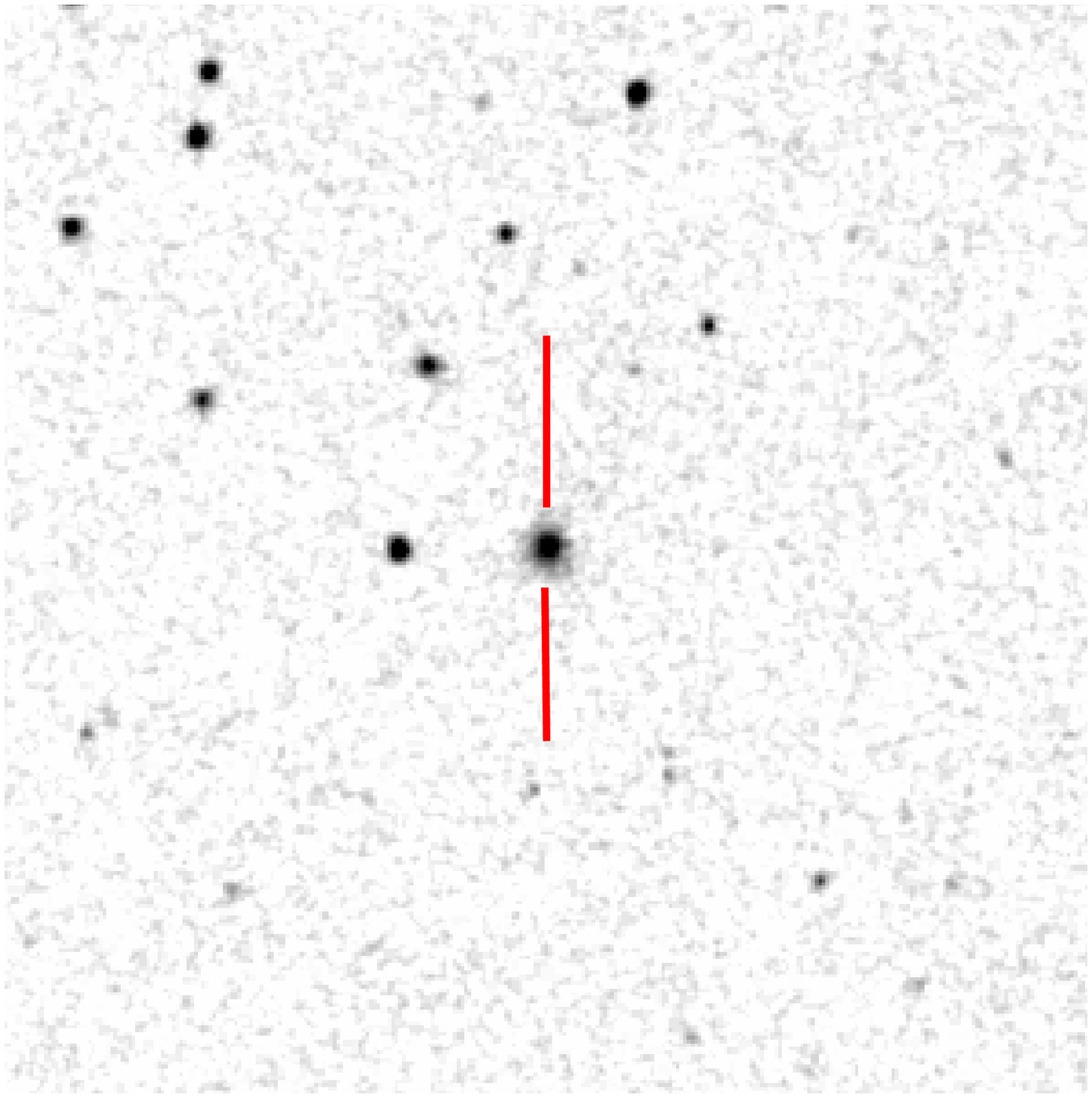,width=5.9cm}}}
\centering{\mbox{\psfig{file=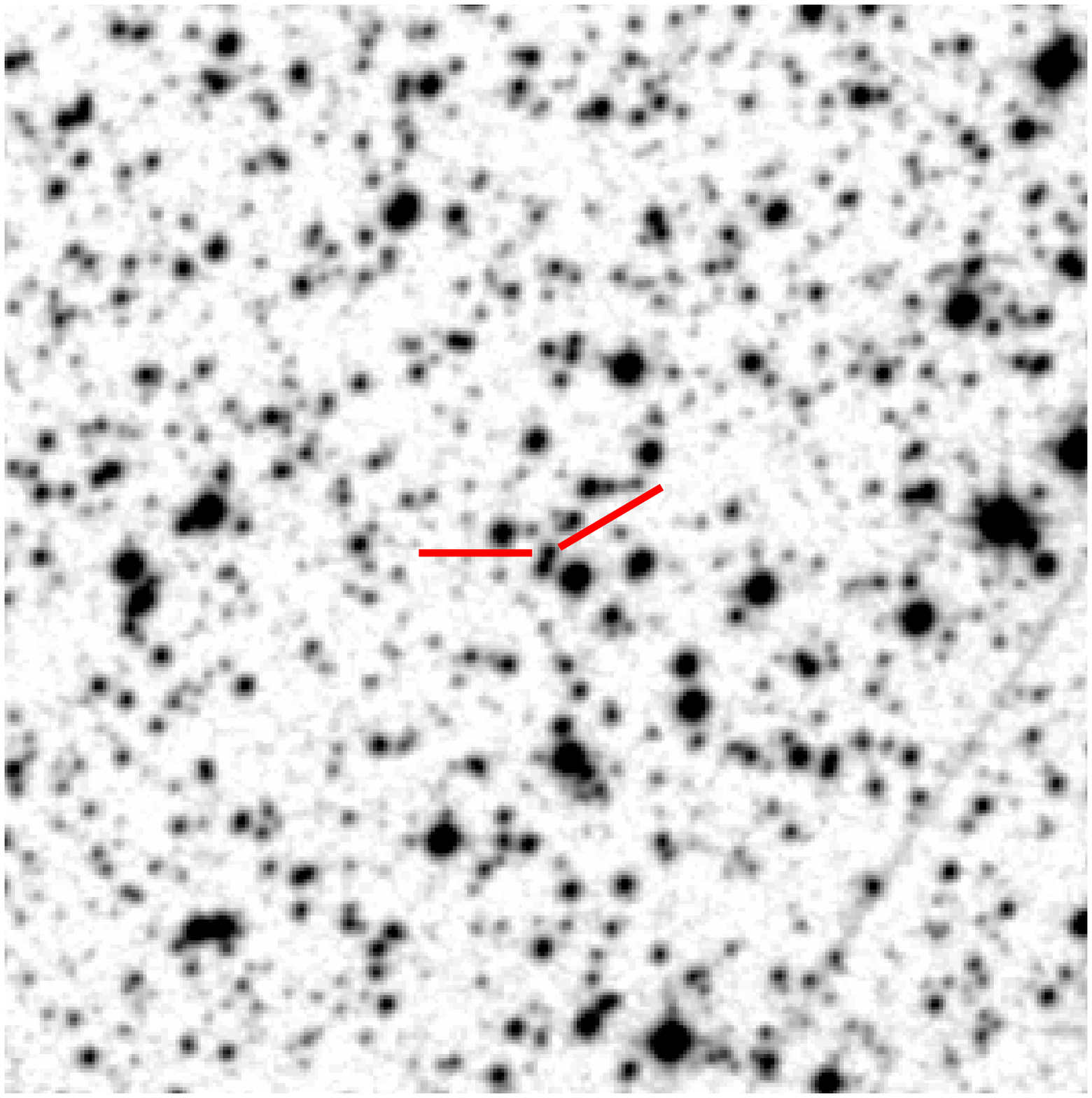,width=5.9cm}}}
\parbox{6cm}{
\psfig{file=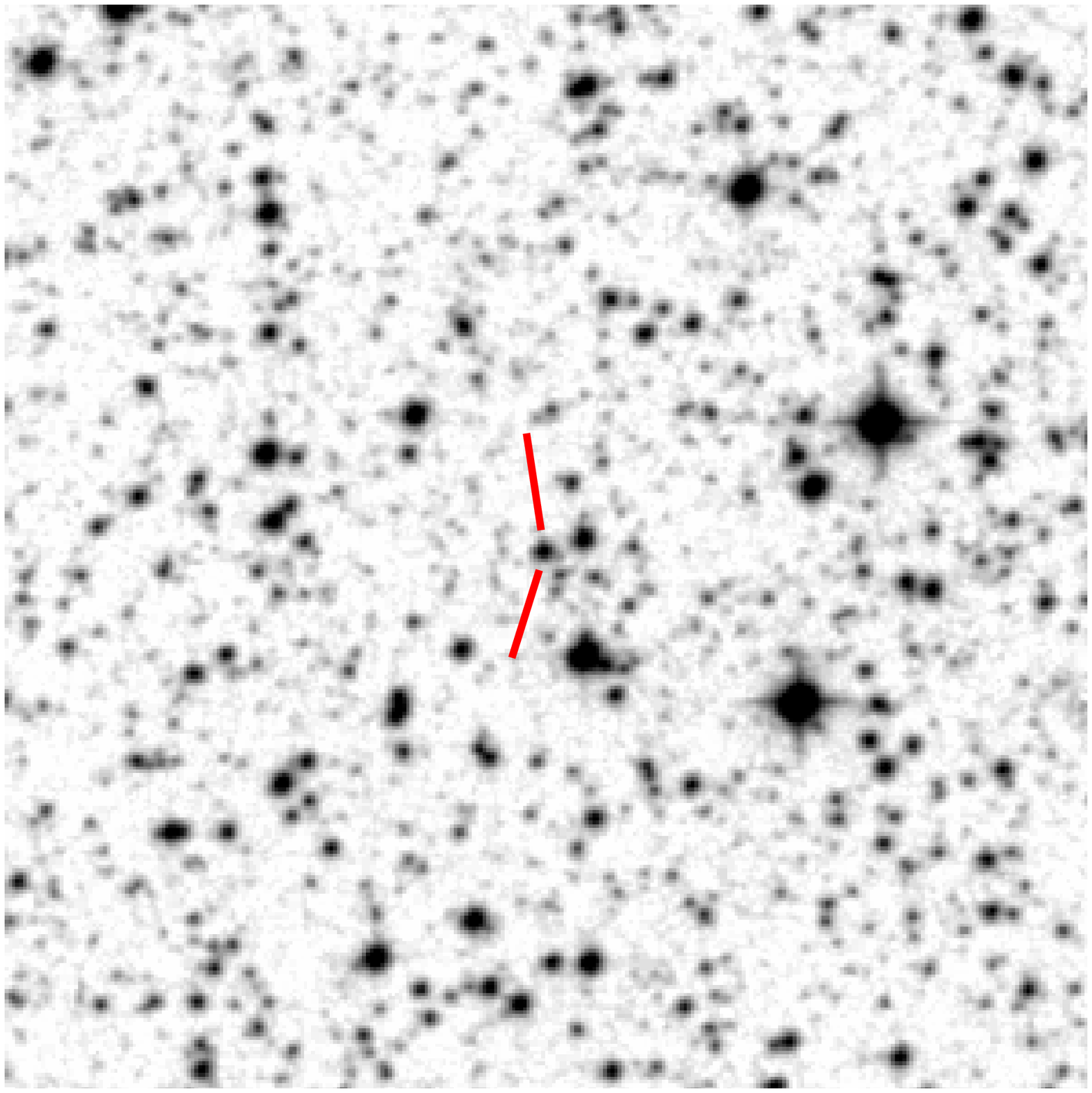,width=5.9cm}
}
\hspace{0.8cm}
\parbox{11cm}{
\vspace{-.5cm}
\caption{From left to  right and top to bottom: optical images of the
fields  of IGR  J00333+6122, Swift J0216.3+5128, IGR J02466$-$4222, IGR
J02524$-$0829, IGR J05270$-$6631, IGR J08390$-$4833, IGR J09025$-$6814, 
IGR J09253+6929, IGR J10147$-$6354 and IGR J11098$-$6457.  The optical 
counterparts  of the {\it INTEGRAL} sources are indicated with tick marks.  
Field sizes are 5$'$$\times$5$'$ and are extracted  from the DSS-II-Red 
survey. In all cases, north is up and east to the left.}}
\end{figure*}

\begin{figure*}[th!]
\hspace{-.1cm}
\centering{\mbox{\psfig{file=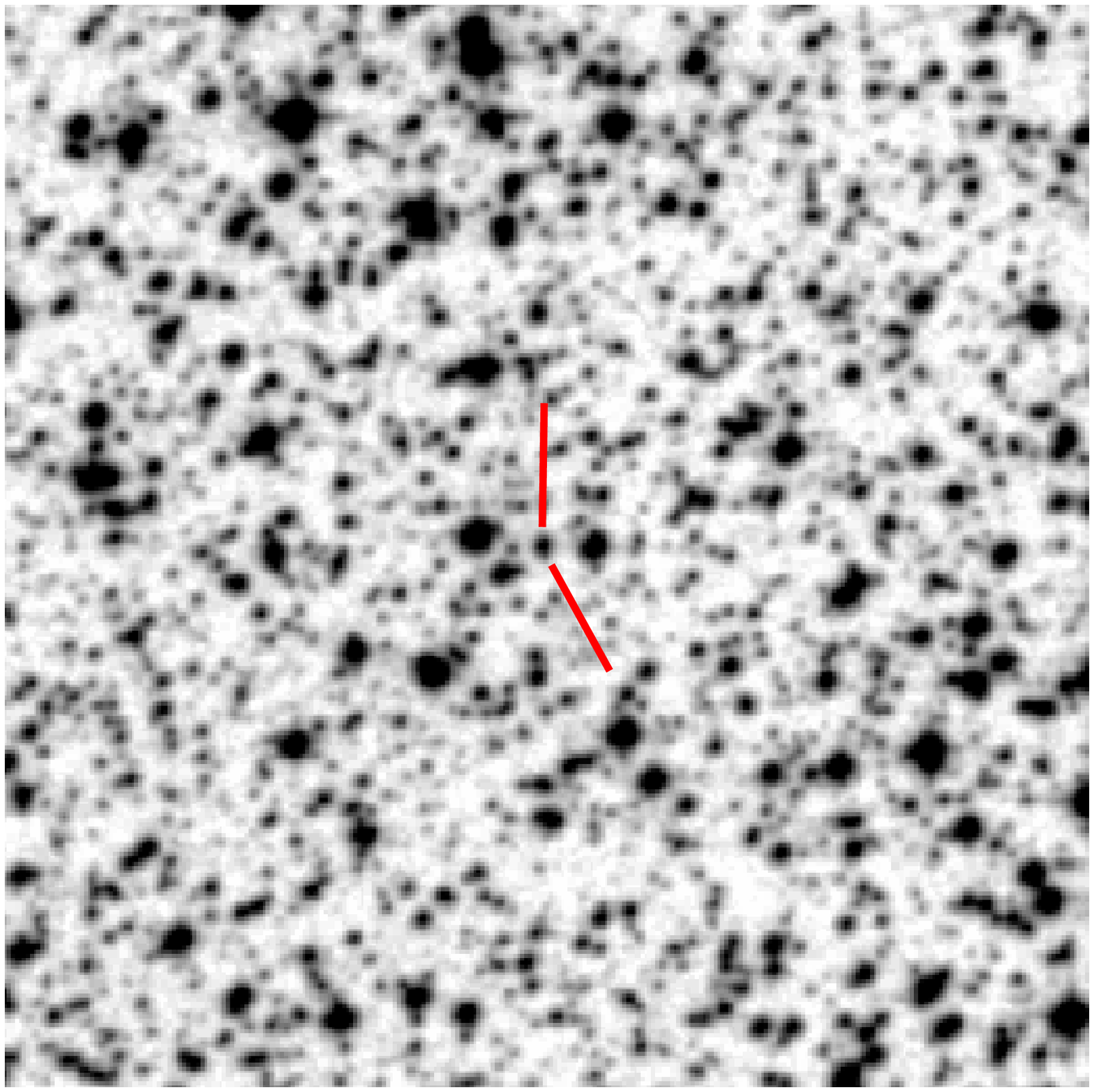,width=5.9cm}}}
\centering{\mbox{\psfig{file=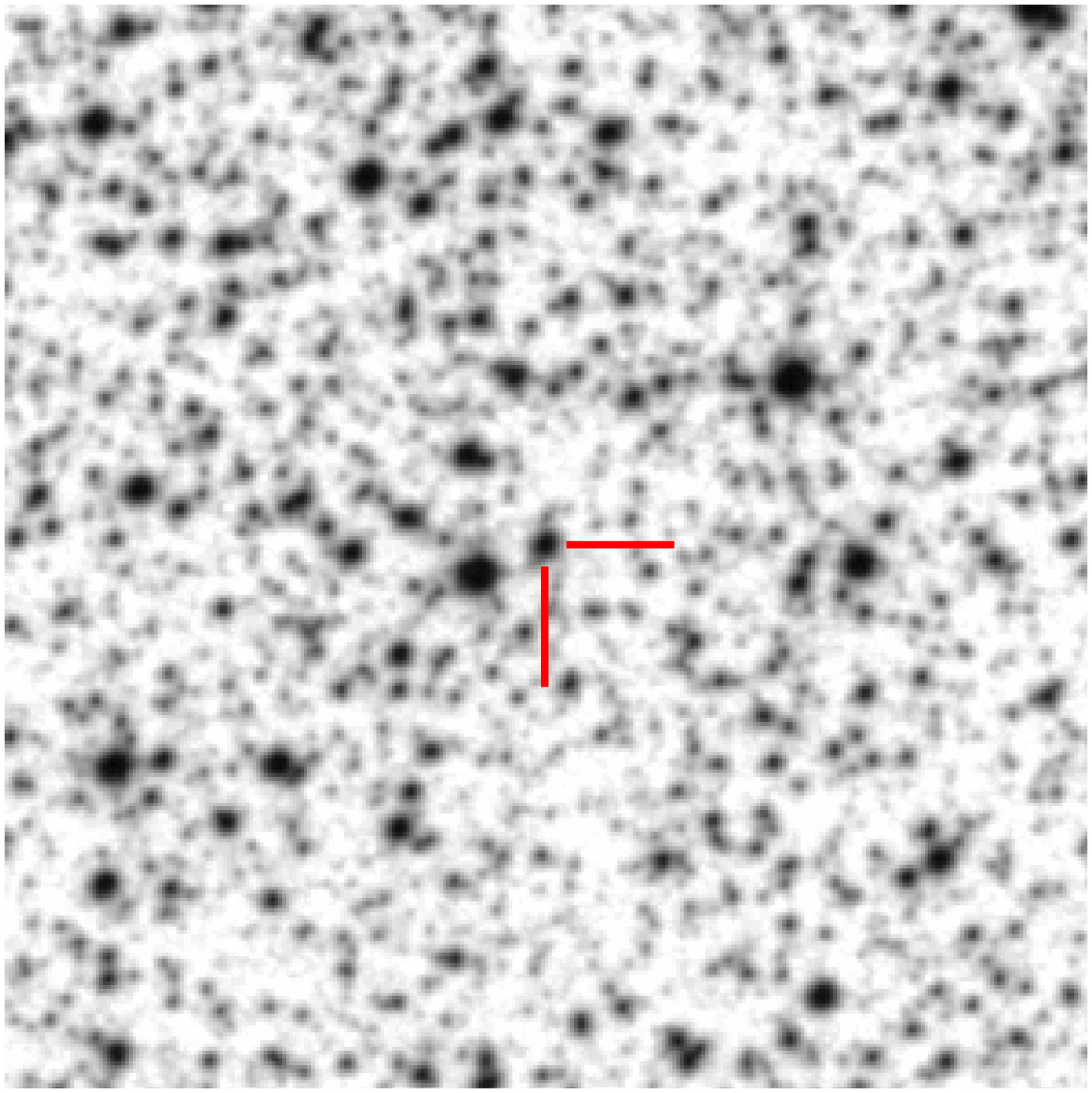,width=5.9cm}}}
\centering{\mbox{\psfig{file=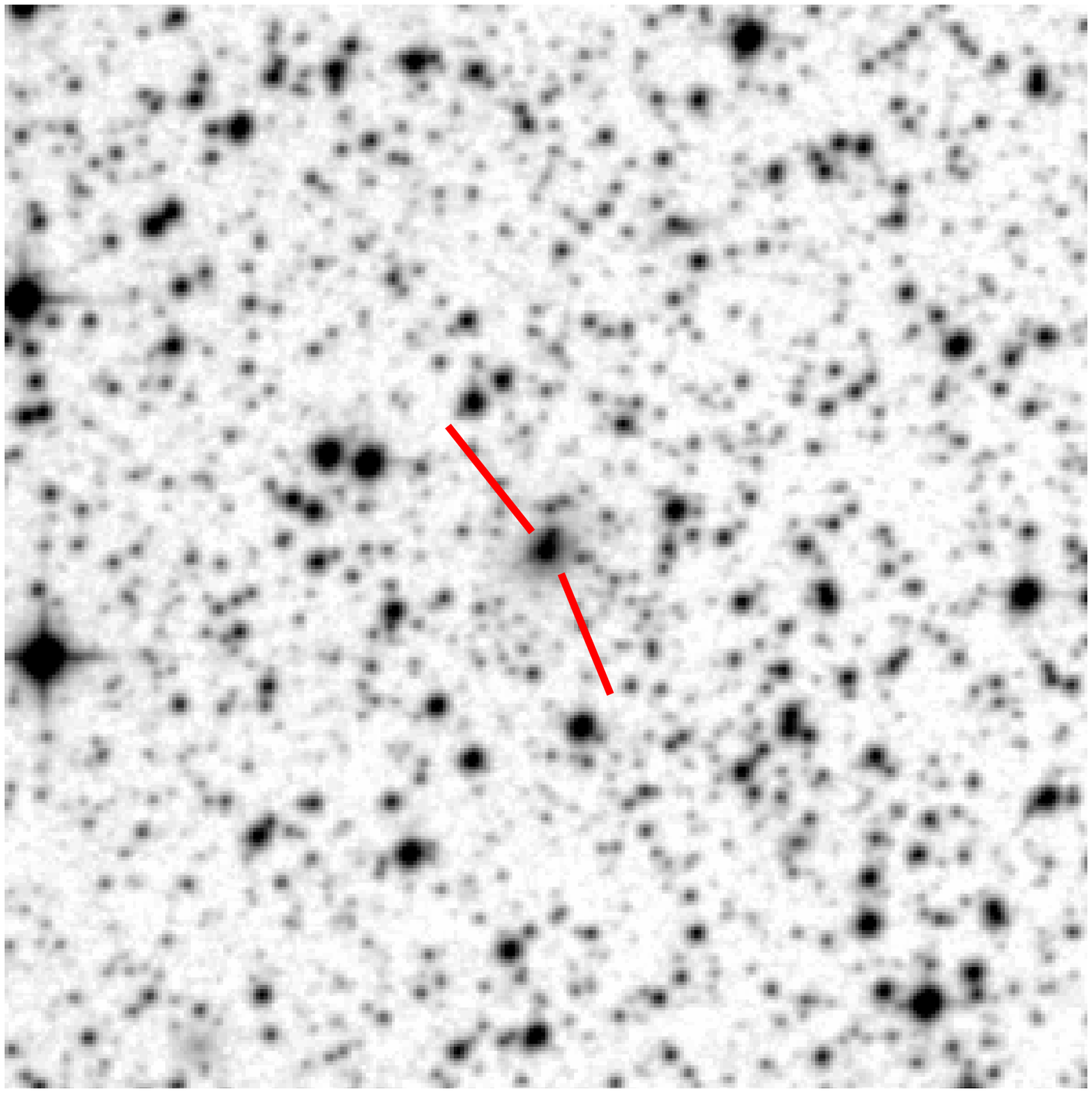,width=5.9cm}}}
\centering{\mbox{\psfig{file=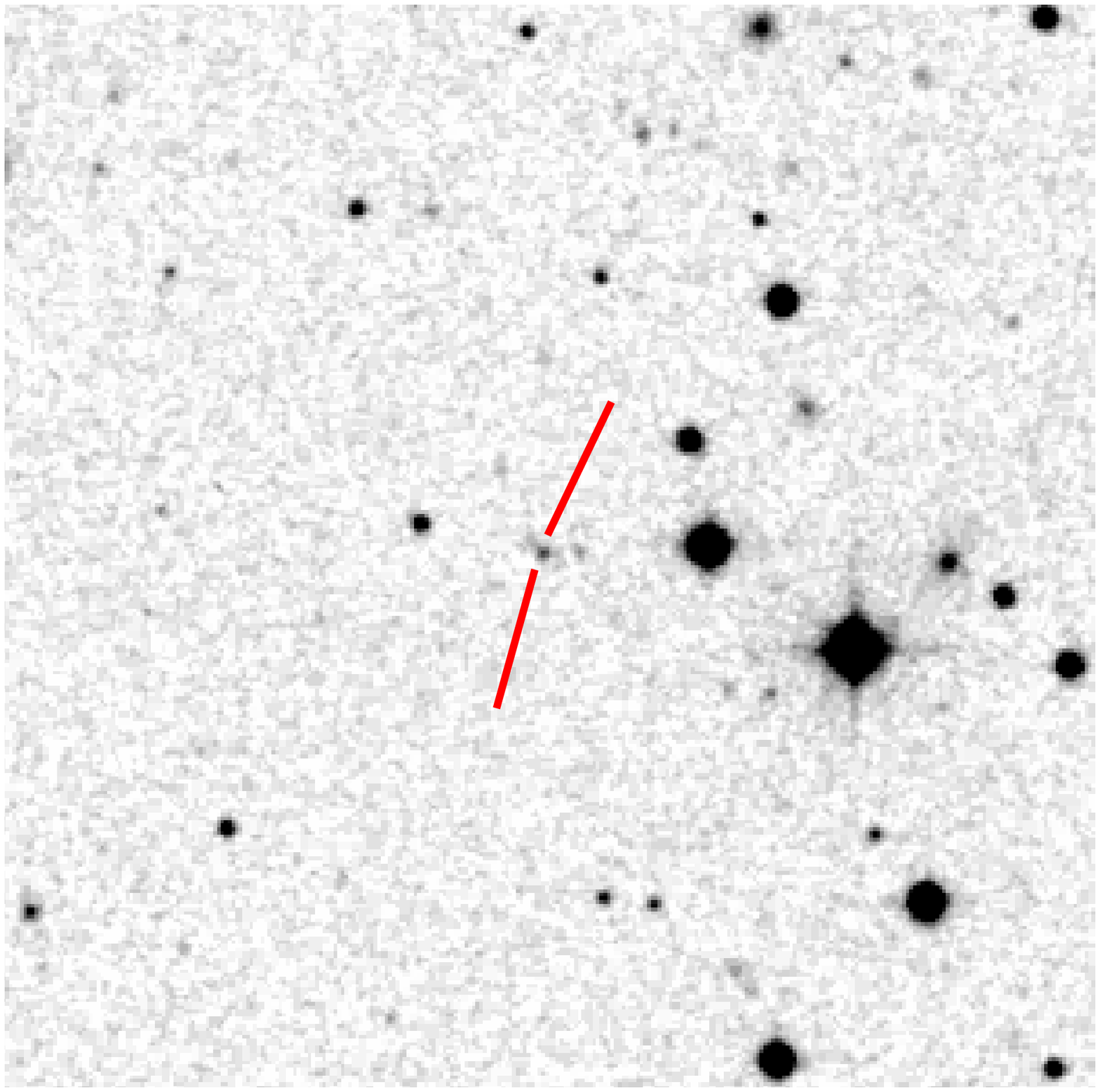,width=5.9cm}}}
\centering{\mbox{\psfig{file=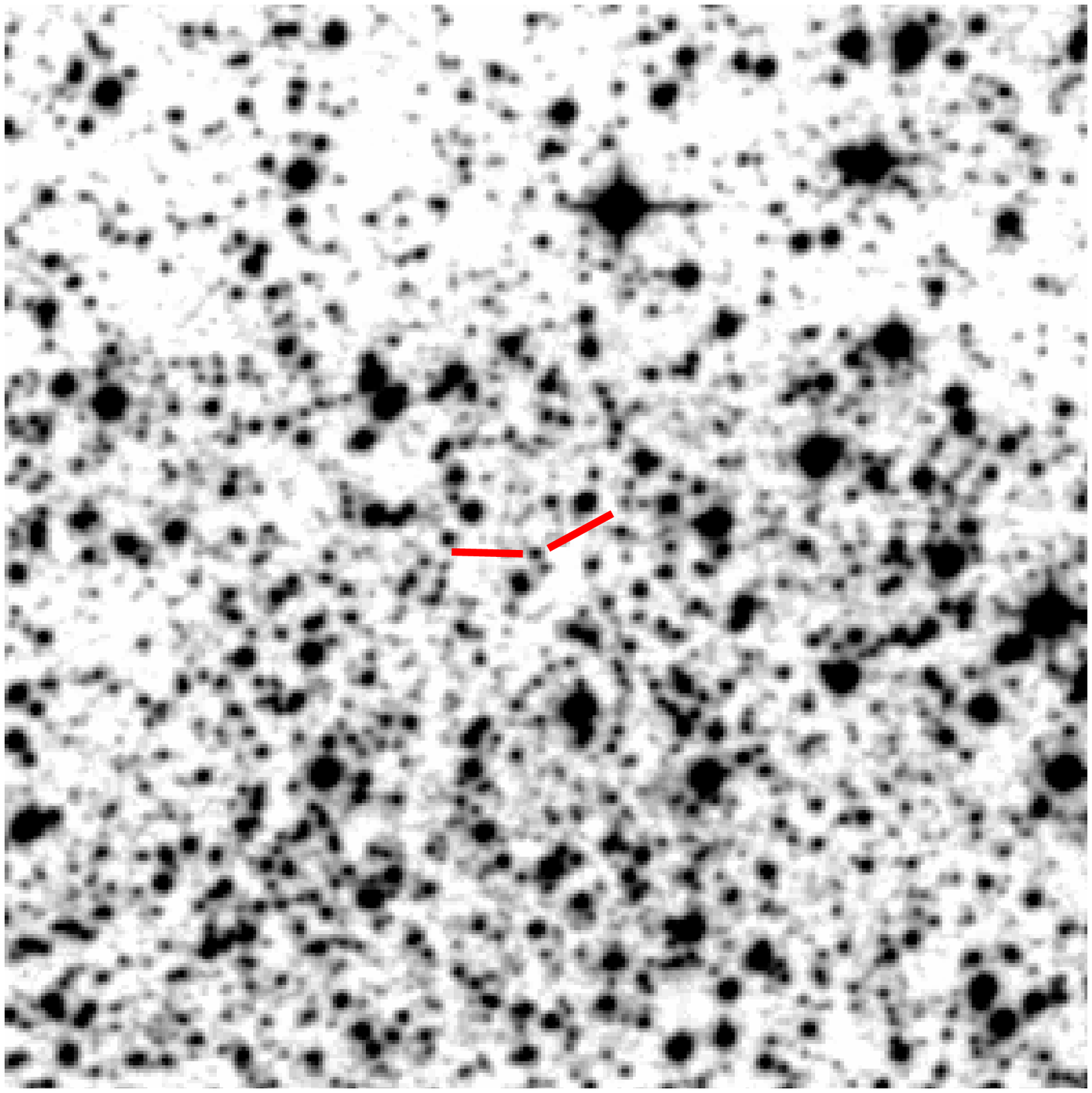,width=5.9cm}}}
\centering{\mbox{\psfig{file=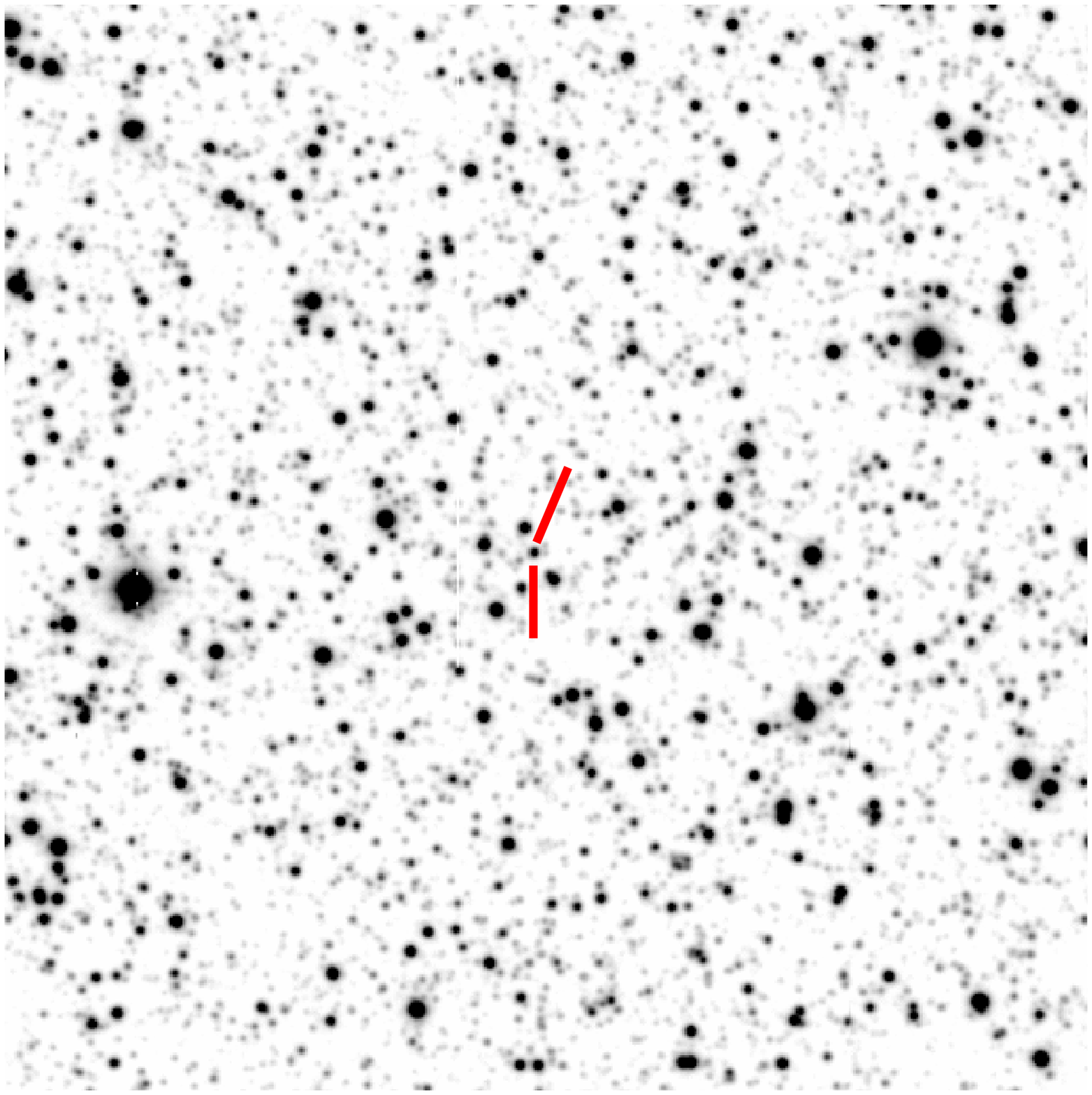,width=5.9cm}}}
\centering{\mbox{\psfig{file=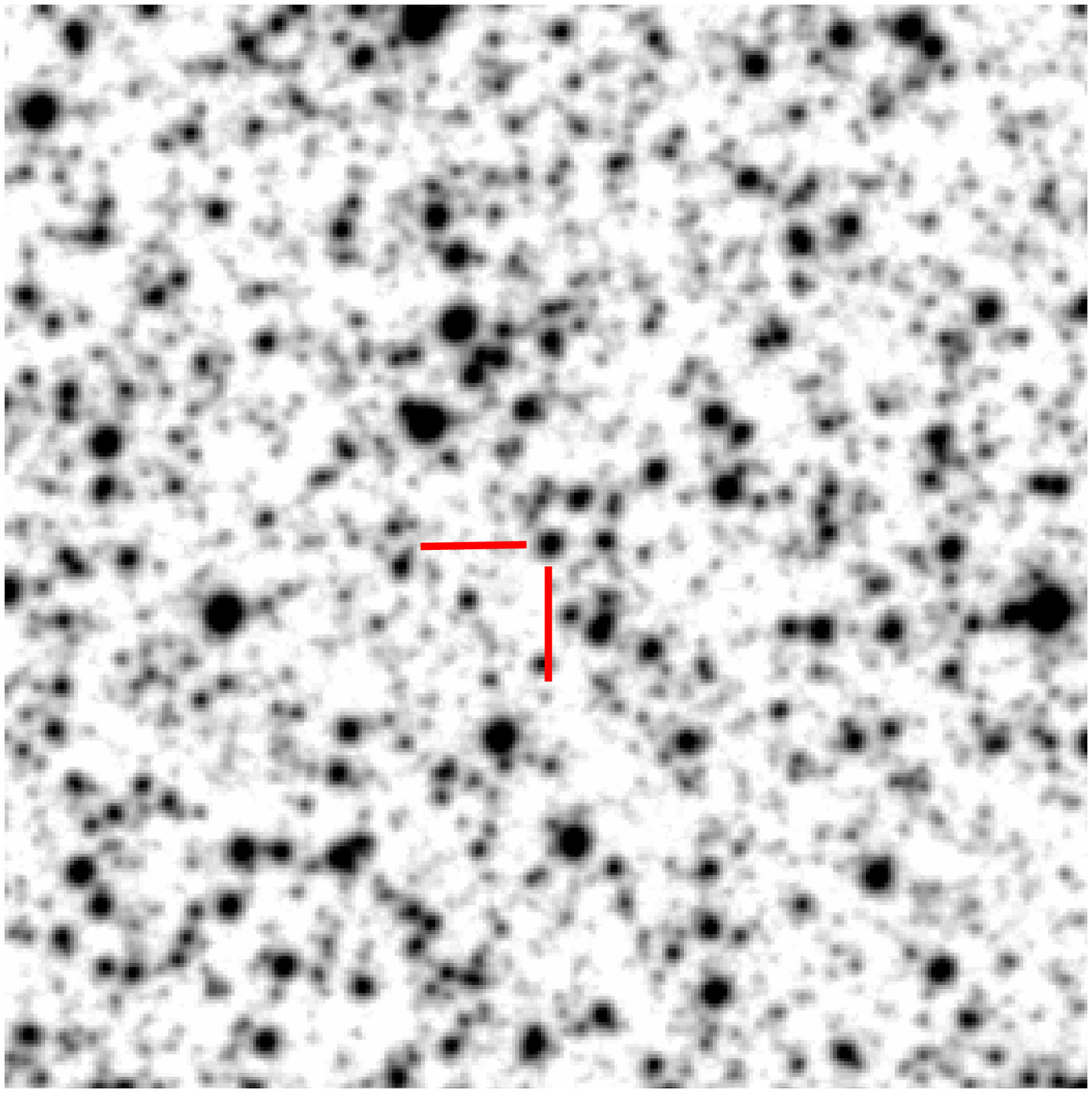,width=5.9cm}}}
\centering{\mbox{\psfig{file=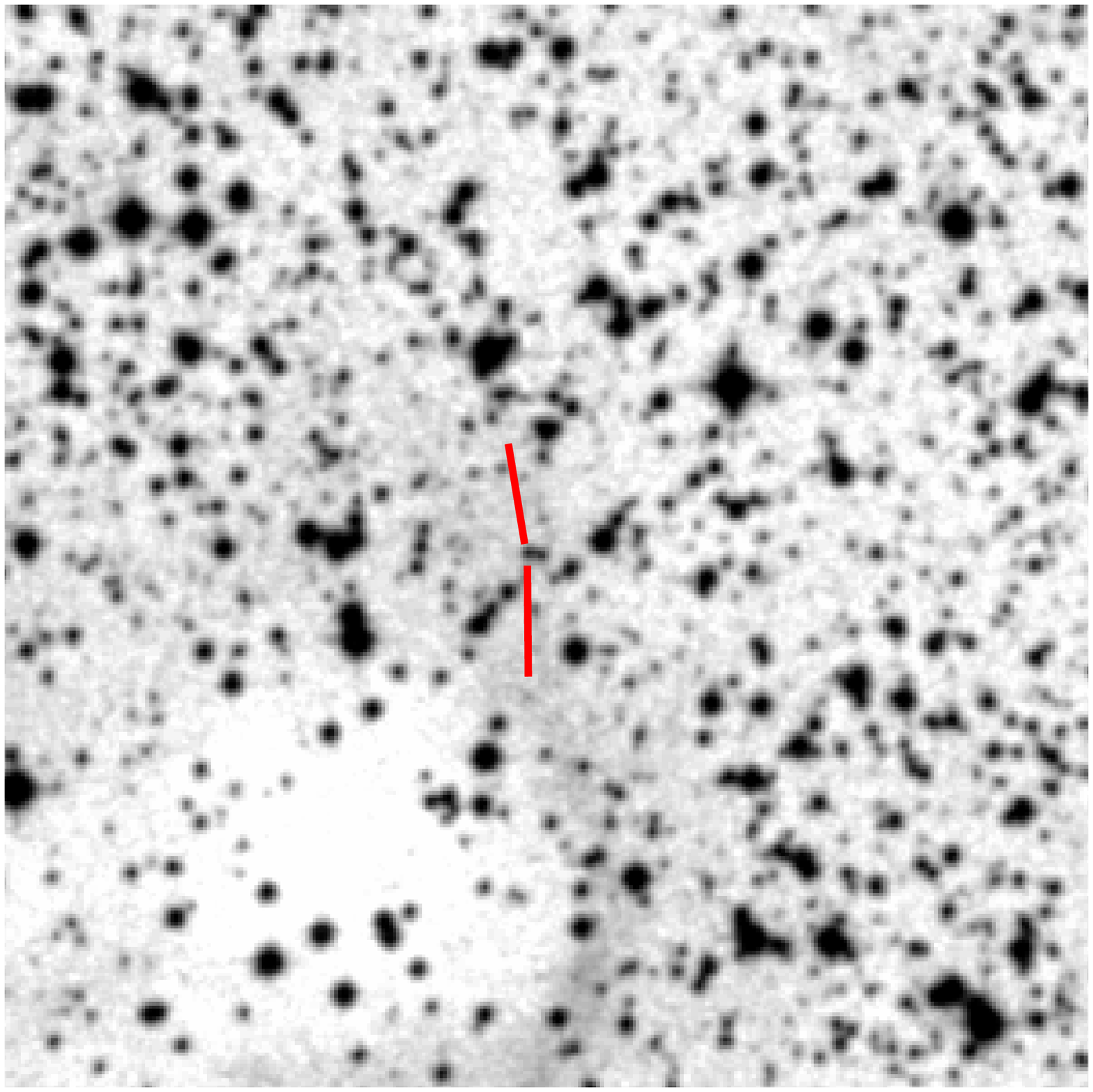,width=5.9cm}}}
\centering{\mbox{\psfig{file=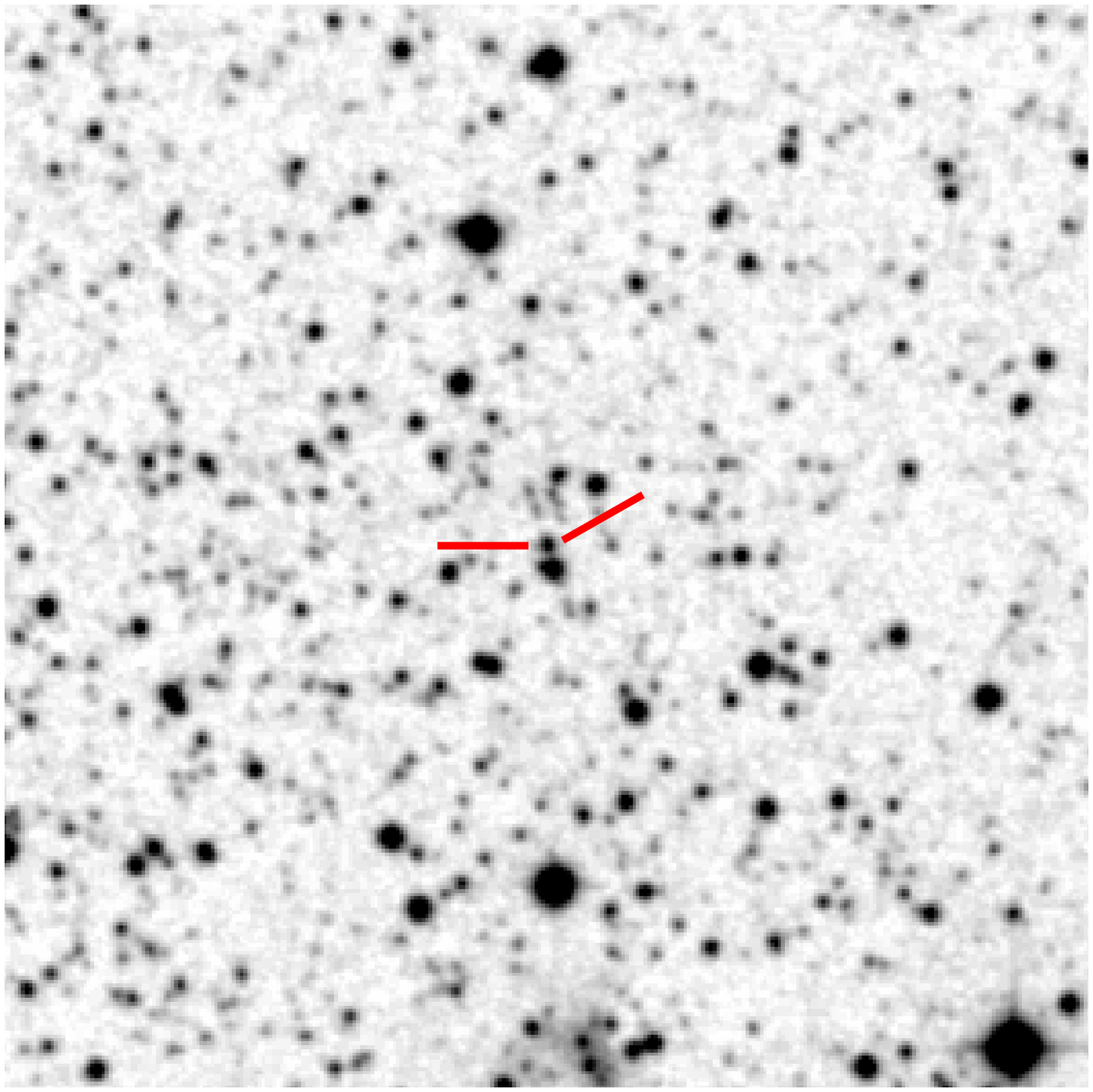,width=5.9cm}}}
\parbox{6cm}{
\psfig{file=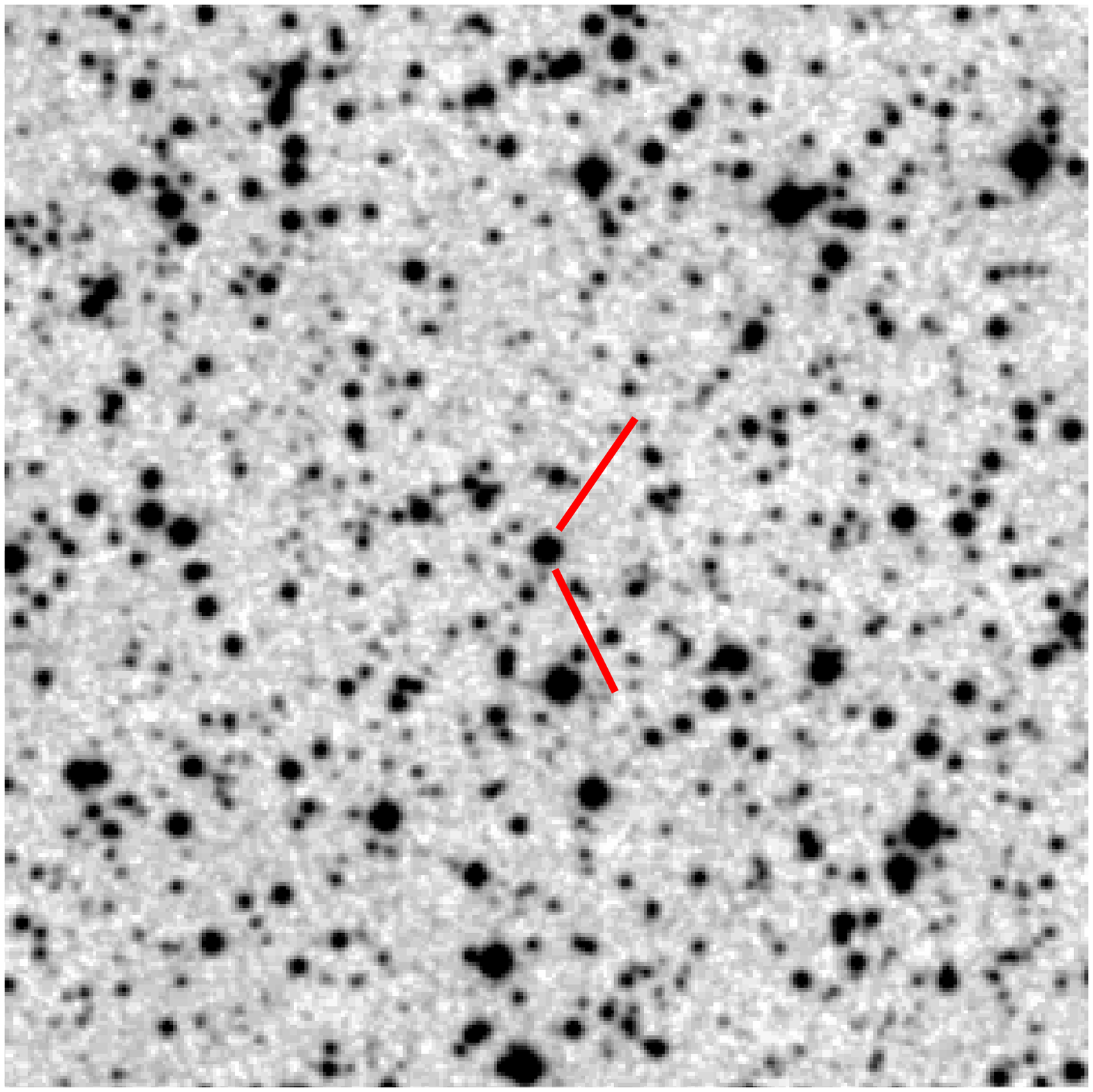,width=5.9cm}
}
\hspace{0.8cm}
\parbox{11cm}{
\vspace{-.5cm}
\caption{As Fig. 1, but for the fields of IGR J11435$-$6109, 1ES 
1210$-$646, IGR J12415$-$5750, IGR J16426+6536, IGR J17404$-$3655, IGR 
J18173$-$2509, IGR J18249$-$3243, IGR J18308$-$1232, IGR J19267+1325
and IGR J21347+4737. The field of IGR J18173$-$2509 is 3$'$$\times$3$'$
in size and was obtained at the ESO 3.6m telescope using EFOSC2 and 
the $R$ filter.}}
\end{figure*}

\begin{table*}[th!]
\caption[]{Log of the spectroscopic observations presented in this paper
(see text for details). If not otherwise indicated, source coordinates
are extracted from the 2MASS catalogue and have an accuracy better than 
0$\farcs$1.}
\scriptsize
\begin{center}
\begin{tabular}{llllcccr}
\noalign{\smallskip}
\hline
\hline
\noalign{\smallskip}
\multicolumn{1}{c}{{\it (1)}} & \multicolumn{1}{c}{{\it (2)}} & \multicolumn{1}{c}{{\it (3)}} & \multicolumn{1}{c}{{\it (4)}} & 
{\it (5)} & {\it (6)} & {\it (7)} & \multicolumn{1}{c}{{\it (8)}} \\
\multicolumn{1}{c}{Object} & \multicolumn{1}{c}{RA} & \multicolumn{1}{c}{Dec} & 
\multicolumn{1}{c}{Telescope+instrument} & $\lambda$ range & Disp. & \multicolumn{1}{c}{UT Date \& Time}  & Exposure \\
 & \multicolumn{1}{c}{(J2000)} & \multicolumn{1}{c}{(J2000)} & & (\AA) & (\AA/pix) & 
\multicolumn{1}{c}{at mid-exposure} & time (s)  \\

\noalign{\smallskip}
\hline
\noalign{\smallskip}

IGR J00333+6122    & 00:33:18.34 & +61:27:43.3 & WHT+ISIS & 5050-10300 & 1.8 & 05 Jan 2008, 22:51 & 2$\times$1000 \\ 
Swift J0216.3+5128 & 02:16:26.73$^\dagger$ & +51:25:25.1$^\dagger$ & WHT+ISIS & 5050-10300 & 1.8 & 06 Jan 2008, 00:30 & 2$\times$1800 \\ 
IGR J02466$-$4222  & 02:46:37.03 & $-$42:22:01.5 & AAT+6dF    & 3900-7600 & 1.6 & 30 Aug 2003, 18:06 & 1200+600 \\ 
IGR J02524-0829    & 02:52:23.39 & $-$08:30:37.5 & SDSS+CCD Spc. & 3800-9200 & 1.0 & 23 Dec 2000, 04:07 & 2700 \\
IGR J05270$-$6631  & 05:26:14.45$^\dagger$ & $-$66:30:45.0$^\dagger$ & 3.6m+EFOSC & 3685-9315 & 2.8 & 02 Jan 2008, 06:19 & 900 \\ 
IGR J08390$-$4833  & 08:38:49.11 & $-$48:31:24.8 & CTIO 1.5m+RC Spec. & 3300-10500 & 5.7 & 04 Mar 2008, 05:19 & 2$\times$1800 \\ 
IGR J09025$-$6814  & 09:02:39.46 & $-$68:13:36.6 & CTIO 1.5m+RC Spec. & 3300-10500 & 5.7 & 19 Feb 2007, 06:11 & 2$\times$1800 \\ 
IGR J09253+6929    & 09:25:47.56 & +69:27:53.6  & Cassini+BFOSC & 3500-8700 & 4.0 & 29 Mar 2008, 21:13 & 2$\times$1800  \\ 
IGR J10147$-$6354  & 10:14:15.55 & $-$63:51:50.1 & 3.6m+EFOSC & 3685-9315 & 2.8 & 02 Jan 2008, 06:44 & 900 \\ 
IGR J11098$-$6457  & 11:09:47.79 & $-$64:52:45.3 & CTIO 1.5m+RC Spec. & 3300-10500 & 5.7 & 04 Mar 2008, 06:44 & 2$\times$1200 \\ 
IGR J11435$-$6109  & 11:44:00.30 & $-$61:07:36.5 & CTIO 1.5m+RC Spec. & 3300-10500 & 5.7 & 01 Jul 2008, 23:13 & 2$\times$1200 \\ 
1ES 1210$-$646     & 12:13:14.79 & $-$64:52:30.5 & CTIO 1.5m+RC Spec. & 3300-10500 & 5.7 & 04 Mar 2008, 07:26 & 2$\times$900 \\ 
IGR J12415$-$5750  & 12:41:25.74 & $-$57:50:03.5 & CTIO 1.5m+RC Spec. & 3300-10500 & 5.7 & 04 Mar 2008, 08:10 & 2$\times$1200 \\ 
IGR J16426+6536    & 16:43:04.07$\dagger$ & +65:32:50.9$\dagger$ & TNG+DOLoRes & 3800-8000 & 2.5 & 04 Feb 2008, 06:11 & 2$\times$900 \\ 
IGR J17404$-$3655  & 17:40:26.86 & $-$36:55:37.4 & CTIO 1.5m+RC Spec. & 3300-10500 & 5.7 & 02 Jul 2008, 05:36 & 2$\times$2400 \\ 
IGR J18173$-$2509  & 18:17:22.3$^*$ & $-$25:08:43$^*$ & CTIO 1.5m+RC Spec. & 3300-10500 & 5.7 & 02 Jul 2008, 04:04 & 2$\times$2400 \\ 
IGR J18249$-$3243  & 18:24:55.92 & $-$32:42:57.7 & CTIO 1.5m+RC Spec. & 3300-10500 & 5.7 & 04 Mar 2008, 09:21 & 2$\times$1200 \\ 
IGR J18308$-$1232  & 18:30:49.88$\dagger$ & $-$12:32:18.7$\dagger$ & SPM 2.1m+B\&C Spec. & 3450-7650 & 4.0 & 29 Jun 2008, 07:49 & 3$\times$2400 \\ 
IGR J19267+1325    & 19:26:27.00 & +13:22:05.0 & SPM 2.1m+B\&C Spec. & 3450-7650 & 4.0 & 30 Jun 2008, 06:15 & 2$\times$2400 \\ 
IGR J21347+4737    & 21:34:20.38 & +47:38:00.2 & Cassini+BFOSC & 3500-8700 & 4.0 & 11 May 2008, 01:01 & 2$\times$1800 \\ 
\noalign{\smallskip}
\hline
\noalign{\smallskip}
\multicolumn{8}{l}{$^\dagger$: coordinates extracted from the USNO catalogues, having 
an accuracy of about 0$\farcs$2 (Deutsch 1999; Assafin et al. 2001; Monet et al. 2003).}\\
\multicolumn{8}{l}{$^*$: coordinates extracted from the DSS-II-Red frames, having an
accuracy of $\sim$1$''$.} \\
\noalign{\smallskip}
\hline
\hline
\noalign{\smallskip}
\end{tabular}
\end{center}
\end{table*}

\section{Optical spectroscopy}

As in the case of Paper VI, the data presented in this paper were 
collected in the course of a campaign that involved observations at the 
following telescopes in the past 2 years:

\begin{itemize}
\item the 1.5m at the Cerro Tololo Interamerican Observatory (CTIO), Chile;
\item the 1.52m ``Cassini'' telescope of the Astronomical Observatory of 
Bologna, in Loiano, Italy; 
\item the 2.1m telescope of the Observatorio Astron\'omico Nacional in San 
Pedro Martir, M\'exico, 
\item the 3.6m telescope at the ESO-La Silla Observatory, Chile;
\item the 3.58m ``Telescopio Nazionale Galileo" (TNG) and the 4.2m 
``William Herschel Telescope'' (WHT) at the Roque de Los Muchachos 
Observatory in La Palma, Spain. 
\end{itemize}

The spectroscopic data secured at these telescopes were optimally 
extracted (Horne 1986) and reduced following standard procedures using 
IRAF\footnote{IRAF is the Image Reduction and Analysis Facility made 
available to the astronomical community by the National Optical Astronomy 
Observatories, which are operated by AURA, Inc., under contract with the 
U.S. National Science Foundation. It is available at {\tt 
http://iraf.noao.edu/}}.  Calibration frames (flat fields and bias) were 
taken on the day preceeding or following the observing night.  The 
wavelength calibration was obtained using lamps acquired soon after each 
on-target spectroscopic acquisiton; uncertainty on this calibration was 
$\sim$0.5~\AA~for all cases: this was checked using the positions of 
background night sky lines. Flux calibration was performed using 
catalogued spectrophotometric standards.

Further spectra were retrieved from two different astronomical archives:
the Sloan Digitized Sky Survey\footnote{{\tt http://www.sdss.org/}} (SDSS, 
Adelman-McCarthy et al. 2007) archive, and the Six-degree Field Galaxy 
Survey\footnote{{\tt http://www.aao.gov.au/local/www/6df/}} (6dFGS) 
archive (Jones et al. 2004). As the 6dFGS archive provides spectra which 
are not calibrated in flux, we used the optical photometric information in 
Jones et al. (2005) to calibrate the 6dFGS data presented here.

We report in Table 1 the detailed log of observations. We list in column 
1 the name of the observed {\it INTEGRAL} sources. In columns 2 and 3 we 
give the object coordinates, extracted from the 2MASS catalogue (with an 
accuracy of $\leq$0$\farcs$1, according to Skrutskie et al. 2006),
from the USNO catalogues (with uncertainties of about 0$\farcs$2: Deutsch 
1999; Assafin et al. 2001; Monet et al. 2003), or from the DSS-II-Red 
astrometry (which has a precision of $\sim$1$''$). In 
column 4 we report the telescope and the instrument used for the 
observations. The characteristics of each spectrograph are presented in 
columns 5 and 6. Column 7 provides the observation date and the UT time at 
mid-exposure, while column 8 reports the exposure times and the number of 
observations of each source.

As a complement to the information on the putative counterpart of IGR 
J18173$-$2509, we analyzed an optical $R$-band frame acquired with the 
ESO 3.6m telescope plus EFOSC2 on 21 
June 2007 (start time: 07:28 UT; duration: 20 s) under a seeing of 
1$\farcs$0; the 2$\times$2-rebinned CCD of EFOSC2 secured a plate scale 
of 0$\farcs$31/pix, and a useful field of 5$\farcm$2$\times$5$\farcm$2.

This imaging frame was corrected for bias and flat-field in the usual 
fashion and was calibrated using nearby USNO-A2.0\footnote{available at \\ 
{\tt http://archive.eso.org/skycat/servers/usnoa/}} stars. Simple aperture 
photometry, within the 
MIDAS\footnote{\texttt{http://www.eso.org/projects/esomidas}} package, was 
then used to measure the $R$-band magnitude of the putative optical 
counterpart of IGR J18173$-$2509.

\section{Results}

Here we present the results of our spectroscopic campaign described in 
Sect. 2. The optical magnitudes quoted below, if not otherwise stated, are 
extracted from the USNO-A2.0 catalogue.

As done in Papers I-VI, we describe here the identification and
classification criteria for the optical spectra of the sources considered
in this work.

In order to infer the reddening along the line of sight of Galactic 
sources, when possible, we considered an intrinsic H$_\alpha$/H$_\beta$ 
line ratio of 2.86 (Osterbrock 1989) and we determined the corresponding 
colour excess, using the Galactic extinction law of Cardelli et al. (1989), 
from the comparison between the intrinsic line ratio and the measured one.

To derive the distance of the compact Galactic X--ray sources of our 
sample, in the case of CVs we assumed an absolute magnitude M$_V \sim$ 9 
and an intrinsic color index $(V-R)_0 \sim$ 0 mag (Warner 1995), whereas 
for HMXBs, when applicable, we used the intrinsic stellar color indices 
and absolute magnitudes as reported in Lang (1992) and Wegner (1994). For 
the single low-mass X-ray binary (LMXB) of our sample, we considered 
$(V-R)_0$ $\sim$ 0 $\sim$ M$_R$ (e.g., van Paradijs \& McClintock 1995).

For the emission-line AGN classification, we used the criteria 
of Veilleux \& Osterbrock (1987) and the line ratio diagnostics 
of Ho et al. (1993, 1997) and of Kauffmann et al. (2003); 
moreover, for the subclass assignation of Seyfert 1 nuclei, we used 
the H$_\beta$/[O {\sc iii}]$\lambda$5007 line flux ratio criterion 
described in Winkler (1992).

The spectra of the galaxies shown here were not corrected for starlight 
contamination (see, e.g., Ho et al. 1993, 1997) given the limited S/N and 
the spectral resolution. We do not consider that this will affect any of 
our main results and conclusions.

In the following we consider a cosmology with $H_{\rm 0}$ = 65 km s$^{-1}$ 
Mpc$^{-1}$, $\Omega_{\Lambda}$ = 0.7 and $\Omega_{\rm m}$ = 0.3; the 
luminosity distances of the extragalactic objects reported in this paper 
are computed with these parameters using the Cosmology Calculator of 
Wright (2006). Moreover, when not explicitly stated otherwise, for our 
X--ray flux estimates we will assume a Crab-like spectrum except for the 
{\it XMM-Newton} sources, for which we considered the 0.2--12 keV flux 
reported in Saxton et al. (2008) or Watson et al. (2008). We also remark 
that the results presented here supersede the preliminary ones reported in 
Masetti et al. (2008c) and in Parisi et al. (2008a,b,c,d).

As in Papers IV-VI, the next subsections will deal with the object 
identifications by dividing them into three broad classes (CVs, X--ray 
Binaries and AGNs) ordered according to their increasing distance from 
Earth.

\subsection{CVs}

\begin{figure*}[th!]
\mbox{\psfig{file=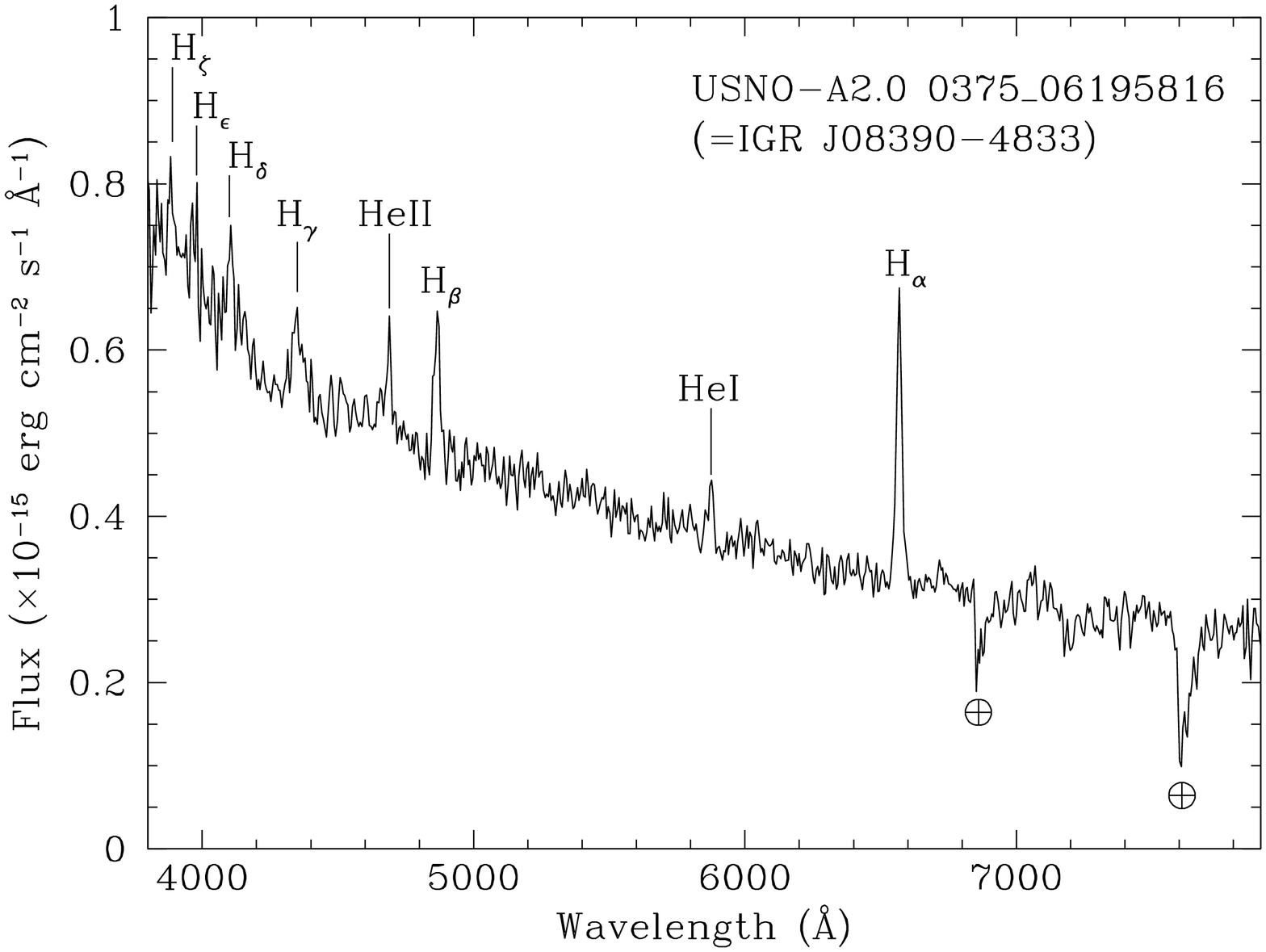,width=9cm,angle=270}}
\mbox{\psfig{file=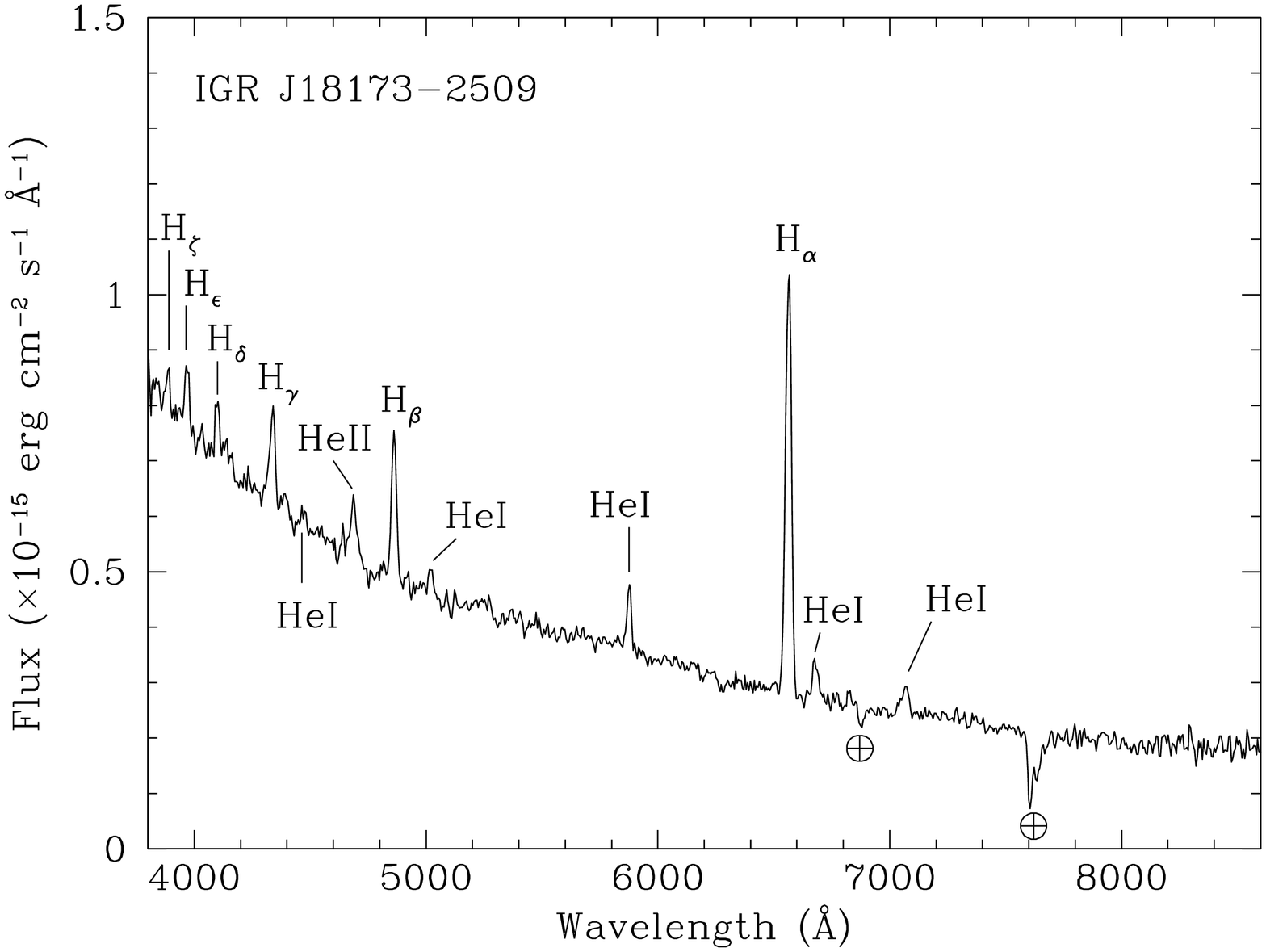,width=9cm,angle=270}}

\vspace{-.9cm}
\mbox{\psfig{file=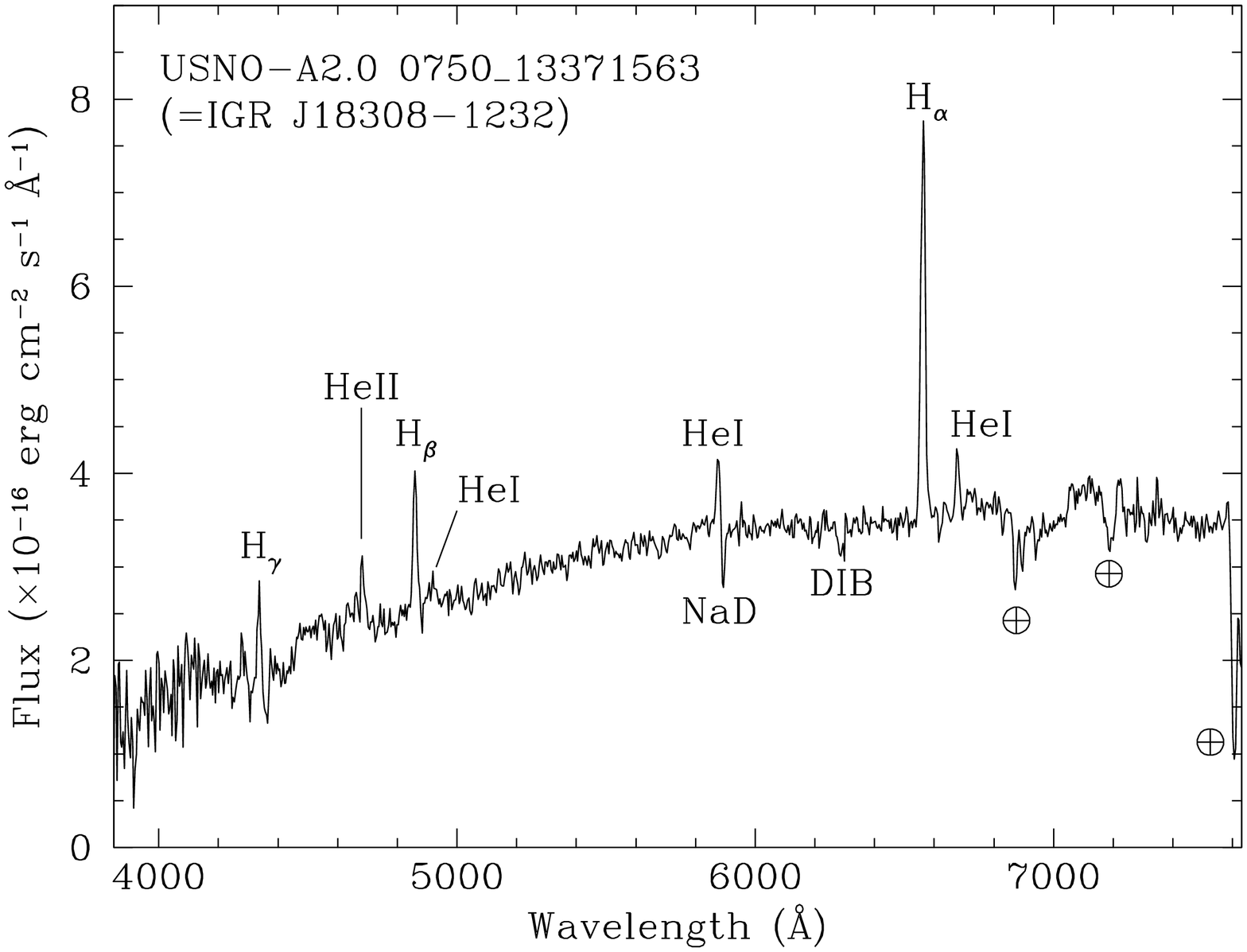,width=9cm,angle=270}}
\mbox{\psfig{file=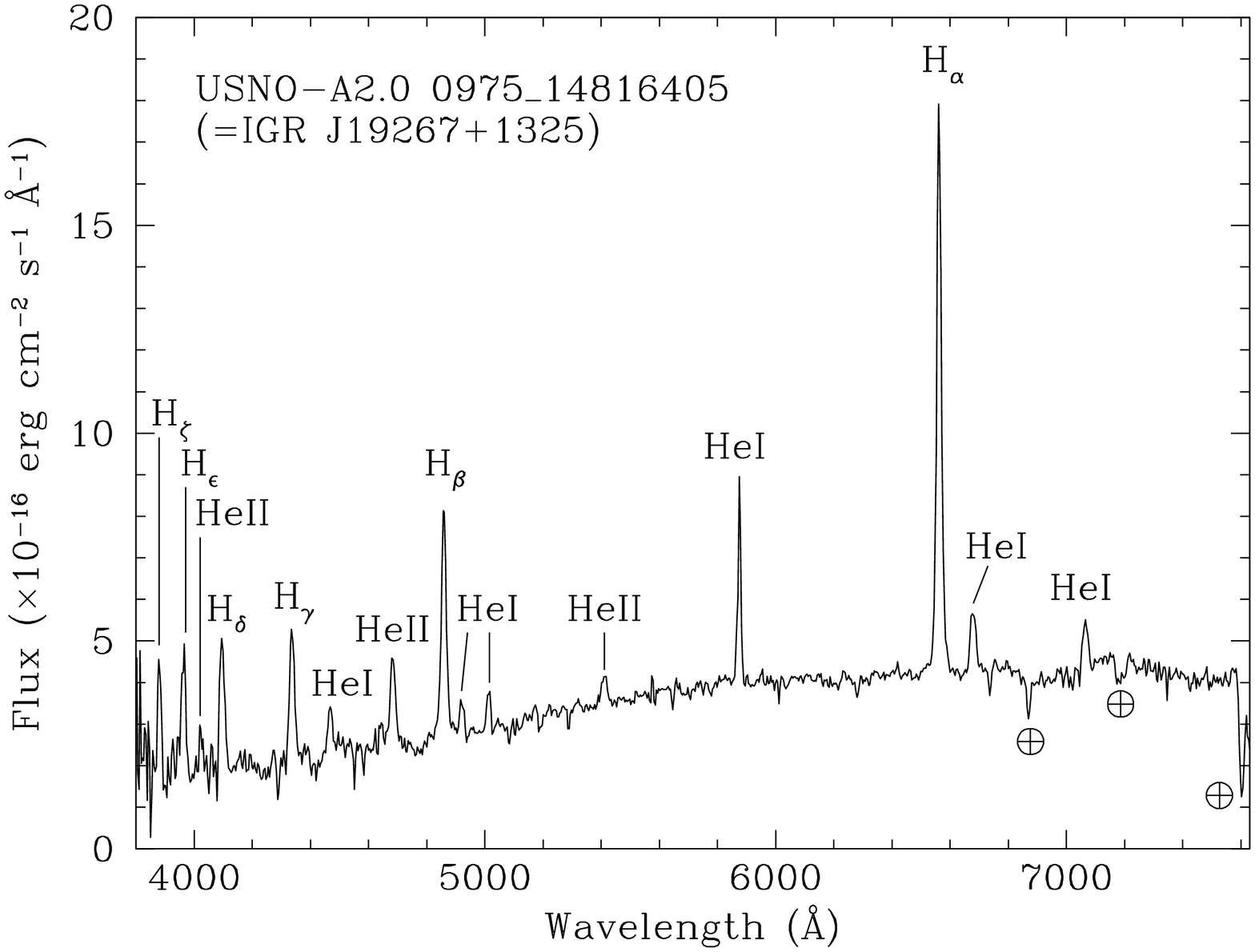,width=9cm,angle=270}}

\vspace{-.9cm}
\parbox{9cm}{
\psfig{file=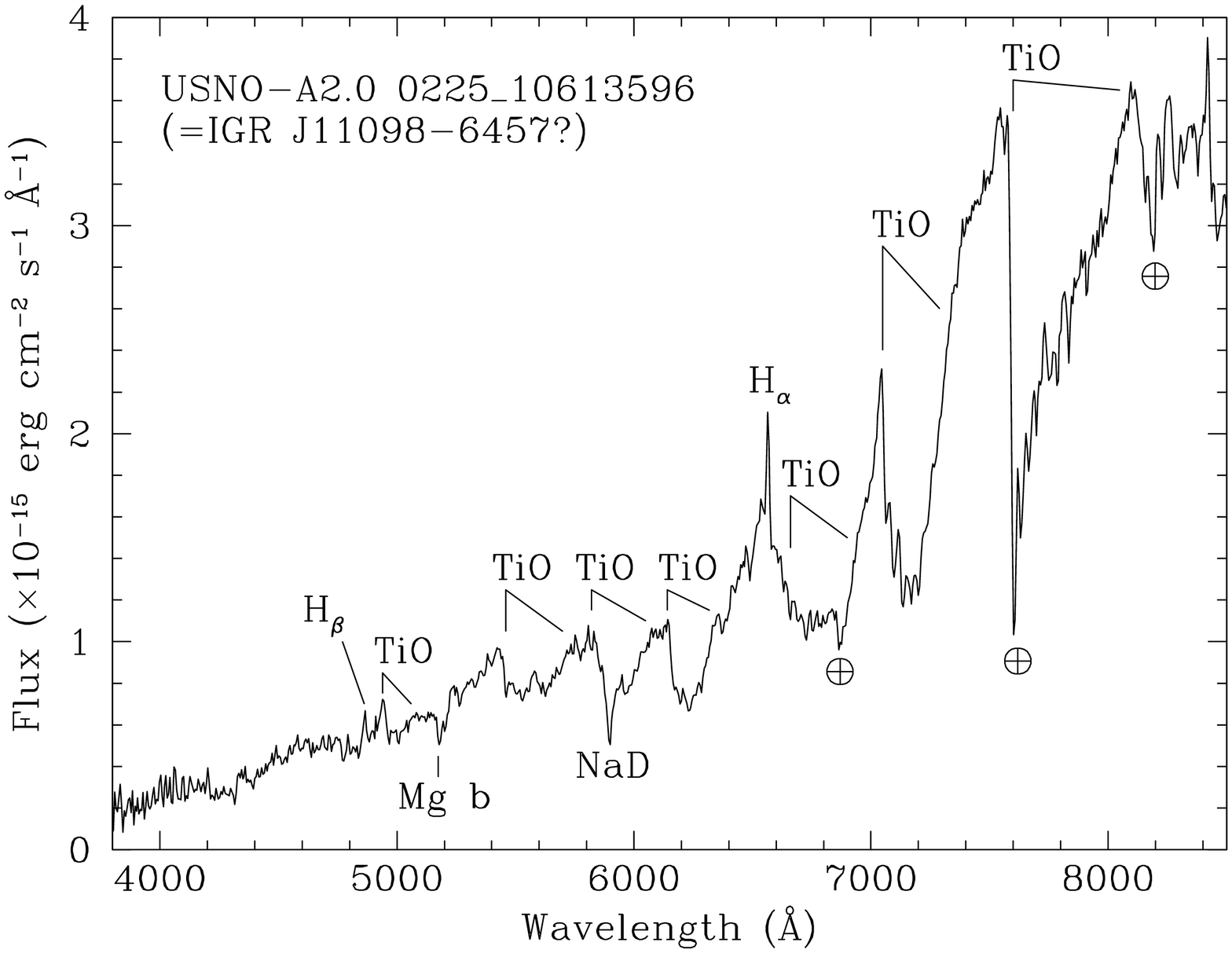,width=9cm,angle=270}
}
\hspace{0.5cm}
\parbox{7.5cm}{
\vspace{-.5cm}
\vspace{-.5cm}
\caption{Spectra (not corrected for the intervening Galactic absorption) 
of the optical counterparts of the CVs belonging to the sample of {\it 
INTEGRAL} sources presented in this paper: the 4 dwarf novae are reported 
in the upper and central panels, while the symbiotic star is shown in the 
lower left panel. For each spectrum the main spectral features are labeled. 
The symbol $\oplus$ indicates atmospheric telluric absorption bands.}
}
\end{figure*}

We identify 4 objects of our sample (IGR J08390$-$4833, IGR J18173$-$2509, 
IGR J18308$-$1232 and IGR J19267+1325) as dwarf nova CVs because of the 
characteristics of their optical spectra (Fig. 3). All of them show Balmer 
emission lines up to at least H$_\gamma$, as well as He {\sc i} and He 
{\sc ii} lines in emission. All of the detected lines are consistent with 
being at $z$ = 0, indicating that these objects lie within our Galaxy. In 
addition, source IGR J11098$-$6457 is identified as a symbiotic star given 
its optical spectral continuum, which shows the typical features of a red 
giant star with superimposed H$_\alpha$ and H$_\beta$ emissions, again at 
$z$ = 0 (Fig. 3, lower left panel). Our $R$-band photometry of the 
counterpart of IGR J18173$-$2509, described in Sect. 2, yields a magnitude 
$R$ = 17.2$\pm$0.1.

The main spectral features of these objects, and the main astrophysical 
parameters which can be inferred from the available optical and X--ray 
observational data, are given in Table 2. The X--ray luminosities listed 
in this Table for the various objects were computed using the fluxes 
reported in Voges et al. (1999), White et al. (2000), Bird et al. (2007), 
Landi et al. (2007b, 2008c,d), Sazonov et al. (2008), Ibarra et al. 
(2008a,b) and Tomsick et al. (2008b).

In the spectra of the sources identified here as dwarf nova CVs the He 
{\sc ii}$\lambda$4686/H$_\beta$ equivalent width (EW) ratio is $\ga$0.5 
and the EWs of He {\sc ii} and H$_\beta$ are around (or larger than) 10 
\AA: this indicates that these sources are quite likely magnetic CVs 
belonging to the Intermediate Polar (IP) subclass (see Warner 1995 and 
references therein). Indeed, for IGR J08390$-$4833 and IGR J19267+1325, 
Sazonov et al. (2008) and Evans et al. (2008), respectively, confirmed 
this hypothesis by measuring the likely spin period of the white dwarf 
(WD) harboured in these systems.

Our findings are consistent with the results reported in Revnivtsev et al. 
(2008) and Steeghs et al. (2008), thus confirming their identification of 
IGR J08390$-$4833 and IGR J19267+1325, respectively, as magnetic dwarf 
nova CVs. For the latter object, however, we measure an H$_\alpha$ EW 
which is a factor of $\sim$2 lower than the value reported by Steeghs et 
al. (2008), obtained from a spectrum acquired about one year before ours; 
this long-term emission line variability is however not unusual for IP CVs 
(e.g., Warner 1995).

For the case of IGR J11098$-$6457, using the 
Bruzual-Persson-Gunn-Stryker\footnote{available at:\\ {\tt 
ftp://ftp.stsci.edu/cdbs/grid/bpgs/}} (Gunn \& Stryker 1983) and 
Jacoby-Hunter-Christian\footnote{available at:\\ {\tt 
ftp://ftp.stsci.edu/cdbs/grid/jacobi/}} (Jacoby et al. 1984) 
spectroscopy atlases, we constrain the spectral type of its optical 
counterpart to be M2\,III. We also note a possible excess on the blue side 
of the optical continuum, which is a common feature in symbiotic stars.

From this spectral information, assuming colours and absolute $V$ 
magnitude of a M2\,III star (Ducati et al. 2001; Lang 1992) and 
considering the measured Balmer line ratio, we obtain a distance of 
$\sim$20 kpc for the source, which would place it well beyond the Carina 
arm of the Galaxy. This suggests that more absorption should occur along 
the line of sight and that this number should rather be used as an 
(admittedly loose) upper limit for the distance to this object. Indeed, if 
one assumes the total Galactic colour excess along the source line of 
sight, $E(B-V)_{\rm Gal}$ = 1.042 mag (Schlegel et al. 1998), we obtian a 
distance of $\sim$5.2 kpc to the source, which is well within the 
aforementioned Galactic arm.

We add that, as reported in Landi et al. (2008d), a further soft X--ray 
source is detected with {\it Swift}/XRT in the field of IGR J11098$-$6457, 
although formally outside of the IBIS error circle. While further 
investigation is needed to understand whether this is the actual soft 
X--ray counterpart of the {\it INTEGRAL} object, the {\it Swift}/XRT 
observation of Landi et al. (2008d) nevertheless indicates that X--ray 
activity up to 10 keV is produced by the symbiotic star identified here.

\begin{table*}[th!]
\caption[]{Synoptic table containing the main results concerning the 5 CVs
(see Fig. 3) identified in the present sample of {\it INTEGRAL} sources.}
\scriptsize
\vspace{-.3cm}
\begin{center}
\begin{tabular}{lcccccccccr}
\noalign{\smallskip}
\hline
\hline
\noalign{\smallskip}
\multicolumn{1}{c}{Object} & \multicolumn{2}{c}{H$_\alpha$} & 
\multicolumn{2}{c}{H$_\beta$} & \multicolumn{2}{c}{He {\sc ii} $\lambda$4686} & 
$R$ & $A_V$ & $d$ & \multicolumn{1}{c}{$L_{\rm X}$} \\
\cline{2-7}
\noalign{\smallskip} 
 & EW & Flux & EW & Flux & EW & Flux & mag & (mag) & (pc) & \\

\noalign{\smallskip}
\hline
\noalign{\smallskip}

IGR J08390$-$4833 & 27$\pm$3 & 8.6$\pm$0.8 & 8.7$\pm$0.9 & 4.1$\pm$0.4 & 5.2$\pm$0.8 & 
2.6$\pm$0.4 & 16.6 & $\sim$0 & $\sim$330 & 0.46 (0.5--8) \\
 & & & & & & & & & & 0.99 (17--60) \\

& & & & & & & & & & \\ 

IGR J11098$-$6457 & 5.2$\pm$0.5 & 8.0$\pm$0.8 & 3.6$\pm$0.7 & 1.9$\pm$0.4 & $<$2.0 & $<$1.0 & 
15.6 & 1.23 & $<$20000 & $<$33 (2--10) \\
 & & & & & & & & & & $<$6200 (20--100) \\

& & & & & & & & & & \\ 

IGR J18173$-$2509 & 92$\pm$3 & 25.0$\pm$0.8 & 14.0$\pm$1.0 & 6.9$\pm$0.5 & 7.0$\pm$1.0 & 
3.6$\pm$0.5 & 17.2 & 0.73 & $\sim$330 & 0.35--0.49 (0.2--12) \\
 & & & & & & & & & & 1.7 (2--10) \\ 
 & & & & & & & & & & 2.1 (20--100) \\ 

& & & & & & & & & \\ 

IGR J18308$-$1232 & 21.9$\pm$1.1 & 7.7$\pm$0.4 & 8.5$\pm$0.9 & 2.2$\pm$0.2 & 3.9$\pm$0.9 & 
9.3$\pm$2.0 & 17.0 & 0.67 & $\sim$320 & 0.32 (0.2--12) \\
 & & & & & & & & & & 0.060 (0.24--2.0) \\ 
 & & & & & & & & & & 1.8 (20--100) \\ 

& & & & & & & & & \\ 

IGR J19267+1325 & 63.6$\pm$1.9 & 26.9$\pm$0.8 & 45$\pm$3 & 11.8$\pm$0.8 & 22$\pm$2 & 
5.3$\pm$0.5 & 17.8 & $\sim$0 & $\sim$580 & 0.13 (0.1--2.4) \\
 & & & & & & & & & & 4.0 (0.3--10) \\ 
 & & & & & & & & & & 3.3 (2--10) \\ 
 & & & & & & & & & & 4.0 (20--100) \\ 

\noalign{\smallskip} 
\hline
\noalign{\smallskip} 
\multicolumn{11}{l}{Note: EWs are expressed in \AA, line fluxes are
in units of 10$^{-15}$ erg cm$^{-2}$ s$^{-1}$, whereas X--ray luminosities
are in units of 10$^{32}$ erg s$^{-1}$ and the} \\
\multicolumn{11}{l}{reference band (between brackets) is expressed in keV.} \\
\noalign{\smallskip} 
\hline
\hline
\noalign{\smallskip} 
\end{tabular} 
\end{center}
\end{table*}

\subsection{X--ray binaries}

\begin{figure*}[th!]
\mbox{\psfig{file=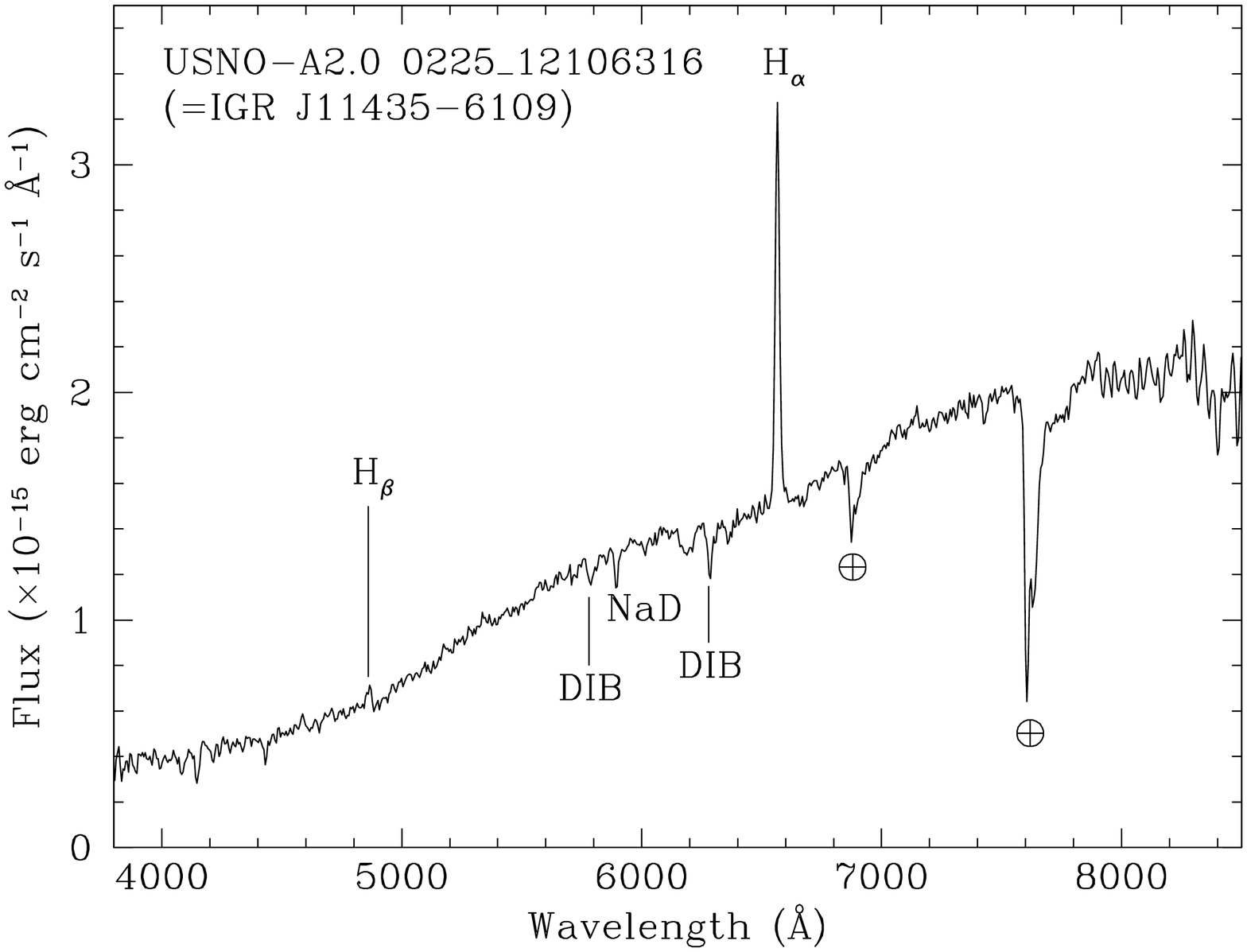,width=9cm,angle=270}}
\mbox{\psfig{file=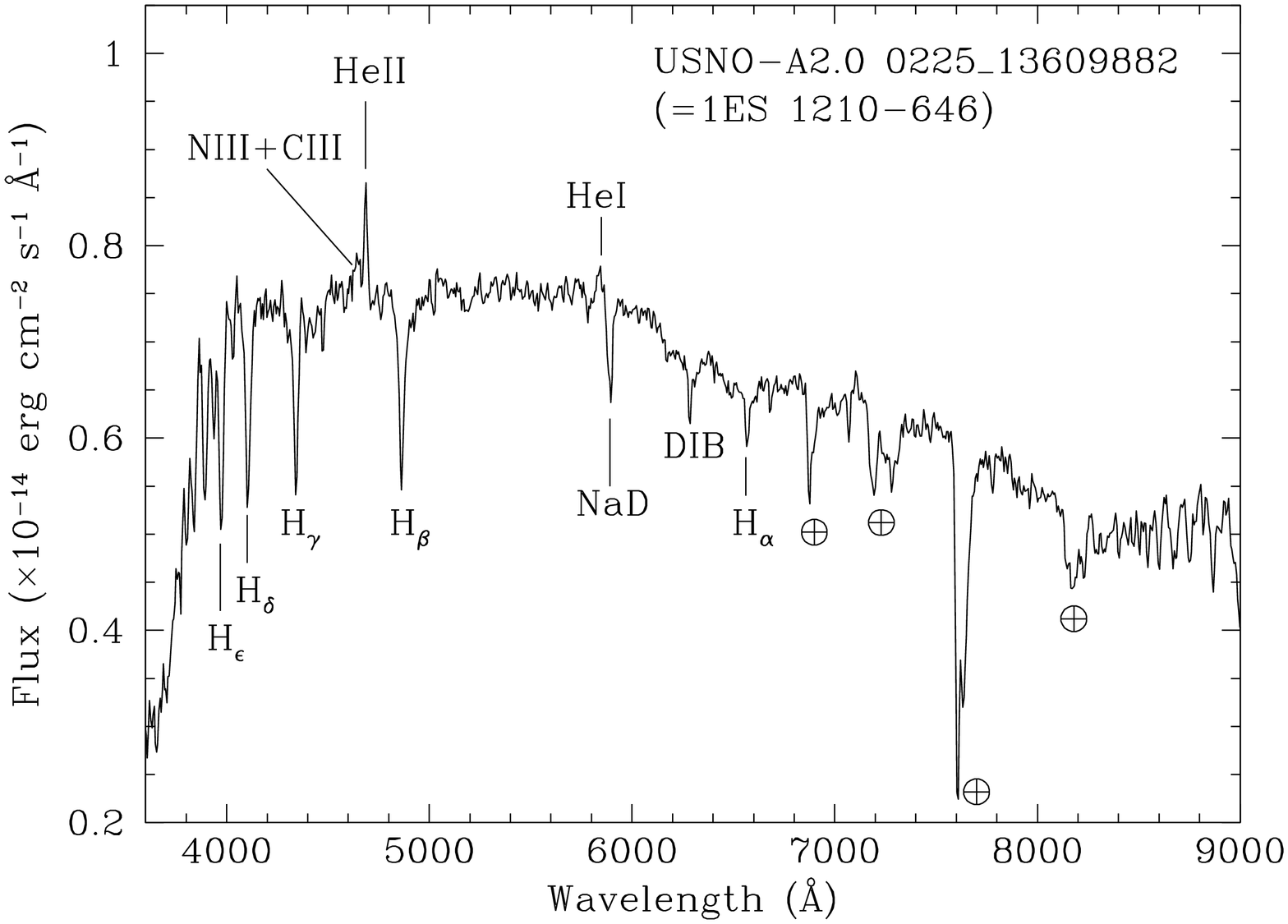,width=9cm,angle=270}}

\vspace{-.9cm}
\mbox{\psfig{file=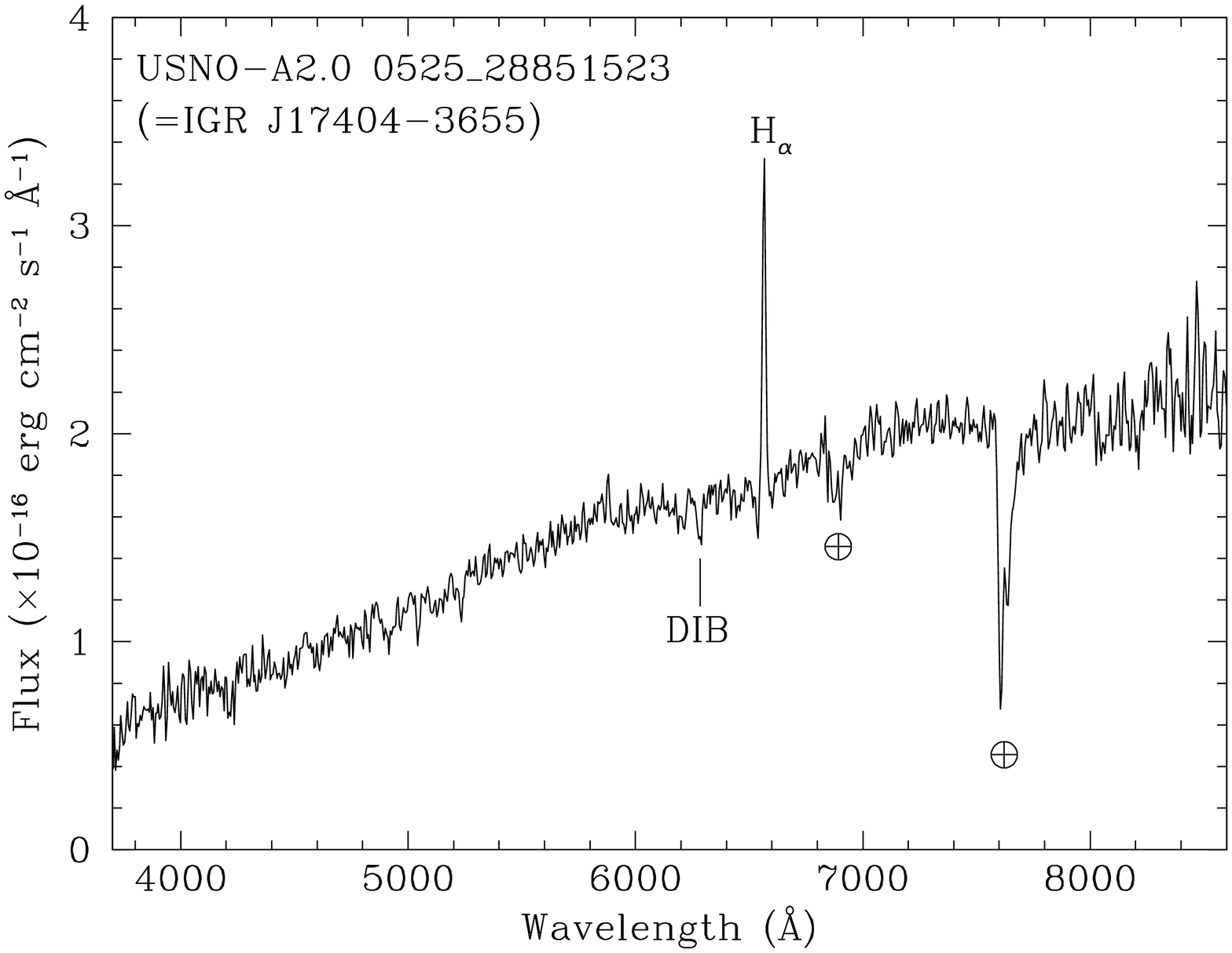,width=9cm,angle=270}}
\mbox{\psfig{file=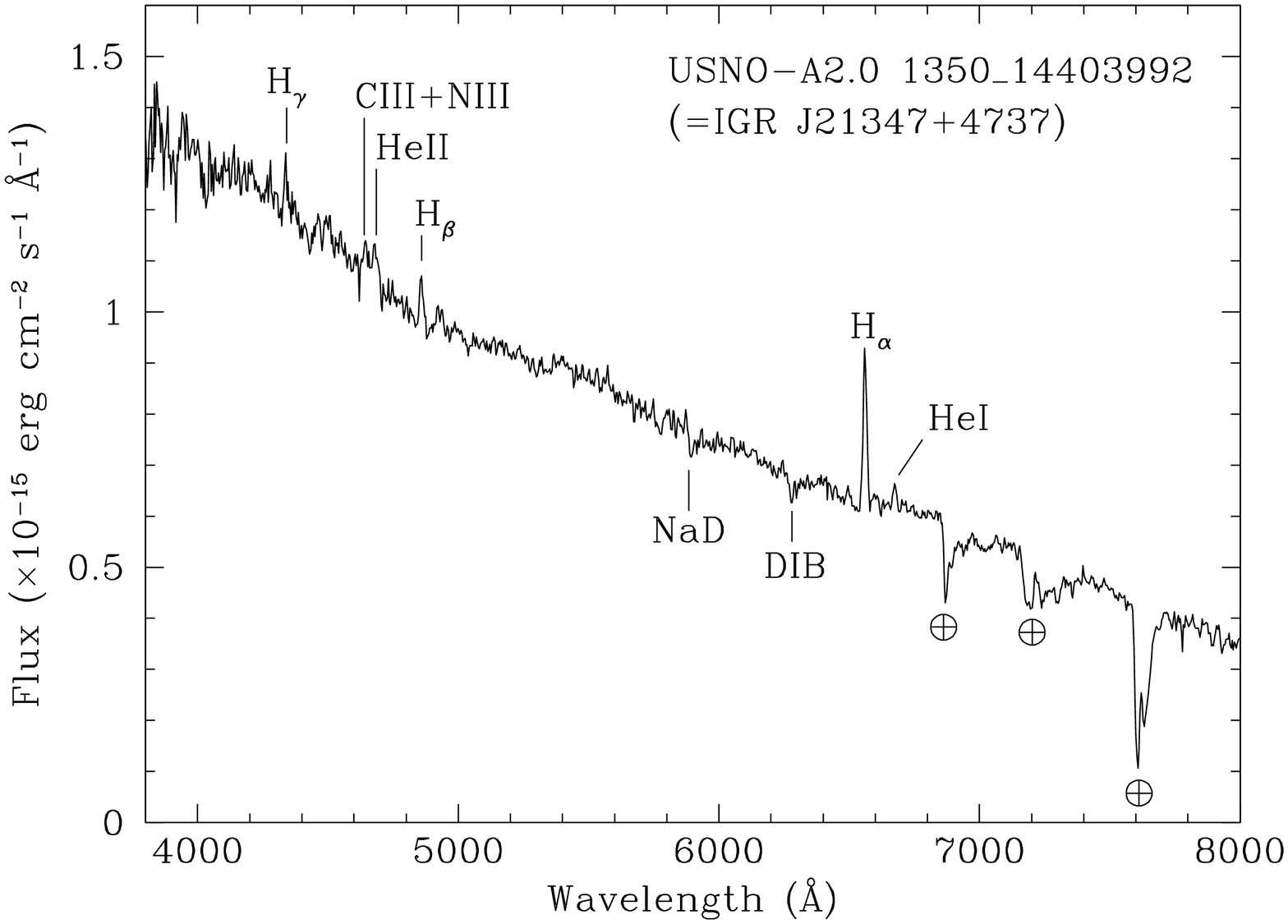,width=9cm,angle=270}}
\vspace{-.5cm}
\caption{Spectra (not corrected for the intervening Galactic absorption) 
of the optical counterparts of the X--ray binaries belonging to
the sample of {\it INTEGRAL} sources presented in this paper.
For each spectrum the main spectral features are labeled. The 
symbol $\oplus$ indicates atmospheric telluric absorption bands.}
\end{figure*}

\begin{table*}[th!]
\caption[]{Synoptic table containing the main results concerning the 4 
X--ray binaries (see Fig. 4) identified or observed in the present sample 
of {\it INTEGRAL} sources.}
\scriptsize
\vspace{-.5cm}
\begin{center}
\begin{tabular}{lccccccccccr}
\noalign{\smallskip}
\hline
\hline
\noalign{\smallskip}
\multicolumn{1}{c}{Object} & \multicolumn{2}{c}{H$_\alpha$} & 
\multicolumn{2}{c}{H$_\beta$} &
\multicolumn{2}{c}{He {\sc ii} $\lambda$4686} &
Optical & $A_V$ & $d$ & Spectral & \multicolumn{1}{c}{$L_{\rm X}$} \\
\cline{2-7}
\noalign{\smallskip} 
 & EW & Flux & EW & Flux & EW & Flux & mag. & (mag) & (kpc) & type & \\

\noalign{\smallskip}
\hline
\noalign{\smallskip}

IGR J11435$-$6109 & 25.8$\pm$1.3 & 39$\pm$2 & 3.5$\pm$1.1 & 2.2$\pm$0.7 & $<$2.2 & $<$1.2 &
16.43$^{\rm a}$ ($V$) & 5.7 & $\sim$8.6 & B2\,III & 2.5 (0.16--3.5) \\
& & & & & & & & & & or B0\,V & 8.2 (0.3--10) \\
& & & & & & & & & & & 16 (20--100) \\

& & & & & & & & & & & \\ 

1ES 1210$-$646 & in abs. & in abs. & in abs. & in abs. & 2.8$\pm$0.3 & 21$\pm$2 &
13.9 ($R$) & 3.3 & $\sim$2.8 & B5\,V & 1.9 (0.16--3.5) \\
& & & & & & & & & & & 64 (1--8) \\
& & & & & & & & & & & 12 (2--10) \\
& & & & & & & & & & & 1.0 (20--100) \\

& & & & & & & & & & & \\ 

IGR J17404$-$3655 & 14.2$\pm$0.14 & 2.5$\pm$0.3 & $<$9 & $<$0.9 & $<$6 & $<$0.6 &
17.3 ($R$) & 3.1 & $\sim$9.1 & --- & 12 (2--10) \\
& & & & & & & & & & & 13 (20--100) \\

& & & & & & & & & & & \\ 

IGR J21347+4737 & 8.1$\pm$0.6 & 5.1$\pm$0.4 & 1.2$\pm$0.2 & 1.2$\pm$0.2 & 1.6$\pm$0.4 & 1.7$\pm$0.4 &
14.2 ($R$) & 2.2 & $\sim$5.8 & B3\,V & 0.08 (0.5--8) \\
& & & & & & & & & & & 6.0 (20--100) \\

\noalign{\smallskip} 
\hline
\noalign{\smallskip}
\multicolumn{12}{l}{Note: EWs are expressed in \AA, line fluxes are
in units of 10$^{-15}$ erg cm$^{-2}$ s$^{-1}$, whereas X--ray luminosities
are in units of 10$^{34}$ erg s$^{-1}$ and the reference} \\
\multicolumn{12}{l}{band (between brackets) is expressed in keV.} \\
\multicolumn{12}{l}{$^{\rm a}$: from Negueruela et al. (2007)} \\
\noalign{\smallskip} 
\hline
\hline
\end{tabular} 
\end{center} 
\end{table*}

Four of the {\it INTEGRAL} sources selected here (IGR J11435$-$6109, 1ES 
1210$-$646, IGR J17404$-$3655 and IGR J21347+4737) can be classified as 
Galactic X--ray binaries by their overall spectral appearance (see Fig. 
4), which is typical of this class of objects (see e.g. Papers I-VI), with 
narrow Balmer and/or helium emission lines at a wavelength consistent with 
that of the laboratory restframe, superimposed on an intrinsically blue 
continuum.

The spectral shape of all of these sources but IGR J21347+4737 appears 
however substantially reddened, implying the presence of interstellar dust 
along the line of sight. This is quite common in X--ray binaries detected 
with {\it INTEGRAL} (e.g., Paper IV-VI) and indicates that these objects 
are relatively far from Earth. The confirmation of the presence of 
reddening along the line of sight towards these sources comes from either 
the observed H$_\alpha$/H$_\beta$ line ratio (see Table 3) or the observed 
optical colours. In all cases we indeed find a reddening compatible with 
the Galactic one along the line of sight of the object (Schlegel et al. 
1998).

Three of the sources of our sample (IGR J11435$-$6109, 1ES 1210$-$646 and 
IGR J21347+4737) appear to be HMXBs due to the optical spectral features 
readily detected, and which point to a early-type star as the optical 
counterpart of these hard X--ray sources. These findings confirm the 
results of Negueruela et al. (2007) regarding IGR J11435$-$6109.

A word of caution should however be spent about the classification of 1ES 
1210$-$646: although our optical spectroscopy excludes that this system is 
a magnetic dwarf nova as suggested by Revnivtsev et al. (2007) on the 
grounds of its {\it Swift}/XRT X--ray spectrum, the detection of an iron 
line at 6.7 keV reported by these authors is rarely seen in accreting 
compact objects (such as neutron stars and black holes) hosted in HMXBs 
(but see Rib\'o et al. 1999, Burderi et al. 2000, Paul et al. 2002 and 
Naik \& Paul 2003 for detections of this line in HMXBs). However, the 
hypothesis of a WD as the accretor also has problems. The only case (to 
the best of our knowledge) of an X--ray binary composed of an early type 
star and a white dwarf, i.e. CI Cam (see e.g. Orlandini et al. 2000, 
Filippova et al. 2008, and references therein) shows a completely 
different behaviour, both in terms of secondary star nature and persistent 
X--ray emission: indeed, CI Cam has a supergiant B[e] star as the optical 
companion (Hynes et al. 2002); moreover, its low-level 2--10 keV 
luminosity (Boirin et al. 2002) is at least 2 orders of magnitude fainter 
than that of 1ES 1210$-$646 (see Table 3). Thus, we still suggest that 
this source is an HMXB, although with peculiarities which certainly 
deserve further multiwavelength studies.

We can instead rule out an HMXB nature for IGR J17404$-$3655 because its 
optical magnitudes do not fit any star of early spectral type, not 
even if absorption along the line of sight is considered (see Table 3). It 
also does not show features of a red giant star (as, for instance, in the 
case of IGR J11098$-$6457); rather, its reddened spectrum suggests that it 
is actually substantially absorbed by the Galactic dust, indicating 
that this object lies far from Earth. Besides, its emission-line spectral 
appearance is quite different from that of dwarf nova CVs in Fig. 3.
Because of all this, we can exclude that it is a CV. We thus identify 
this hard X--ray source as an LMXB.

Table 3 collects the relevant optical spectral information on these 4 
sources, along with their main parameters inferred from the available 
X--ray and optical data. X--ray luminosities in Table 3 were calculated 
using the fluxes in Forman et al. (1978), Elvis et al. (1992), Thompson 
et al. (1998), Reynolds et al. (1999), Bird et al. (2007), Landi et al. 
(2008e), Sazonov et al. (2008) and Tomsick et al. (2008a).

For the HMXBs detected in our sample we obtained the constraints 
for distance, reddening, spectral type and X--ray luminosity shown 
in Table 3 by considering the absolute magnitudes of early-type stars 
and by applying the method described in Paper III for the classification 
of source 2RXP J130159.6$-$635806.
Given the information on the optical magnitudes and the line-of-sight
reddening (see Table 3), we can state that none of these 3 objects hosts
a blue supergiant as a secondary star, as this would place them far
outside of the Galaxy. 

Going now into detail for some of these sources, we note that inspection 
of the optical spectrum of 1ES 1210$-$646 shows that the entire Balmer 
series appears in absorption. However, the presence of He {\sc i}, 
He {\sc ii} and N {\sc iii} + C {\sc iii} (Bowen blend) emissions makes us 
confident that this object is the optical counterpart of this X--ray 
source. We also note that the H$_\alpha$ absorption is not very pronounced 
with respect to the other Balmer absorption lines: indeed, its EW is 
around 2.5 \AA, which is less than half of the value expected from a B5\,V 
star (Jaschek \& Jaschek 1987). This may hint at the presence of a hidden 
H$_\alpha$ emission which partially fills the corresponding absorption 
line. The dereddened optical spectral continuum of 1ES 1210$-$646 is 
nevertheless consistent with that of an intermediate-type, main-sequence B 
star.

Concerning the counterpart of IGR J21347+4737, we detect a number of 
emission lines in its optical spectrum: while this is the confirmation 
that this object is the optical counterpart of the aforementioned {\it 
INTEGRAL} source, the optical spectral appearance is completely different 
from the findings of Bikmaev et al. (2008a), who observed this source 
about one year earlier than we did. This suggests that the circumstellar 
disk around the optical companion formed again during this time lapse. 
This behaviour is not surprising, given the marked X--ray variability 
of this HMXB (Sazonov et al. 2008; Bikmaev et al. 2008a). Due to the large 
variability of the Balmer lines, for this source we used its USNO-A2.0 $B$ 
and $R$ magnitudes to determine its reddening, as we do not expect them to 
vary substantially with time given the HMXB nature of this source.

It is moreover noted that IGR J11435$-$6109 has a hydrogen column density 
$N_{\rm H}$, inferred by Tomsick et al. (2008a) from the source X--ray 
spectrum, which is $\sim$15 times higher than the one derived from the 
optical reddening using the empirical formula of Predehl \& Schmitt 
(1995). This is often observed in absorbed HMXBs detected with {\it 
INTEGRAL} (e.g., Chaty 2008) and suggests the presence of additional 
absorbing material in the vicinity of the X--ray source, likely due to the 
accretion stream flowing onto the compact object in this X--ray system.
Regarding this source, we also point out that both of our possible 
optical spectral classifications for the companion star fit with the 
position occupied by this object in the Be/X locus of the Corbet diagram 
(Corbet 1986) on the basis of its spin and orbital periods (Swank \& 
Markwardt 2004; Corbet \& Remillard 2005).

None of these systems is associated with a radio source. This means that 
none of them is likely to be a jet-emitting X--ray binary (i.e., a 
microquasar).

\subsection{AGNs}

\begin{figure*}
\mbox{\psfig{file=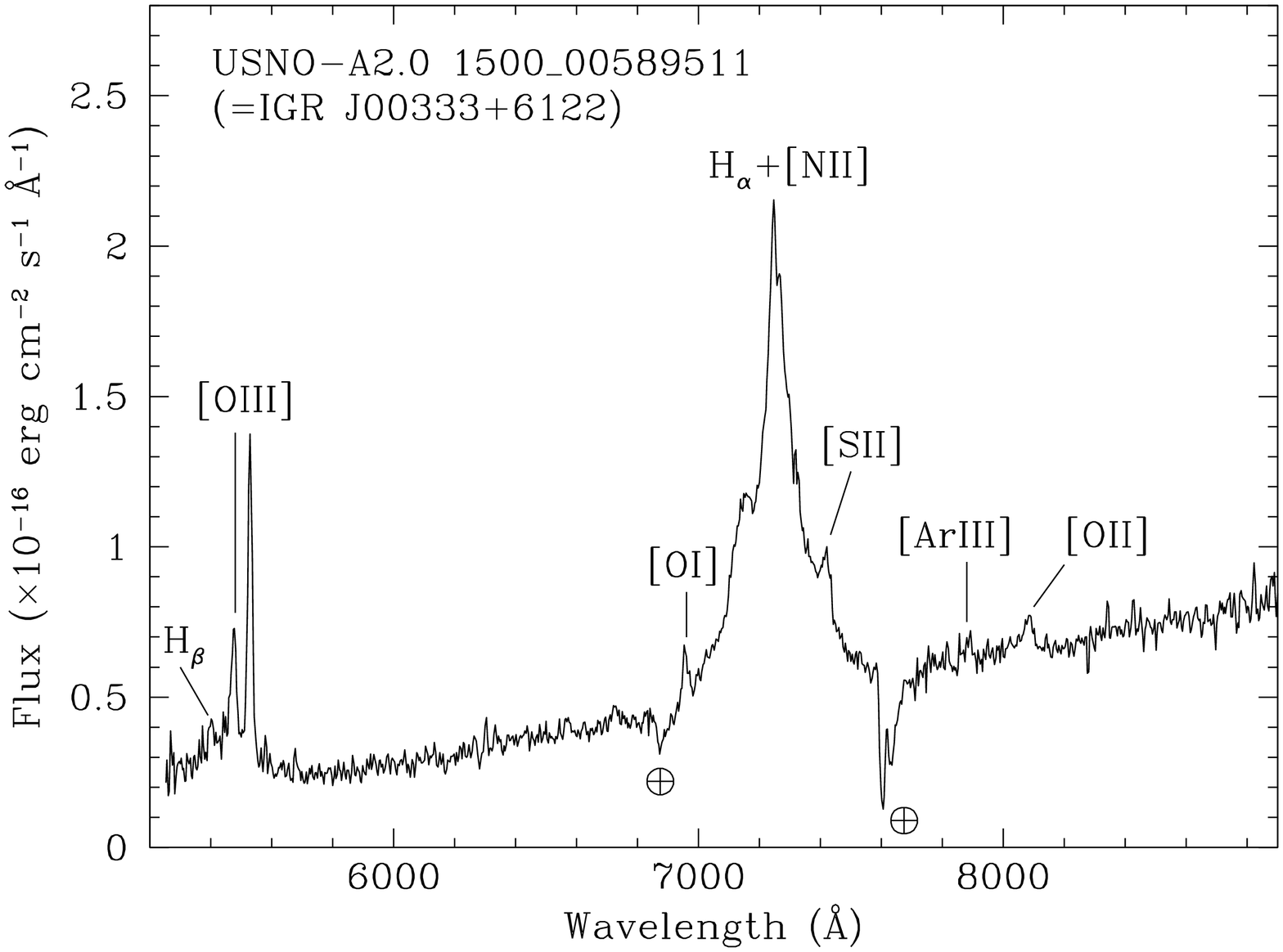,width=9cm,angle=270}}
\mbox{\psfig{file=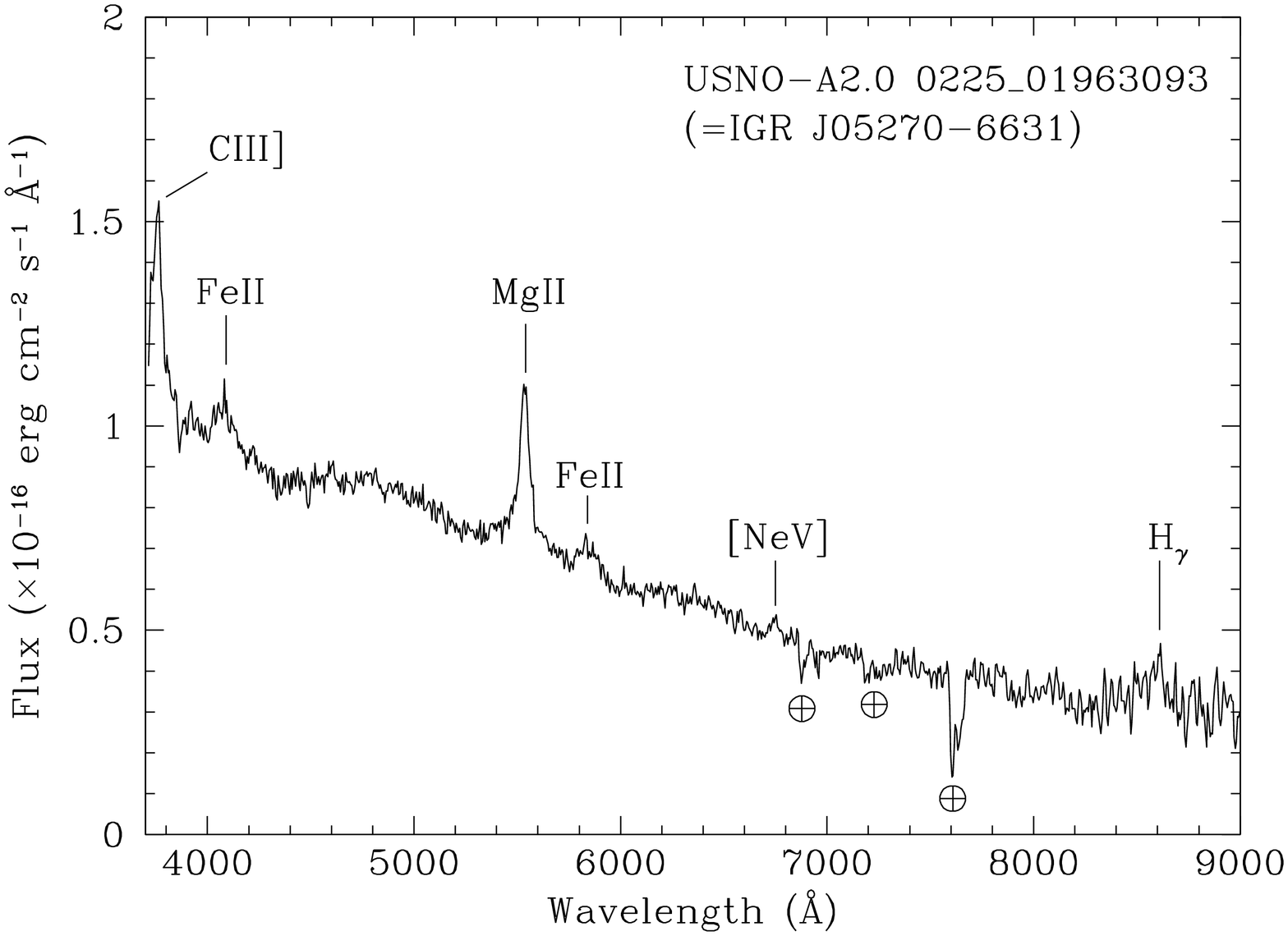,width=9cm,angle=270}}

\vspace{-.9cm}
\mbox{\psfig{file=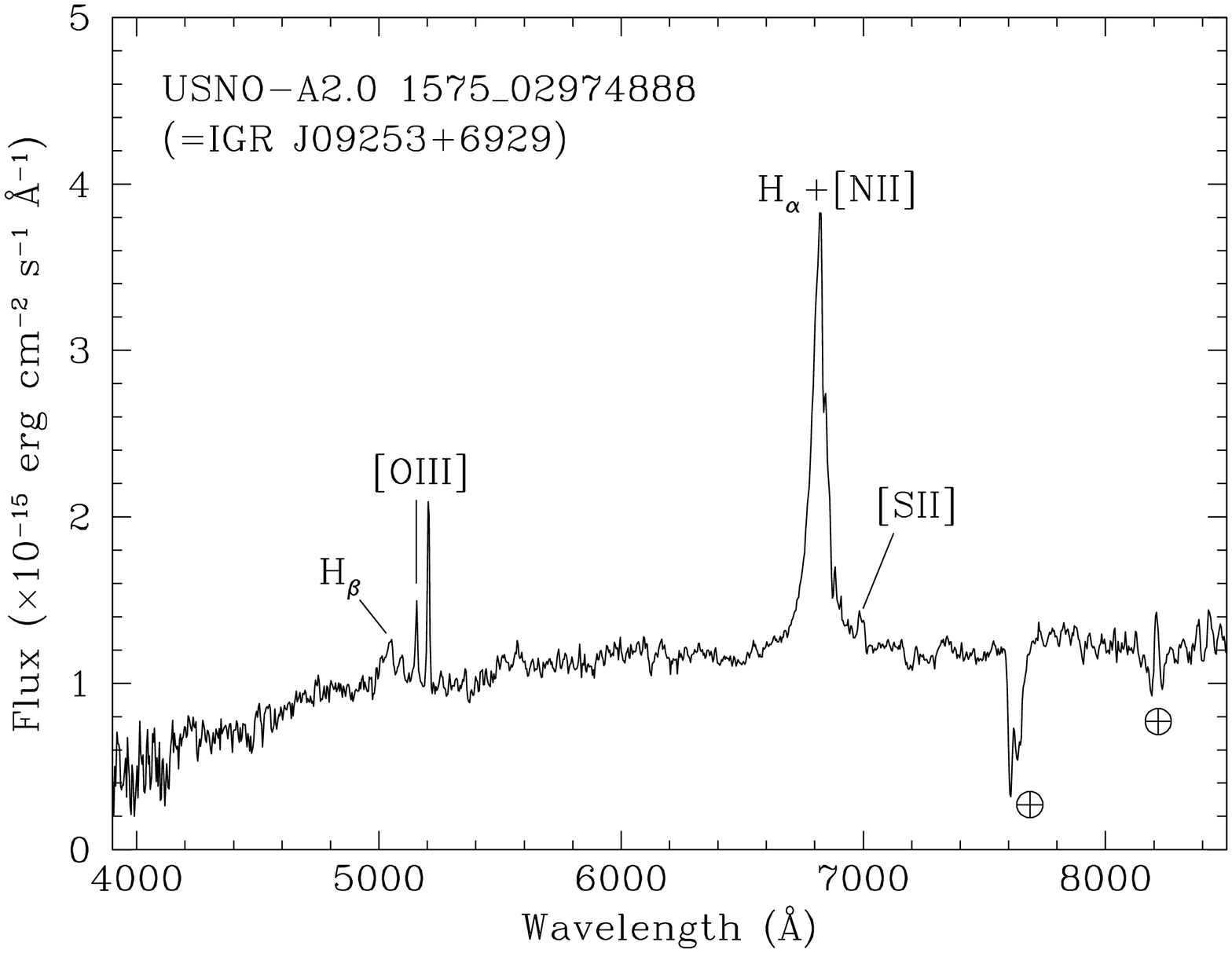,width=9cm,angle=270}}
\mbox{\psfig{file=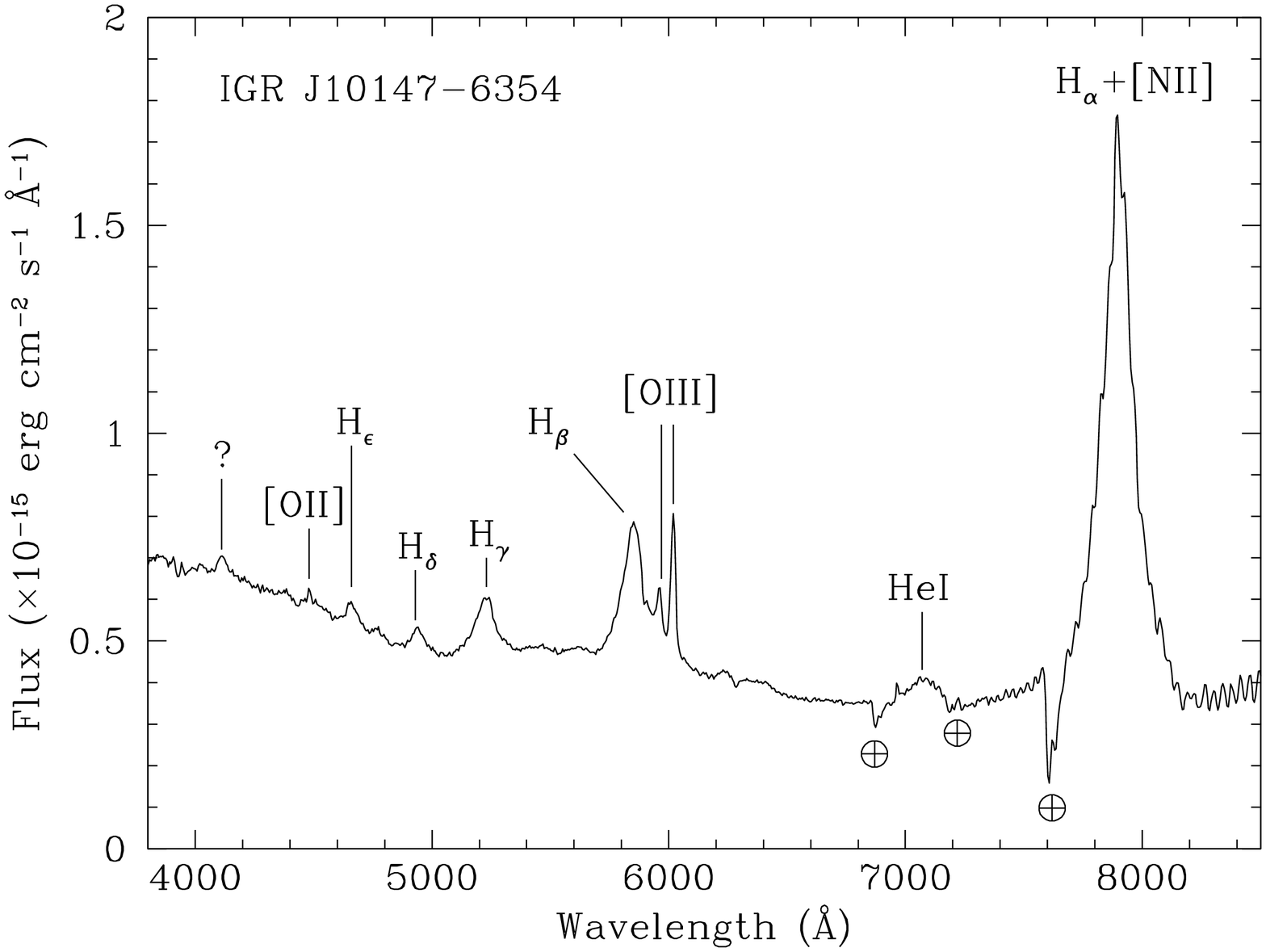,width=9cm,angle=270}}

\vspace{-.9cm}
\mbox{\psfig{file=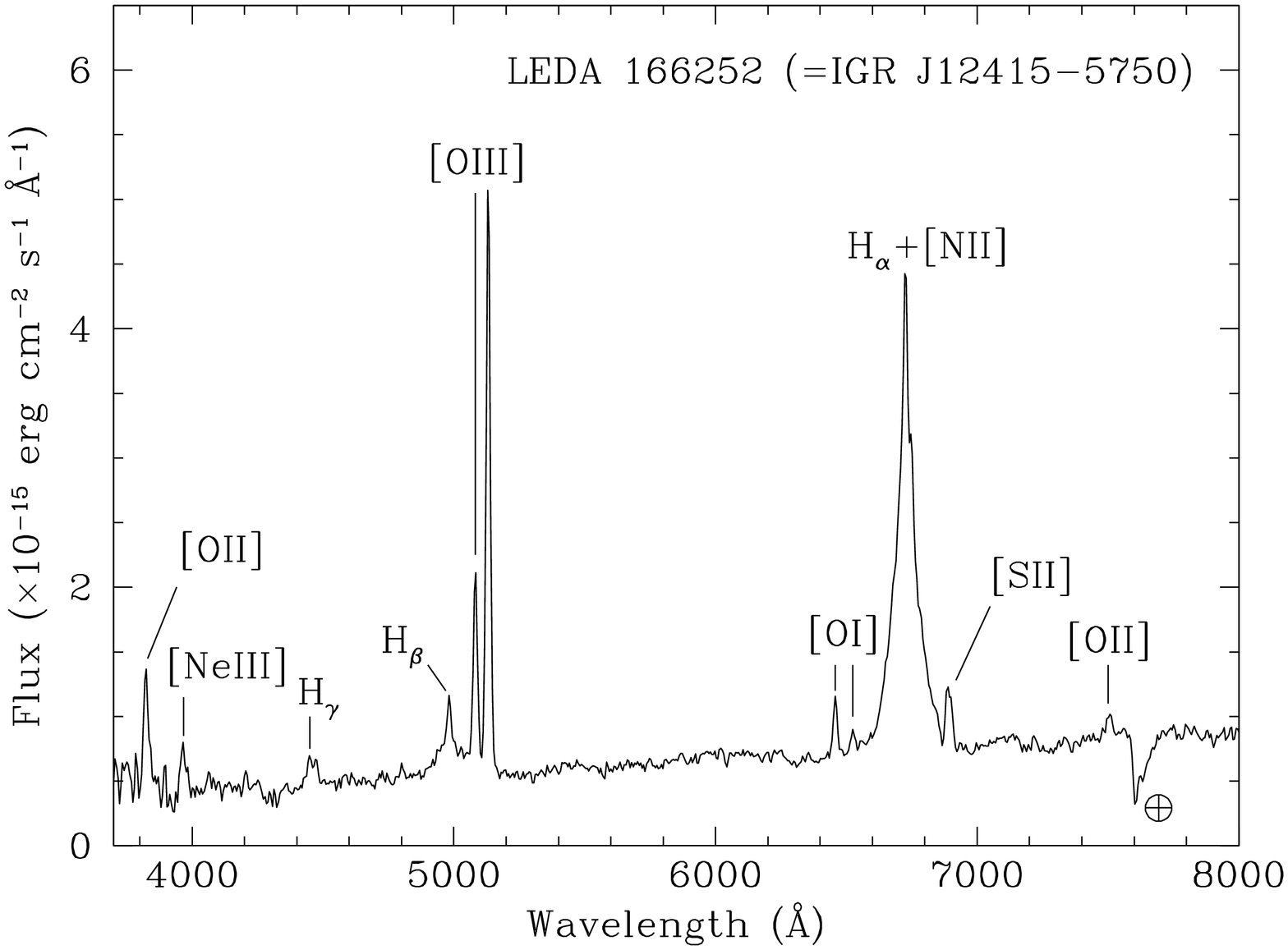,width=9cm,angle=270}}
\mbox{\psfig{file=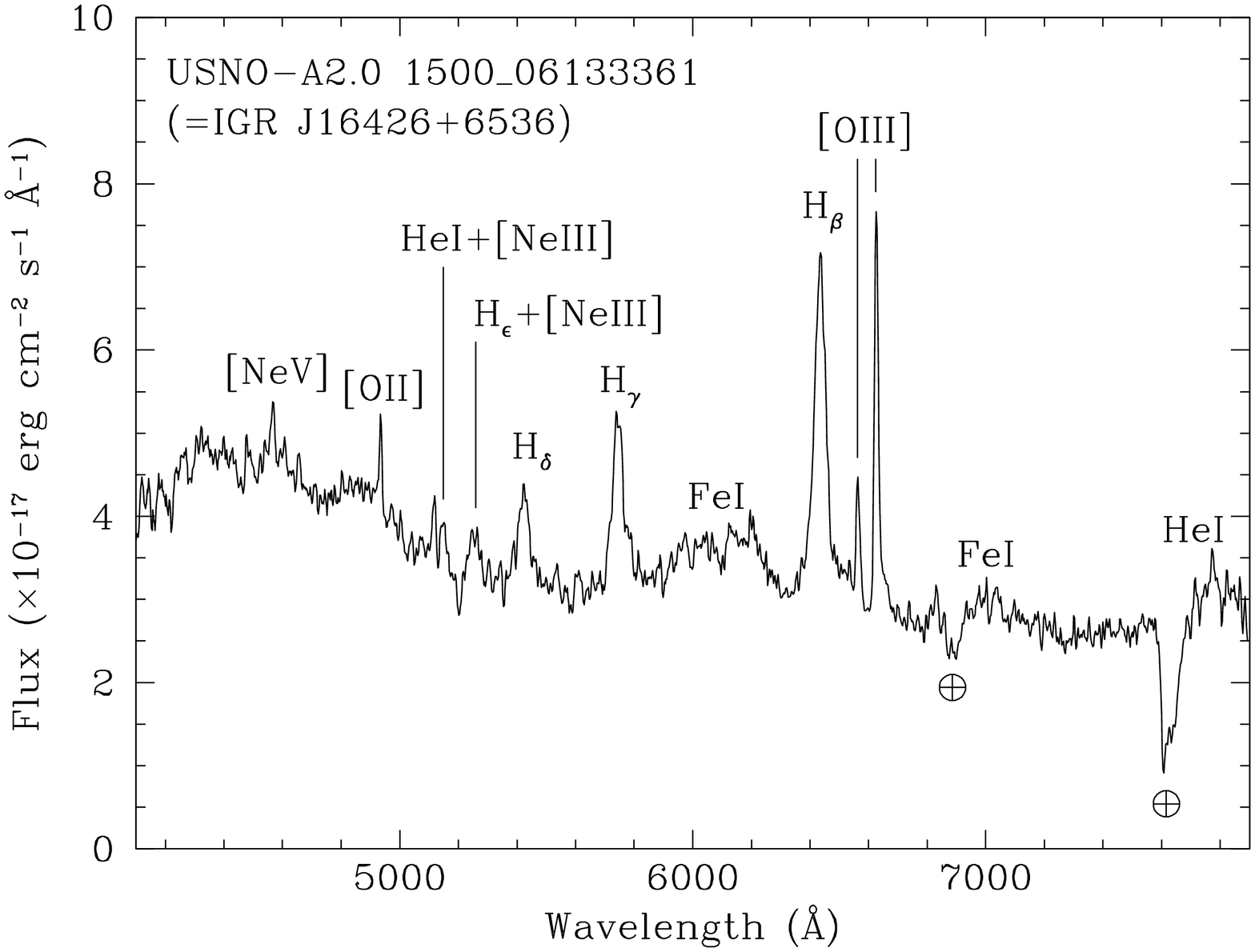,width=9cm,angle=270}}

\vspace{-.9cm}
\parbox{9cm}{
\psfig{file=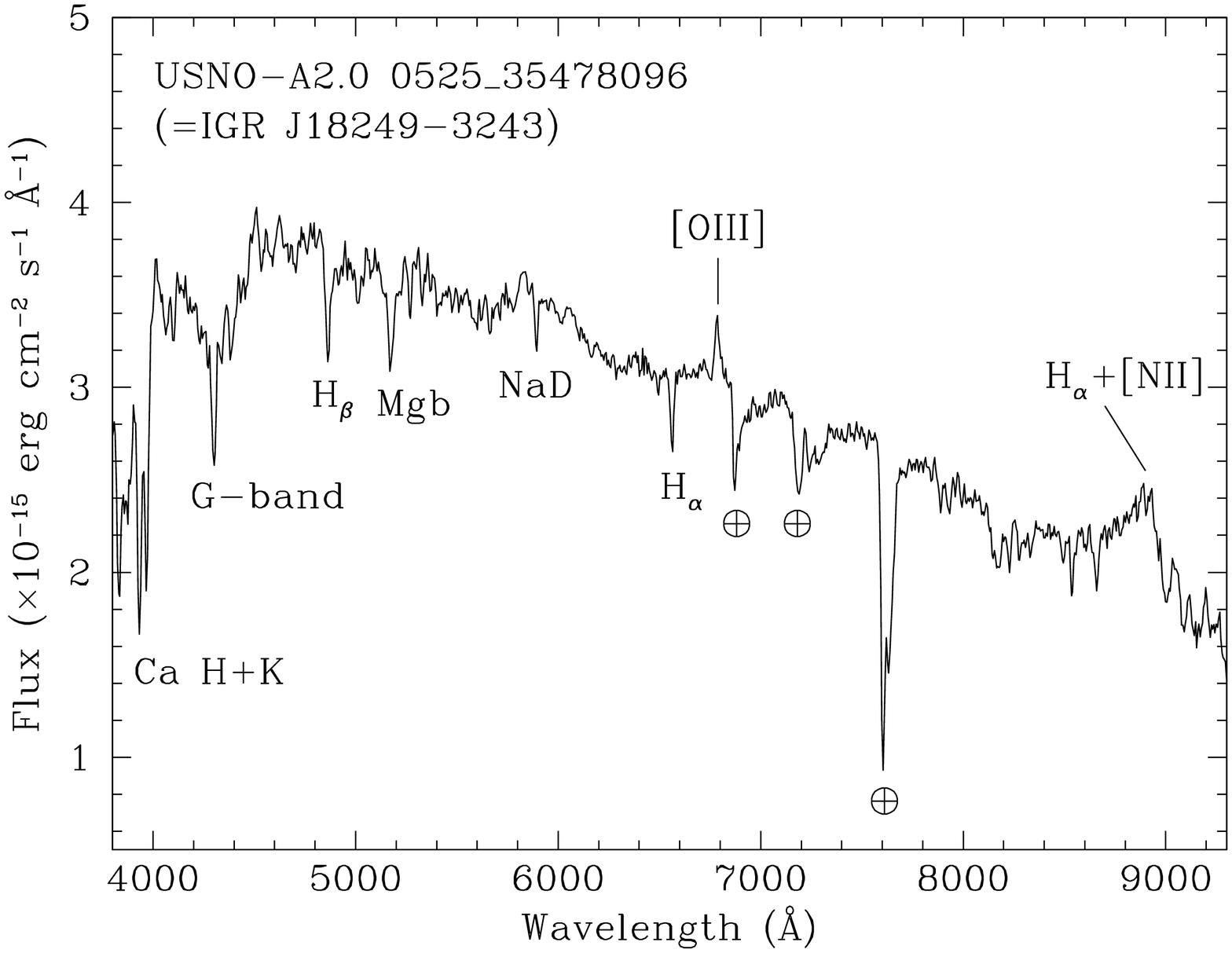,width=9cm,angle=270}
}
\hspace{.5cm}
\parbox{7.5cm}{
\vspace{1.5cm}
\caption{Spectra (not corrected for the intervening Galactic absorption) 
of the optical counterparts of the 7 broad emission line AGNs belonging to 
the sample of {\it INTEGRAL} sources presented in this paper.
For each spectrum the main spectral features are labeled. The 
symbol $\oplus$ indicates atmospheric telluric absorption bands.
The TNG spectrum has been smoothed using a Gaussian filter with
$\sigma$ = 3 \AA.
\vspace{1.5cm}
}
}
\end{figure*}

\begin{figure*}
\mbox{\psfig{file=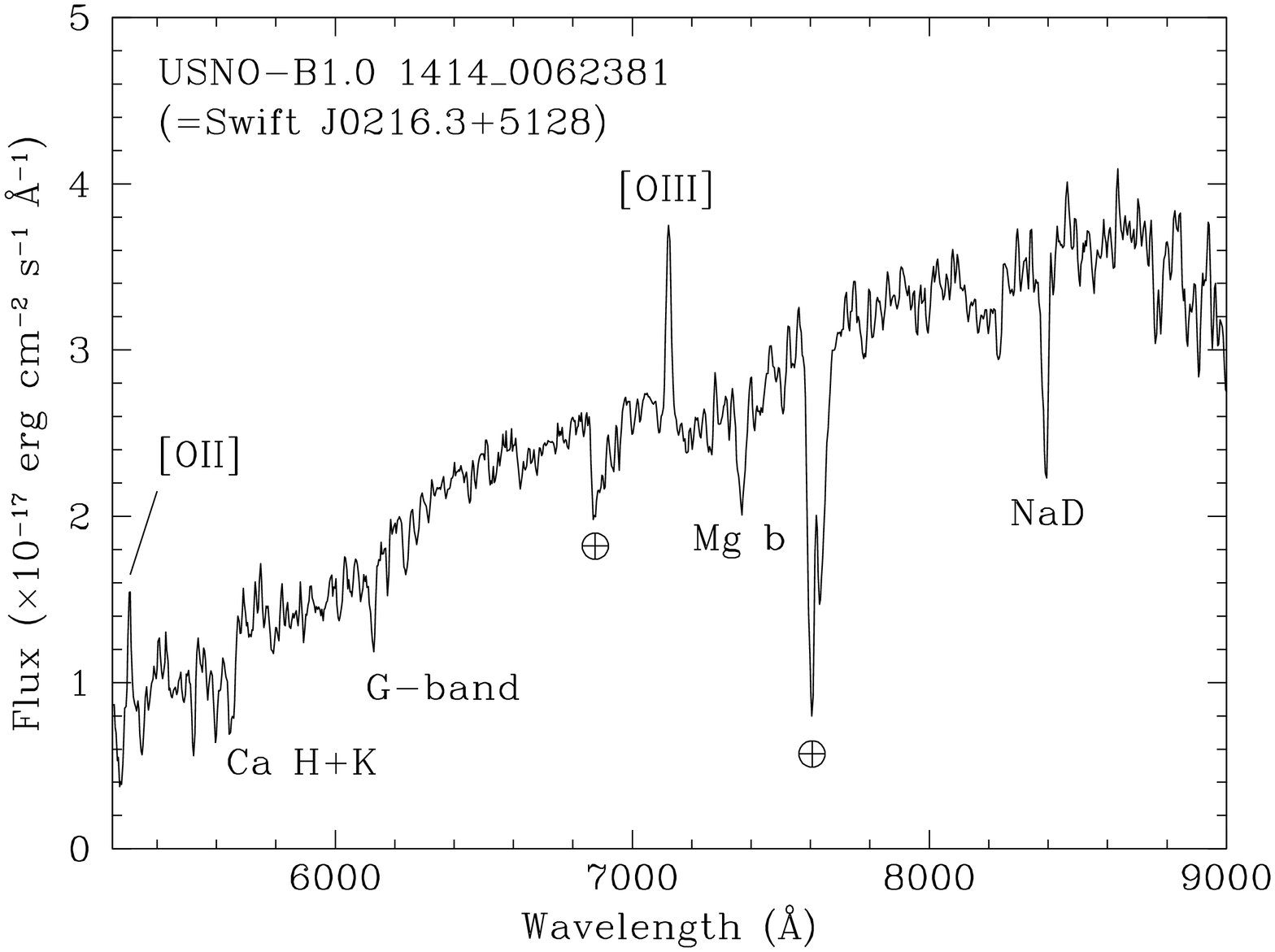,width=9cm,angle=270}}
\mbox{\psfig{file=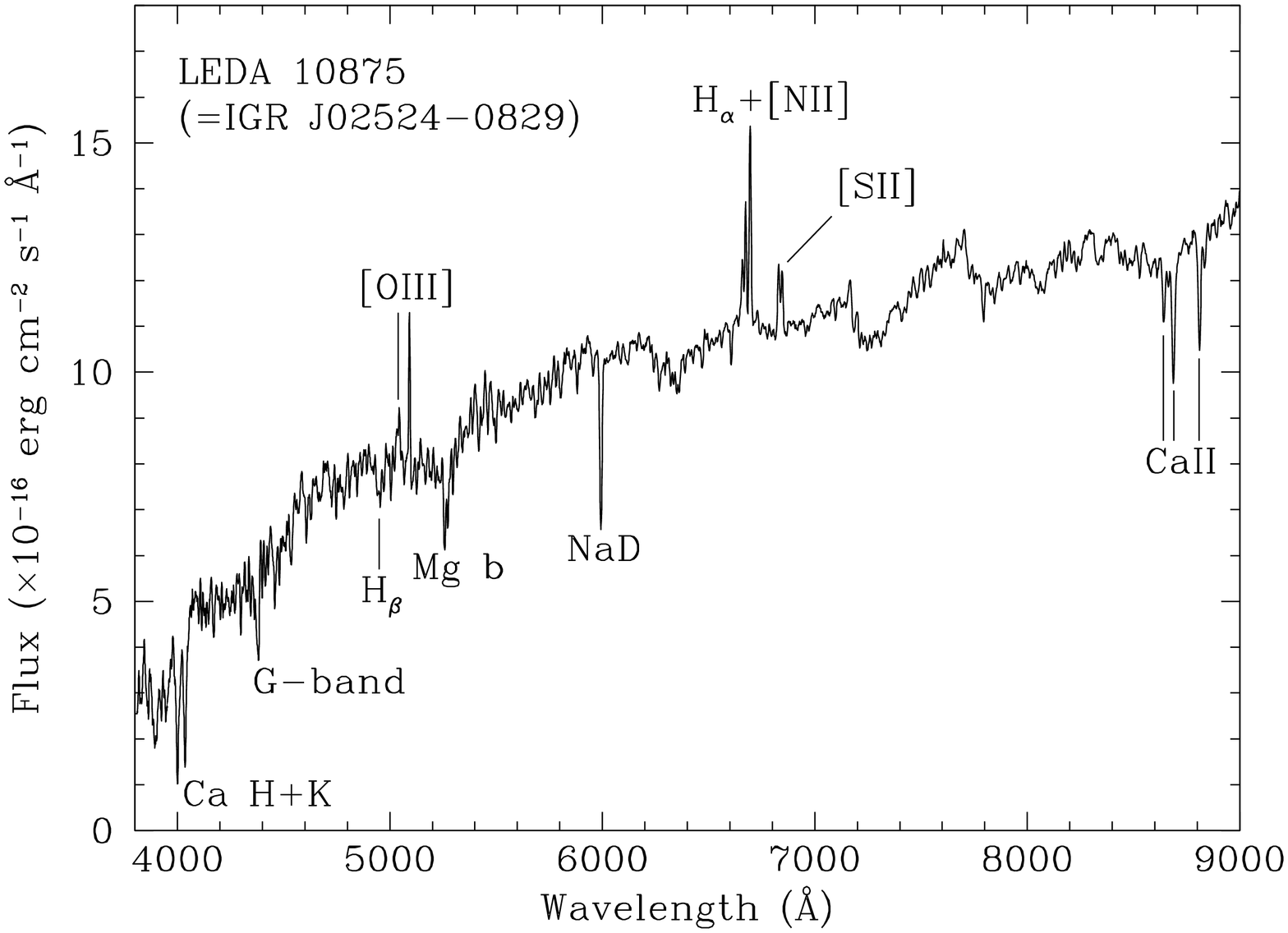,width=9cm,angle=270}}
\vspace{-.5cm}
\caption{Spectra (not corrected for the intervening Galactic absorption)
of the optical counterparts of the 2 narrow emission line AGNs belonging 
to the sample of {\it INTEGRAL} sources presented in this paper.
For each spectrum the main spectral features are labeled. The
symbol $\oplus$ indicates atmospheric telluric absorption bands.
The WHT spectrum has been smoothed using a Gaussian filter with
$\sigma$ = 3 \AA.}
\end{figure*}

\begin{figure*}[th!]
\mbox{\psfig{file=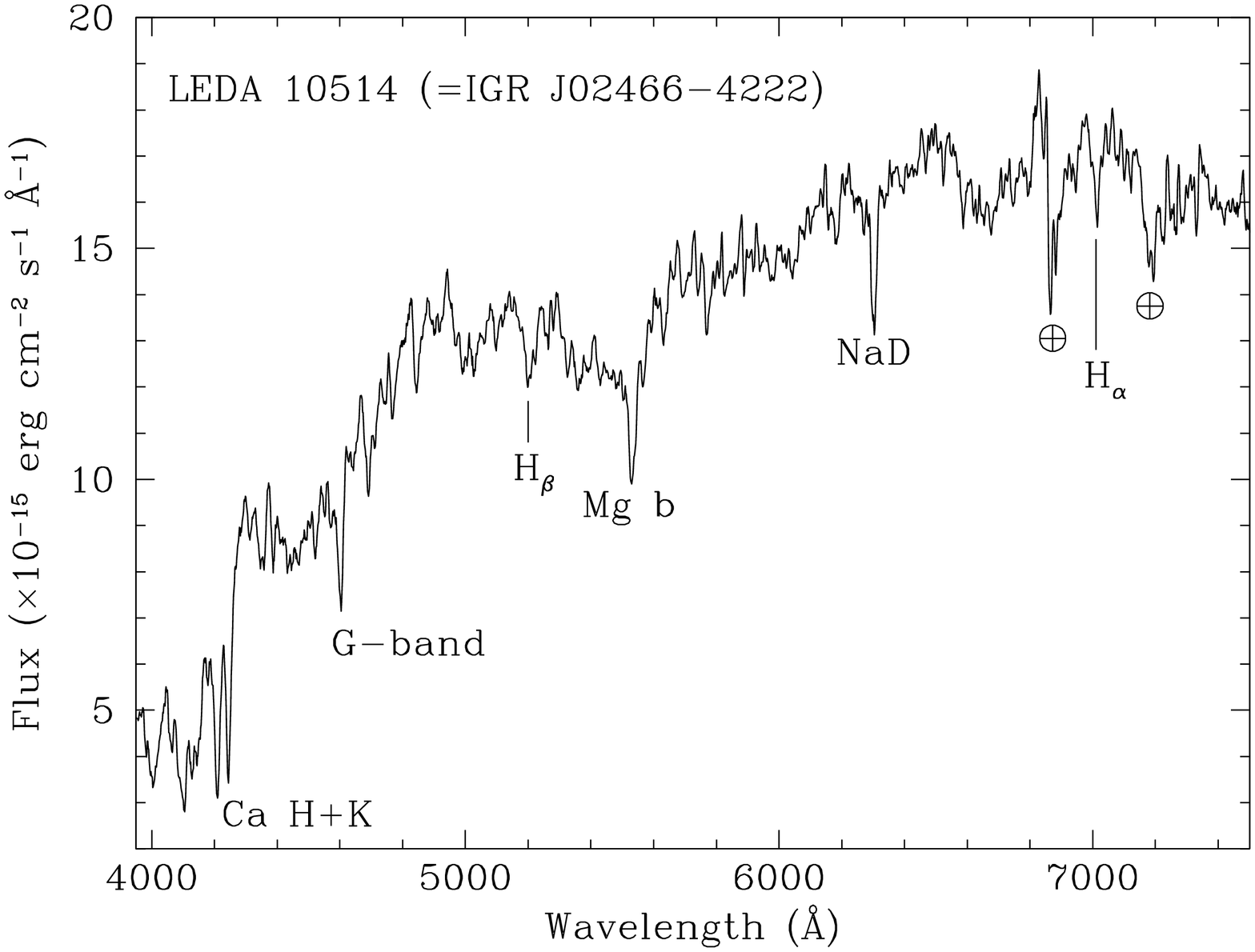,width=9cm,angle=270}}
\mbox{\psfig{file=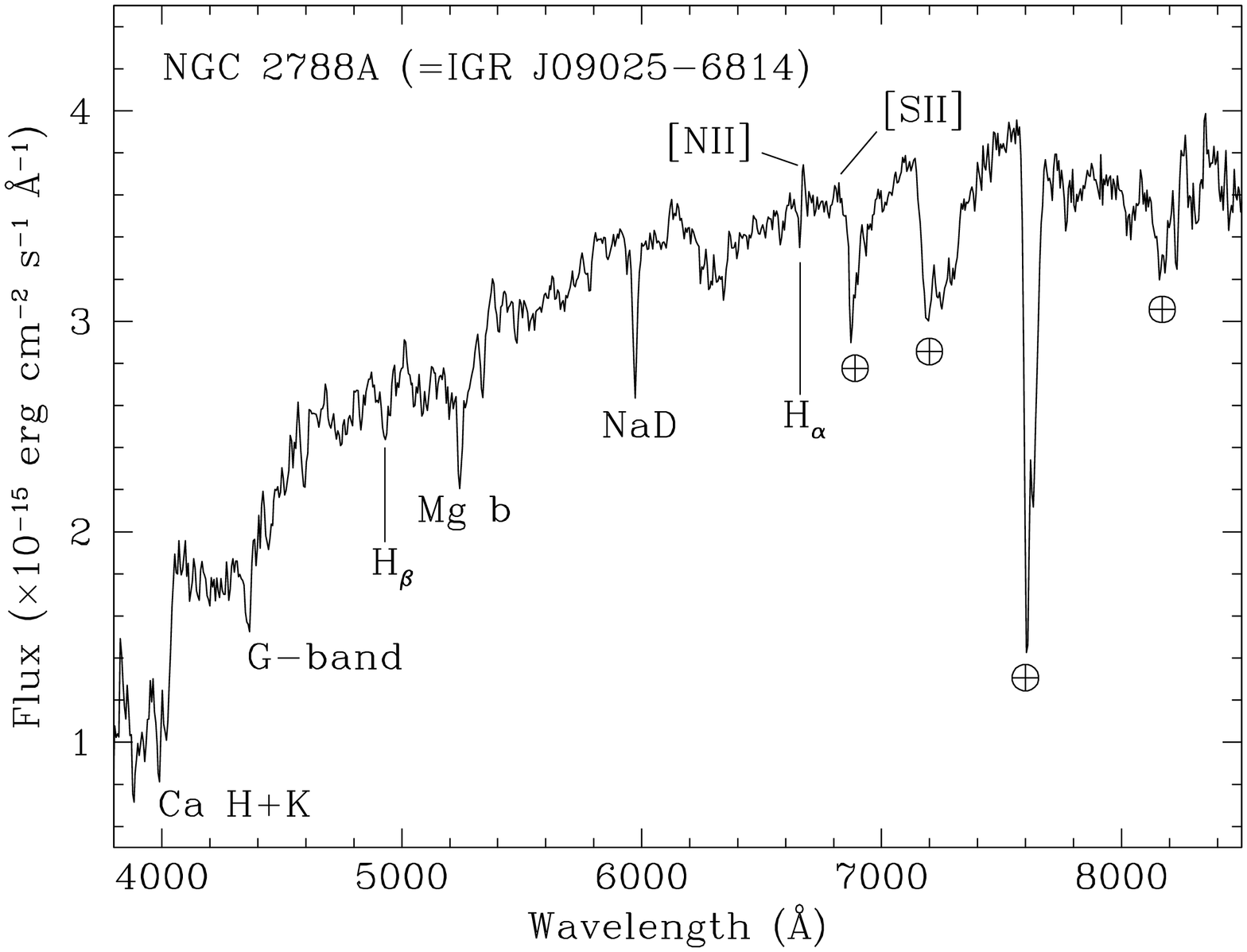,width=9cm,angle=270}}
\vspace{-.5cm}
\caption{Spectra (not corrected for the intervening Galactic absorption)
of the optical counterparts of the 2 XBONGs belonging to the sample of 
{\it INTEGRAL} sources presented in this paper.
For each spectrum the main spectral features are labeled. The
symbol $\oplus$ indicates atmospheric telluric absorption bands.}
\end{figure*}

\begin{table*}[th!]
\caption[]{Synoptic table containing the main results concerning the 11
AGNs (Figs. 5-7) identified or observed in the present sample of 
{\it INTEGRAL} sources.}
\begin{center}
\begin{tabular}{lcccccrcr}
\noalign{\smallskip}
\hline
\hline
\noalign{\smallskip}
\multicolumn{1}{c}{Object} & $F_{\rm H_\alpha}$ & $F_{\rm H_\beta}$ &
$F_{\rm [OIII]}$ & Class & $z$ &
\multicolumn{1}{c}{$D_L$ (Mpc)} & $E(B-V)_{\rm Gal}$ & \multicolumn{1}{c}{$L_{\rm X}$} \\
\noalign{\smallskip}
\hline
\noalign{\smallskip}

IGR J00333+6122 & * & 2.5$\pm$0.6 & 1.76$\pm$0.09 & Sy1.5 & 0.105 & 522.2 & 1.231 & 22 (2--10) \\
 & * & [90$\pm$20] & [58$\pm$3] & & & & & 39 (20--100) \\

 & & & & & & & & \\

Swift J0216.3+5128 & --- & --- & 0.24$\pm$0.03 & likely Sy2 & 0.422 & 2492.1 & 0.187 & 890 (0.2--12) \\
 & --- & --- & [0.36$\pm$0.05] & & & & & 970 (2--10) \\
 & & & & & & & & 1100 (20--100) \\

 & & & & & & & & \\

IGR J02466$-$4222 & in abs. & in abs. & $<$5.0 & XBONG & 0.0695 & 337.4 & 0.018 & $\sim$0.1 (0.5--8) \\
 & [in abs.] & [in abs.] & [$<$5.6] & & & & & 42 (17--60) \\

 & & & & & & & & \\

IGR J02524$-$0829 & 2.6$\pm$0.3 & in abs. & 2.4$\pm$0.2 & Sy2 & 0.0168 & 78.5 & 0.054 & 0.35 (2--10) \\
 & [2.8$\pm$0.3] & [in abs.] & [2.7$\pm$0.3] & & & & & 2.3 (17--60) \\

 & & & & & & & & \\

IGR J05270$-$6631 & --- & --- & --- & Type 1 & 0.978 & 6922.4 & 0.075 & 280 (0.1--2) \\
 & --- & --- & --- & QSO & & & & 11 (2--10) \\
 & & & & & & & & 5600 (20--40) \\
 & & & & & & & & $<$6300 (40--100) \\

 & & & & & & & & \\

IGR J09025$-$6814 & in abs. & in abs. & $<$2.6 & XBONG & 0.014 & 63.4 & 0.101 & 0.0072 (2--10) \\
 & [in abs.] & [in abs.] & [$<$3.8] & & & & & 1.2 (20--100) \\

 & & & & & & & & \\

IGR J09253+6929 & * & 23$\pm$5 & 13.2$\pm$0.7 & Sy1.5 & 0.039 & 185.2 & 0.290 & 0.25 (2--10) \\
 & * & [55$\pm$11] & [31.6$\pm$1.6] & & & & & 4.3 (20--40) \\
 & & & & & & & & $<$4.3 (40--100) \\

 & & & & & & & & \\

IGR J10147$-$6354 & * & 37$\pm$2 & 9.8$\pm$0.5 & Sy1.2 & 0.202 & 1067.2 & 0.314 & 29 (2--10) \\
 & * & [81$\pm$4] & [21.8$\pm$1.1] & & & & & $<$31 (20--40) \\
 & & & & & & & & 160 (40--100) \\

 & & & & & & & & \\

IGR J12415$-$5750 & * & 42$\pm$7 & 88$\pm$3 & Sy1.5 & 0.024 & 113.7 & 0.609 & 0.52 (0.1--2.4) \\
 & * & [300$\pm$70] & [588$\pm$18] & & & & & 2.3 (0.2--12) \\
 & & & & & & & & 3.5 (20--100) \\

 & & & & & & & & \\

IGR J16426+6536 & --- & 1.89$\pm$0.09 & 0.78$\pm$0.04 & NLSy1 & 0.323 & 1821.3 & 0.021 & 44 (0.2--12) \\
 & --- & [2.0$\pm$0.1] & [0.81$\pm$0.04] & & & & & 910 (20--40) \\
 & & & & & & & & $<$790 (40--100) \\

 & & & & & & & & \\

IGR J18249$-$3243 & * & $<$0.9 & 9.6$\pm$0.9 & Sy1 & 0.355 & 2033.2 & 0.233 & 38 (0.1--2.4) \\
 & * & [$<$1.5] & [14.5$\pm$1.5] & & & & & 260 (2--10) \\
 & & & & & & & & 390 (20--100) \\

\noalign{\smallskip} 
\hline
\noalign{\smallskip} 
\multicolumn{9}{l}{Note: emission line fluxes are reported both as 
observed and (between square brackets) corrected for the intervening Galactic} \\ 
\multicolumn{9}{l}{absorption $E(B-V)_{\rm Gal}$ along the object line of sight 
(from Schlegel et al. 1998). Line fluxes are in units of 10$^{-15}$ erg cm$^{-2}$ s$^{-1}$,} \\
\multicolumn{9}{l}{whereas X--ray luminosities are in units of 10$^{43}$ erg s$^{-1}$ 
and the reference band (between round brackets) is expressed in keV.} \\ 
\multicolumn{9}{l}{The typical error on the redshift measurement is $\pm$0.001 
but for the SDSS and 6dFGS spectra, for which an uncertainty} \\
\multicolumn{9}{l}{of $\pm$0.0003 can be assumed.} \\
\multicolumn{9}{l}{$^*$: heavily blended with [N {\sc ii}] lines} \\
\noalign{\smallskip} 
\hline
\hline
\end{tabular} 
\end{center} 
\end{table*}

It is found that 11 objects of our sample show optical spectra that allow 
us to classify them as AGNs (see Table 4). Nine of these objects present 
strong, redshifted broad and/or narrow emission lines typical of nuclear 
galactic activity: seven of them can be classified as Type 1 (broad-line) 
and two as Type 2 (narrow-line) AGNs. Among Type 1 AGNs, we find three 
Seyfert 1.5 galaxies, one Sefert 1.2 galaxy, one narrow-line (NL) Seyfert 
1 galaxy and one high-redshift Type 1 QSO; for the case of IGR 
J18249$-$3243, only a general Seyfert 1 classification can be given due to 
the contamination of its spectrum by an interloping Galactic star 
(see Fig. 5, lower left panel): this is similar to other cases that we 
previously found (Paper I and V). We moreover revise here to Seyfert 1.5 
the AGN class of galaxy LEDA 166252, the optical counterpart of IGR 
J12415$-$5750. We can therefore exclude the Seyfert 2 classification 
reported in the literature and in on-line catalogues such as 
SIMBAD\footnote{available at {\tt
http://simbad.u-strasbg.fr}} and NED\footnote{available at {\tt 
http://nedwww.ipac.caltech.edu}}.
This classification also confirms the X--ray spectral description given by 
Malizia et al. (2007) and Winter et al. (2008) for this source and deemed 
too peculiar for a Seyfert 2 AGN according to these authors.

For the remaining two objects, the nuclear activity was detected only 
thanks to soft X--ray emission detected with {\it Chandra} or {\it Swift}, 
given that no or unremarkable emission lines were apparent in their 
optical spectra: we therefore classify them as X--ray bright, optically 
normal galaxies (XBONGs; see Comastri et al. 2002).

The main observed and inferred parameters for each of these objects are 
reported in Table 4. In this table, X--ray luminosities were computed from 
the fluxes reported in Voges et al. (1999), {\it ROSAT} Team (2000), Bird 
et al. (2007), Krivonos et al. (2007), Landi et al. (2007a,b, 2008b,d), 
Ibarra et al. (2008a), Malizia et al. (2007), Rodriguez et al. (2008), 
Sazonov et al. (2008) and Watson et al. (2008). We also used the 2--10 
keV fluxes of IGR J02524$-$0829 and IGR J09025$-$6814 which are 
4.8$\times$10$^{-12}$ and 1.5$\times$10$^{-13}$ erg cm$^{-2}$ s$^{-1}$,
repsectively, according to Landi et al. (in preparation).

For 7 out of 11 reported AGNs, the redshift value was determined in this 
work for the first time. The redshifts of the remaining 4 sources are 
consistent with those reported in the literature, e.g., in the Hyperleda 
archive (Prugniel 2005).

Going into detail for selected sources, we see that, according to Jonker 
\& Kuiper (2007), two soft X--ray emitters were found with {\it Chandra} 
within the IBIS error circle of IGR J00333+6122; however (see also Parisi 
et al. 2008a), we discard source 2 of Jonker \& Kuiper (2007) as the 
possible counterpart of this {\it INTEGRAL} source given its relative 
faintness, its softness and its association with a Galactic star showing 
no peculiarities in its optical spectrum.

We notice here that the optical position of Swift J0216.3+5128
is consistent with the soft X--ray localization of Malizia et al. (2007) 
but not with that of Winter et al. (2008), both obtained with the same 
{\it Swift}/XRT observations. We checked these XRT pointings again and we
found only one source within the {\it INTEGRAL} error circle of Swift 
J0216.3+5128, again at the coordinates given by Malizia et al. (2007).
Moreover, given that the {\it XMM-Newton} position of Watson et al. (2008) 
is fully consistent with that reported by Malizia et al. (2007), we are
confident that this is indeed the true localization of this source, and 
that we identified its correct optical counterpart.

Looking at the optical spectrum of the counterpart of Swift J0216.3+5128 
we see that, at the redshift of this source, the H$_\beta$ emission falls 
right in the O$_2$ telluric band at 6870 \AA; so, no reliable measurements 
of this line are possible. Nevertheless, the overall optical spectral 
appearance and the lack of detection of wide H$_\beta$ emission wings 
allow us to classify this source as a likely Seyfert 2 AGN.

Here we confirm the identification of IGR J02524$-$0829 as a Seyfert 2 AGN 
hosted in the galaxy LEDA 10875 (Bikmaev et al. 2008b). This is done by 
inspecting its optical spectrum (Fig. 6, right panel) and supported by 
the finding of a single {\it Swift}/XRT soft X--ray source, positionally 
consistent with the nucleus of this galaxy, within the IBIS error box of 
this {\it INTEGRAL} object.

The sources IGR J02466$-$4222 (see also Sazonov et al. 2008) and IGR 
J09025$-$6814 could be associated with galaxies LEDA 10514 (=MCG 
$-$07$-$06$-$018) and NGC 2788A only thanks to the {\it Chandra} and {\it 
Swift} positions of their soft X--ray counterparts. Indeed, as it can be 
seen in Fig. 7, both galaxies show no peculiarities in their optical 
spectra: following the approach of Laurent-Muehleisen et al. (1998), we 
can classify both of them as `normal' galaxies. Thus, as noted above, we 
can state that these hard X--ray sources are XBONGs similar to NGC 4992 
(=IGR J13091+1137; Paper IV). The association of NGC 2788A with a 
far-infrared IRAS source (IRAS 1988) further suggests this classification.

As already mentioned, we classify IGR J16426+6536 as an NL Seyfert 1 AGN, 
because its optical spectrum complies with the criteria of Osterbrock \& 
Pogge (1985) concerning the presence of the Fe {\sc ii} bump and the 
narrowness of the Full Width at Half Maximum (FWHM) of the H$_\beta$ 
emission ($\sim$2000 km s$^{-1}$).

It is noted that the optical counterpart of source IGR J18249$-$3243 lies 
$\sim$15 arcsec outside the nominal {\it ROSAT} error circle. However, 
the {\it Swift}/XRT pointing (Landi et al. 2008b) shows that its actual 
soft X--ray position is definitely consistent with the proposed optical 
counterpart.

Unfortunately, due to the impossibility of a simultaneous detection of 
H$_\alpha$ and H$_\beta$ emission lines in the Seyfert 2 galaxies and the 
XBONGs of our sample, we are not able to provide an estimate of the 
absorption local to their AGN and, in turn, a reliable assessment of the 
Compton nature of these galactic nuclei with the method of Bassani et al. 
(1999). Thus, to explore this issue we can apply the soft-to-hard X--ray 
flux ratio diagnostic of Malizia et al. (2007). We find that this 
diagnostic has values 0.89, 0.0024, 0.15 and 0.006 for Swift J0216.3+5128, 
IGR J02466$-$4222, IGR J02524$-$0829 and IGR J09025$-$6814, respectively. 
As a cautionary note, we remark that Malizia et al. (2007) used X--ray 
flux measurements in the 2--10 keV and 20--100 keV bands to compute the 
aforementioned ratio, while only the (0.5--8 keV)/(17--60 keV) and the 
(2--10 keV)/(17--60 keV) flux ratios are avaliable in the literature for 
IGR J02466$-$4222 and IGR J02524$-$0829, respectively (see Table 4).

Nevertheless, if one compares these figures with those of the sample of 
Malizia et al. (2007, their Fig. 5), one sees that the two XBONGs of our 
sample fall in the low-values tail of this flux ratio distribution. This, 
according to Malizia et al. (2007), strongly suggests that these two 
sources are Compton thick AGNs and confirms the indication of Sazonov et 
al. (2008) concerning the nature of IGR J02466$-$4222. The two Seyfert 2 
galaxies of our sample, instead, are very likely Compton thin AGNs.

Following Wu et al. (2004) and Kaspi et al. (2000), and McLure \& Jarvis 
(2002) for the case of IGR J05270$-$6631, we can eventually provide an 
estimate of the mass of the central black hole in 6 of the 7 objects 
classified as Type 1 AGN (this procedure could not be applied to IGR 
J18249$-$3243 as no broad H$_\beta$ emission component was detected for 
it). The corresponding black hole masses for these 6 cases are reported in 
Table 5. Here we assumed a null local absorption for all Type 1 AGNs.

\begin{table}
\caption[]{BLR gas velocities (in km s$^{-1}$) and 
central black hole masses (in units of 10$^7$ $M_\odot$) for 6
Seyfert 1 AGNs belonging to the sample presented in this paper.}
\begin{center}
\begin{tabular}{lcc}
\noalign{\smallskip}
\hline
\hline
\noalign{\smallskip}
\multicolumn{1}{c}{Object} & $v_{\rm BLR}$ & $M_{\rm BH}$ \\
\noalign{\smallskip}
\hline
\noalign{\smallskip}

IGR J00333+6122   & 8100 & 35 \\
IGR J05270$-$6631 & 3900 & 24 \\
IGR J09253+6929   & 4900 & 4.1 \\
IGR J10147$-$6354 & 4800 & 42  \\
IGR J12415$-$5750 & 6700 & 9.2 \\
IGR J16426+6536   & 1900 & 1.1 \\

\noalign{\smallskip}
\hline
\hline
\noalign{\smallskip}
\end{tabular}
\end{center}
\end{table}

\subsection{Statistical considerations}

\begin{figure}[h!]
\hspace{-0.5cm}
\psfig{file=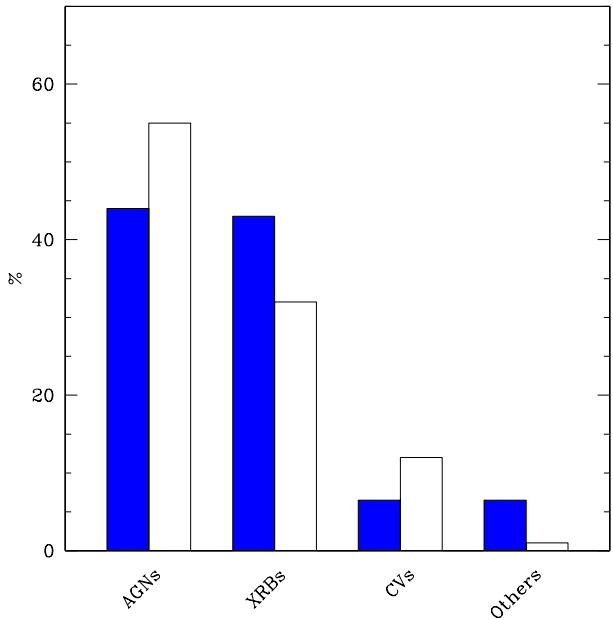,width=9.5cm}
\vspace{-0.7cm}
\caption{Histogram, subdivided into source types, showing the percentage 
of {\it INTEGRAL} objects of known nature in the compilation of Bodaghee 
et al. (2007; left-side, darker columns), and {\it INTEGRAL} sources 
from various surveys and identified through optical or NIR spectroscopy 
(right-side, lighter columns).}
\end{figure}

As is customary in this series of papers, we now update the broad 
statistical approach used in Papers V and VI by including the results 
presented here, along with recent spectroscopic optical and NIR 
identifications of {\it INTEGRAL} sources as HMXBs (Chaty et al. 2008, 
but see Nespoli et al. 2008b; Nespoli et al. 2008a,c), LMXBs (Torres et 
al. 2006; Cadolle Bel et al. 2007; Nespoli et al. 2008b; Torres et al. 
2008) and AGNs (Bikmaev et al. 2008a; Burenin et al. 2008; Gon\c{c}alves 
et al. 2008).

It is found that, of the 129 {\it INTEGRAL} sources identified up to now 
through optical or NIR spectroscopy, 71 (55\%) are AGNs, with the 
following breakdown: 33 (i.e, 45\% of the AGN identifications) are Seyfert 
1 galaxies, 29 (41\%) are Seyfert 2 galaxies, while the QSO, XBONG and BL 
Lac subclasses are populated by 3 objects (4\%) each.
There are 41 X--ray binaries, that is, 32\% of all optical/NIR identifications,
with a large majority, i.e. about 80\%, of HMXBs; 16 objects (12\%) 
are CVs, with 12 of them classified as definitely or likely IPs (see Papers 
IV-V and the present work) and 4 of symbiotic star type.

A comparison with the numbers and percentages of spectroscopically 
identified {\it INTEGRAL} sources in the past years (Papers V-VI) 
illustrates an evolution with time of the population of the classes 
described above. First, one can see that the number of these 
identifications increased by a factor of 2.4 in two years; this indicates
the interest of the community in this work of systematic identification of 
hard X--ray sources, and stresses the utility of soft X--ray satellites
(especially {\it Swift} and {\it Chandra}) for the task of refining the
high-energy error box of these sources. Indeed, our experience tells 
us that follow-up soft X--ray observations of the IBIS error circle of 
unidentified {\it INTEGRAL} objects clearly boosted our identification 
record (Papers V-VI and this work).

Second, in terms of percentage population of the aforementioned classes, 
one can see that AGNs always constituted the largest class, increasing up 
to containing now the absolute majority of new identifications. They are 
followed by the X--ray binaries group, which on the contrary became 
smaller: now it contains one third of all identifications; among these, the 
fraction of HMXBs still by far dominates over LMXBs, although with a 
reduction over time. A fair share of objects is made of (magnetic) 
CVs, which are constantly the third largest group of {\it INTEGRAL} 
sources identified through optical spectroscopy.

Third, while the percentage population of the three classes shows some 
differences when one makes a comparison between Paper V and Paper VI, it 
basically did not vary between Paper VI and the present work. This may 
suggest that the relative sizes of these three largest groups of sources,
when identified through optical and NIR spectroscopy, reached a `stability 
point'.

One can moreover compare the latest percentages (see Fig. 8) with those of 
the 370 identified objects belonging to the collection of {\it INTEGRAL} 
sources (Bodaghee et al. 2007) detected between 2002 and 2006. In 
this catalogue there are 163 (44\%) AGNs, 160 (43\%) X--ray binaries (of 
which 82 are LMXBs and 78 are HMXBs, i.e., 51\% and 49\% of this class), 
and 23 (6.5\%) CVs, nearly all of them being of magnetic nature 
(IPs or Polars; see Barlow et al. 2006 and Landi et al. 2008a).

This confirms once more the effectiveness of the method of catalogue 
cross-correlation plus optical and/or NIR spectroscopy follow-up in 
revealing the nature of unidentified {\it INTEGRAL} sources. The above is 
true especially for AGNs and CVs, even if in Paper VI we suggested that 
optical spectroscopy searches may introduce a bias against the detection 
of (heavily absorbed) Galactic X--ray binaries. Indeed, by looking at the 
results of, e.g., Hannikainen et al. (2007), Chaty et al. (2008) and 
Nespoli et al. (2008a,c), all of the {\it INTEGRAL} objects 
spectroscopically identified in the NIR bands by these authors are HMXBs.

We conclude by stressing that, of the 129 optical and NIR spectroscopic 
identifications mentioned in this subsection, 104 were obtained or 
refined within the framework of our spectroscopic follow-up program 
(Papers I-VI, the present work, and references therein).

\section{Conclusions}

Within our continuing identification program of {\it INTEGRAL} sources by 
means of optical spectroscopy (Papers I-VI) running at various telescopes 
around the world, in this paper we have identified and studied 20 hard 
X--ray objects of unknown or poorly explored nature. This was performed
with the use of 6 different telescopes and using the archival data of 2 
spectroscopic surveys.

We found that the selected sample is made of 11 AGNs (6 of which are 
Seyfert 1 galaxies, 2 are Seyfert 2 galaxies, 2 are XBONGs and one is a 
Type 1 QSO), 4 HMXBs (all of them likely belonging to the Be/X subclass), 
4 CVs, one LMXB and one symbiotic star. Again in the present sample, as in 
the case of our past papers within the framework of this research, we note 
that the absolute majority of identified sources belongs to the AGN 
class, and we see a non-negligible presence of (possibly magnetic) CVs.

The results presented in this work once again demonstrate the high 
effectiveness of the method of catalogue cross-correlation and/or 
follow-up observations with soft X--ray satellites affording arcsec-sized 
localizations (such as {\it Chandra}, {\it XMM-Newton} or {\it Swift}) 
plus optical (and NIR) spectroscopy to pinpoint the actual nature of the 
still unidentified {\it INTEGRAL} sources. Further insights can be given 
by deep optical or NIR Galactic surveys such as the VISTA Variables in the 
Via Lactea public NIR survey (VVV; Minniti et al. 2006, 2008).

\begin{acknowledgements}

We thank Silvia Galleti for Service Mode observations at the Loiano 
telescope; Pablo Rodr\'{i}guez-Gil and Andrew Cardwell for Service Mode 
observations at the WHT; Francesca Ghinassi for Service Mode observations 
at the TNG; Antonio De Blasi and Ivan Bruni for night assistance at the 
Loiano telescope; Edgardo Cosgrove, Manuel Hern\'andez and Jos\'e 
Vel\'asquez for day and night assistance at the CTIO telescope; 
Alessandro Ederoclite for support at the ESO 3.6m telescope; 
Ariel S\'anchez for night assistance at the ESO 3.6m telescope; Mauro 
Orlandini for comments and suggestions. We also thank the anonymous 
referee for useful remarks which helped us to improve the quality of this 
paper. 
This research has made use of the ASI Science Data Center Multimission 
Archive; it also used the NASA Astrophysics Data System Abstract Service, 
the NASA/IPAC Extragalactic Database (NED), and the NASA/IPAC Infrared 
Science Archive, which are operated by the Jet Propulsion Laboratory, 
California Institute of Technology, under contract with the National 
Aeronautics and Space Administration. 
This publication made use of data products from the Two Micron All 
Sky Survey (2MASS), which is a joint project of the University of 
Massachusetts and the Infrared Processing and Analysis Center/California 
Institute of Technology, funded by the National Aeronautics and Space 
Administration and the National Science Foundation.
This research has also made use of data extracted from the 6dF 
Galaxy Survey and the Sloan Digitized Sky Survey archives;
it has also made use of the ESO Science Archive operated at Garching bei 
M\"unchen, Germany, of the SIMBAD database operated at CDS, Strasbourg, 
France, and of the HyperLeda catalogue operated at the Observatoire de 
Lyon, France.
The authors acknowledge the ASI and INAF financial support via grant 
No. I/023/05/0.
PP is supported by the ASI-INTEGRAL grant No. I/008/07.
LM is supported by the University of Padua through grant No. 
CPDR061795/06. 
VC is supported by the CONACYT research grant 54480-F (M\'exico).
DM is supported by the Basal CATA PFB 06/09, and FONDAP Center for 
Astrophysics grant No. 15010003.
NM and LM thank the Pontificia Universidad Cat\'olica de Chile for the 
pleasant hospitality in Santiago de Chile during the preparation of this 
paper.
\end{acknowledgements}

\end{document}